\documentclass[review]{elsarticle}

\usepackage{lineno,hyperref}

\journal{Journal of \LaTeX\ Templates}
\usepackage{amsmath, chemarr, url, tabularx, units}
\usepackage{indentfirst}
\usepackage{graphicx, subfigure}
\usepackage{epsfig}
\usepackage{lineno}
\usepackage{amssymb}
\usepackage{tikz}
\usepackage{setspace}
\usetikzlibrary{arrows,positioning} 

\tikzstyle{process} = [rectangle, minimum width=3cm, minimum height=2.5cm, text 
centered, text width=3cm, draw=black, fill=green!30]
\tikzstyle{region} = [circle, minimum width=3.0cm, minimum height=2.5cm, 
text centered, draw=black, fill=white!30]
\tikzstyle{arrow} = [thick,->,>=stealth]

\newif\ifSubmittal

\ifSubmittal
\graphicspath{
               {.}
             }
\else
\graphicspath{
               {./figs/}
             }
\fi

\usepackage[final]{changes}
\definechangesauthor[name={Bernhard Peters}, color={blue}]{BP}
\definechangesauthor[name={Gabriele Pozzetti}, color={red}]{GP}
\definechangesauthor[name={Maryam Baniasadi}, color={green}]{MB1}
\definechangesauthor[name={Mehdi Baniasadi}, color={orange}]{MB2}
\definechangesauthor[name={Mohammad Mohseni}, color={brown}]{MM}
\definechangesauthor[name={Alvaro Estupinan}, color={purple}]{AE}

\begin{document}

\begin{frontmatter}

\title{The XDEM Multi-physics and Multi-scale Simulation Technology: Review on DEM-CFD Coupling, Methodology 
and Engineering Applications}

\author{Bernhard Peters, Maryam Baniasadi, Mehdi Baniasadi, \\Xavier Besseron, Alvaro Estupinan \added[id=AE]{Donoso}, \\Mohammad Mohseni, Gabriele Pozzetti }
\address{Universit\'e du Luxembourg, 6, rue Coudenhove-Kalergi, L-1359 Luxembourg}
\fntext[myfootnote]{Corrsponding author, bernhard.peters@uni.lu}

\begin{abstract}
The XDEM multi-physics and multi-scale simulation platform roots in the 
Extended Discrete Element Method (XDEM) and
is being developed at the Institute of Computational Engineering at the University of Luxembourg. The platform is an advanced multi-
physics simulation technology that combines flexibility and versatility to
establish the next generation of multi-physics and multi-scale simulation
tools. For this purpose the simulation framework relies on coupling various
predictive tools based on both an Eulerian and Lagrangian approach. Eulerian
approaches represent the wide field of continuum models while the Lagrange
approach is perfectly suited to characterise discrete phases. Thus, continuum
models include classical simulation tools such as Computational Fluid Dynamics 
(CFD) or Finite Element Analysis (FEA) while an extended configuration of
the classical Discrete Element Method (DEM) addresses the discrete 
e.g.~particulate phase. Apart from predicting the trajectories of individual
particles, XDEM extends the application to estimating the thermodynamic state 
of each particle by advanced and optimised algorithms. The thermodynamic state
may include temperature and species distributions due to chemical reaction and 
external heat sources. Hence, coupling these extended features with either CFD
or FEA opens up a wide range of applications as diverse as pharmaceutical
industry e.g. drug production, agriculture food and processing industry, mining, 
construction and agricultural machinery, metals manufacturing, energy production 
and systems biology.\end{abstract}


\end{frontmatter}


\author{Bernhard Peters, Maryam Baniasadi, Mehdi Baniasadi, \\ Alvaro Estupinan, Mohammad Mohseni, Gabriele Pozzetti}

\section{Introduction}
This section reviews major approaches in particular focussing on the combined 
approach of the Euler and Lagrange concept in a multi-phase flow context. For 
classical stand-alone concepts such as Computational Fluid Dynamics (CFD) or 
Finite Element Analysis (FEA) the reader is refereed to standard text books 
\cite{blevins84,Peric96,Bathe96,Zienkiewicz84}.

Numerical approaches to model multi-phase flow  
phenomena including a solid e.g. particulate phase may basically be 
classified into two categories: All phases are treated as a continuum 
on a macroscopic level of which the two fluid model is the most well-known 
representative \cite{multiphase.Gidaspow94}. It is well suited to process 
modelling due to its 
computational convenience and efficiency. However, all the data concerning
size distribution, shape or material properties of individual particles is 
lost to a large extent due to the averaging concept. Therefore, this loss of 
information on small scales has to be compensated for by additional constitutive 
or closure relations. Packed beds are frequently employed in engineering applications such as for
thermal conversion of solid fuels e.g.~biomass \cite{Peters03c, comb.Spliethoff10} 
or blast furnaces for metallurgical processes \cite{multiphase.Omori87}. 
Common to these engineering applications is a complex interaction of processes 
consisting of flow of a solid, liquid and gas phase with heat, mass and momentum  transfer 
between them and including chemical conversion. 

In the following sections numerical approaches to represent multiphase flow as
gas-liquid, gas-powder and solid-fluid configurations are reviewed. It is 
followed by the technological concept of the XDEM suite and concludes with 
relevant and validated applications.

\subsection{Gas-Liquid Multi-phase Flow}

A potential flow model to describe the gas-liquid flow were employed by
Szekely et al.~\cite{multiphase.Szekely79}, Sugiyama et 
al.~\cite{multiphase.Sugiyama87b} and Austin \cite{multiphase.Austin97b}.
Another category of models is based on probability and statistical methods
as employed by Ohno et al.~\cite{multiphase.Ohno88}, Wang et 
al.~\cite{multiphase.Wang97a, multiphase.Wang97b}
and Liu et al.~\cite{multiphase.Liu02}. The velocity of the liquid phase is
determined stochastically in conjunction with a distribution function.
Eto et 
al.~\cite{multiphase.Eto93} employed a tube-network dynamic model for
simulation of the liquid flow in the dripping zone of a blast furnace.

Wang et al.~\cite{multiphase.Wang91} coupled continuum and probability 
approaches to describe the gas-liquid flow in  packed beds.  
Within a continuum approach based on Darcy's equation the 
gas-liquid velocity is predicted by solving the coupled continuity and 
momentum equations as carried out by Takahashi et 
al.~\cite{multiphase.Takahashi89}, Takahashi et al.~\cite{multiphase.Takahashi93}, 
Kuwabara et al.~\cite{multiphase.Kuwabara91}, Chen et 
al.~\cite{multiphase.Chen93}, Takatani et al.~\cite{multiphase.Takatani94}, 
Szekely et al.~\cite{multiphase.Szekely79}, 
Kajiwara et al.~\cite{multiphase.Kajiwara79}, Yagi et al.~\cite{multiphase.Yagi93}, Castro et al.~\cite{multiphase.Castro00},
Sugiyama et al.~\cite{multiphase.Sugiyama87a}, Zhang et al.~\cite{multiphase.Zhang98}, Austin et 
al.~\cite{multiphase.Austin98}, Zaimi et al.~\cite{multiphase.Zaimi00}, Castro et al.~\cite{multiphase.Castro02a, multiphase.Castro02b}, Eto et al.~\cite{multiphase.Eto93} and Chu et al.~\cite{multiphase.Chu02}. Due to the 
continuum approach detailed information on the structure of the packed bed 
is lost, that could be recovered by a fine grid taking into account the 
morphology such as porosity and was undertaken by Li et al.~\cite{multiphase.Li01}
and Wen et al.~\cite{multiphase.Wen01}. However, resolving the structure 
requires high CPU times, which was addressed by Wang et 
al.~\cite{multiphase.Wang08} and Wang et 
al.~\cite{multiphase.Wang97a}. Wang et 
al.~\cite{multiphase.Wang08, multiphase.Wang09} developed a model for which the numerical 
grid must not correspond to the structure of the packed bed e.g.~size of the
particles. Motion of the liquid phase is described by conservation 
equations for mass and momentum in two dimensions. The model was validated 
by data of Sugiyama \cite{multiphase.Sugiyama87a} and yielded reasonable
good agreement.

Based on their recent model developments \cite{multiphase.Zhang98},
Zhang et al.~\cite{multiphase.Zhang02} simulated the flow of solids in a blast
furnace because the layered burden and consumption of solids effects 
the flow of gases in a blast furnaces as pointed out by Takahashi et 
al.~\cite{multiphase.Takahashi96}. Increasing the consumption rate of the solids 
reduces the size of the stagnant zone.

Das et al.~\cite{multiphase.Das2011} applied a continuous model to predict 
silicon transport in the dripping zone of a blast furnace. They distinguished
two mechanisms for hold-up for the liquid in the dripping zone through a 
multi-physics coupling of the heat transfer, liquid flow, species transport 
and chemical kinetics processes: Hold-up in the absence of gas flow and in the 
presence of counter current gas flow.  Owing to the complexity of liquid hold-
up, they applied a semi-empirical approach \cite{multiphase.Tsuchiya76, 
multiphase.Wang97c, multiphase.Austin98b} including both static
and dynamic components. They concluded that the counter current gas flow 
has no significant effect on the liquid hold-up under a stable blast furnace 
operating conditions.

Jin et al.~\cite{multiphase.Jin10a, multiphase.Jin10b} 
described dripping of liquid metal through a packed bed taking into 
account counter-current flow and reactions between  gas, solid and
liquid phases. The latter were described by differential conservation equations
for mass, momentum and energy and solved with COMSOL Multiphysics 3.4.
The predicted results include both dynamic and static hold-up of liquid and showed the 
dependence of the temperature distribution due to endothermic reduction of 
silica.  

Danloy et al.~\cite{multiphase.Danloy2011} developed a modular model to describe
the performance of a blast furnace under steady state conditions. The modules
incorporated represent a burden distribution according to Hamilius 
\cite{multiphase.Hamilius78}, flow of gases, liquids and solids including heat transfer
between the phases and a cohesive zone sub-model. The flow of gases, liquids and 
solids is based on differential conservation equations. The cohesive zone 
sub-model starts at a solid temperature of 1200 $^\circ$C and ends where ore 
is liquid accompanied by a set of five reactions for reduction of ore. After 
validation with vertical probe measurements of the blast furnace B at Sidmar,
they predicted the influence of operating parameters such as burden distribution 
on gas distribution, pressure drop, productivity, shape and position of the
cohesive zone and top gas temperature profile. Nogami et 
al.~\cite{multiphase.Nogami06} undertook a numerical analysis of the blast
furnace performance with novel feed material as carbon composite agglomerates
(CCB). Their multi-fluid approach included reduction kinetics for ore
\cite{multiphase.Takahashi84, multiphase.Omori87, multiphase.Fun70, general.Levenspiel76}  and silicon \cite{multiphase.Turkdogan80, multiphase.Inoue87, multiphase.Ozturk86}. Additional charging of carbon composite agglomerates lowered melting and hot metal
temperatures, that contributes to an improved efficiency of the blast furnace. 

Yu et al.~\cite{multiphase.Yu04} also modelled the gas-liquid 
flow by a continuous approach in conjunction with a 
force balance approach and stochastic treatment of the packed bed structure.
Good agreement between measurements and predictions allowed studying localised
liquid flow and the effect of the cohesive zone on the gas flow. 
 
However, experimental observations \cite{multiphase.Gupta94, multiphase.Gupta96, 
multiphase.Gupta97, multiphase.Liu02, multiphase.Mackey73, 
multiphase.Standish68, multiphase.Wang97a, multiphase.Wang97b, 
multiphase.Wang97c, multiphase.Warner59, multiphase.Yagi93, multiphase.Liu98} 
show that an above-mentioned pure 
continuous approach is inaccurate to represent the flow 
of discrete rivulets or droplets \cite{granular.Bridgewater94} for the configurations 
addressed. Therefore, a 
discrete approach seems to be more appropriate that describes the trajectories 
of individual liquid droplets or solid particles 
under the influence of gas drag forces, resistance from the solid phase possibly 
including its properties and gravity acting on discrete particles. Ohno and 
Schneider \cite{multiphase.Ohno88}
were the first to represent liquid flow as spherical droplets moving through
the porous space of a packed bed. Gupta et al.~\cite{multiphase.Gupta96, 
multiphase.Gupta97} applied
a balance of all major forces acting on a liquid mass to model liquid flow
in a packed bed and has been used by Liu et al.~\cite{multiphase.Liu02} and Wang 
et al.~\cite{multiphase.Wang97a, multiphase.Wang97b, multiphase.Wang97c, 
multiphase.Wang00}. Xu et al.~\cite{multiphase.Xu00} and Chew et 
al.~\cite{multiphase.Chew01a, multiphase.Chew01b, multiphase.Chew01c} also 
investigated into the discrete behaviour of the fluid phase.
Singh and Gupta \cite{multiphase.Singh06} investigated into the non-wetting 
liquid flow from a point source in a packed by a discrete model.

These models lack dispersion of the fluid phase, 
that is compensated by the statistical approaches of by Wang et 
al.~\cite{multiphase.Wang97a, multiphase.Wang97b} and Liu et 
al.~\cite{multiphase.Liu02} to some extent. Chew et al.~\cite{cohesivezone.Chew97}
reported from their experimental study that a flow pattern including horizontal
and even upward flow
exists. 

Within a discrete approach both size and shape of liquid masses have to be 
determined for which various studies have been carried out 
\cite{multiphase.Aussillous04, multiphase.Dimitrakopoulos97, multiphase.Durbin88,
multiphase.Dussan87, multiphase.Dussan74, multiphase.King93, 
multiphase.Wilson88}. However, due to the complexity to determine the shape 
Singh and Gupta \cite{multiphase.Singh06} assumed a spherical 
shape for sufficiently large pore spaces and a rivulet otherwise. While the gas 
flow was described as a turbulent flow with the $k-\epsilon$-model, a global 
force balance determined the flow of droplets. Their approach was validated by 
measurements in a uniformly packed bed and agreement was reported to be 
reasonable good.

\subsection{Gas-Solid Multi-phase Flow}  

Gas-Solid multi-phase flow is distinguished by the size of the particulate flow.
For very small sizes of the particles it is classified as powder while larger 
particle dimensions lead to the domain of packed, bubbling or fluidised beds.
Hence, the following sections review individually these two multi-phase flow 
regimes. 

\subsubsection{Gas-Powder Flow}  

Fines may be entrained by ascending gas (dynamic holdup) whereas remaining 
fines are caught in the packed bed (static holdup). Hence, the fines 
effect the gas flow to a large extent as pointed out by
Shibata et al.~\cite{multiphase.Shibata91} and Chen et 
al.~\cite{multiphase.Chen94}. Shibata et al.~\cite{multiphase.Shibata91} 
observed that static powder hold-up as a blockage takes place for gas velocities
lower than a critical value that corresponds well with the transport velocity
of one-dimensional models. Their model took into account convective transport, 
however, neglected viscous stress and body forces and was employed by 
Austin et al.~\cite{multiphase.Austin97a, multiphase.Austin97c}, Castro et 
al.~\cite{multiphase.Castro02} and Nogami et al.~\cite{multiphase.Nogami04}.
Chen et al.~\cite{multiphase.Chen94} extended the model of Shibata et al.~\cite{multiphase.Shibata91} by static and dynamic powder hold-up. Pintowantoro
et al.~\cite{multiphase.Pintowantoro04} investigated the static hold-up as 
proposed by Hikado et al.~\cite{multiphase.Hidaka00}. 
These findings were 
enhanced by modelling the effect of impermeable obstacles in the flow and powder accumulation in a packed bed
by Dong et al.~\cite{multiphase.Dong04a}. The strong influence of permeability on 
the gas-liquid flow was also emphasised by 
Horio \cite{multiphase.Horio92}. They employed the continuous 
two-fluid approach that has been successfully applied by Jackson 
\cite{multiphase.Jackson63},  Soo \cite{multiphase.Soo90}, Garg et 
al.~\cite{multiphase.Garg75}, Gidaspow et al.~\cite{multiphase.Gidaspow83}, 
Kuipers et al.~\cite{multiphase.Kuipers91} and Levy \cite{multiphase.Levy00}.
They concluded that differences in physical properties determine the flow 
behaviour to a large extent including the dependency of dust accumulation
on gas velocity, powder diameter and shape. 
The latter
and its effect on gas-powder flow in a blast furnace in particular was investigated 
by Dong et al.~\cite{cohesivezone.Dong06}. They identified 
numerically powder hold-up regions in the central part of a blast furnace and 
lower part of the cohesive zone. Sugiyama et al.~\cite{multiphase.Sugiyama96}
analysed accumulation in the deadman and dripping zone of blast furnace
through experimental investigation of a two-phase flow through a parallel 
packed bed. 
 
Yamaoka \cite{multiphase.Yamaoka86} reported that
these regions of blockage also occur in two- or three-dimensional packed beds.
He proposed a simple model for which the collision between fines and the particles 
of the packed bed dominate the behaviour. These findings were confirmed by 
Ichida et al.~\cite{multiphase.Ichida92}
who observed that fines were deposited near 
walls 
due to low gas velocities. Similarly, Yamaoka \cite{multiphase.Yamaoka91}
found that dust may stick to particle surfaces close to injection, and thus
reducing the outflow, however, not caused by low velocities. 

In order to contribute to clarification of these phenomena observed 
Hidaka et al.~\cite{multiphase.Hidaka98} investigated experimentally into the
entrainment of fine particles in a packed bed with a counter-current gas flow.
Both dynamic and static hold-up were correlated with empirical equations.  
Furthermore, the pressure drop for a gas flow with entrained fines was slightly 
higher than for a pure gas flow. Similar experiments were carried out by 
Dong et al.~\cite{multiphase.Dong04a} and they found that accumulation regions 
increase with high dust rates and low velocities. Furthermore, the structure 
of a packed bed in particular impermeable blocks in a packed bed  
promote accumulation of fines. The existence of impermeable
fused layers was also confirmed by Steiler et al.~\cite{multiphase.Steiler91}.
This phenomenon of impermeable fused layers was investigated numerically by
Wang et al.~\cite{multiphase.Wang00}. Their results indicate that a complex 
and non-uniform flow with horizontal and even upward flow takes place in the 
cohesive zone and in front of the raceways. Furthermore, liquids penetrating the
void space in the coke layers leads to an increased hold-up.

Yagi \cite{multiphase.Yagi93} investigated the 
flow of four phases namely gas, liquid, solids and fines 
by differential conservation equations. Four phase flow was also addressed by 
Dong et al.~\cite{multiphase.Dong09}. They found that motion and structure of 
the solids effects the powder and liquid holdup to a large extent. Thus, 
they predicted a solid stagnant zone with a hold-up region for powder and 
liquid. They also stressed the fact 
that a flow model has to be coupled to a heat transfer and a chemical conversion 
model. Similar investigations were carried
out by Aoki et al.~\cite{multiphase.Aoki93} who predicted the behaviour of fine 
particles in the vicinity of the injection area. In the injection region 
Yuu et al.~\cite{multiphase.Yuu05} predicted stable and unstable flows and 
was carried out similarly by Takeda and Lockwood 
\cite{multiphase.Takeda96, multiphase.Takeda97}.

\subsubsection{Gas-Solid Flow}

An alternative approach considers
the solid phase as discrete, while the flow of liquids or gases is treated
as a continuum phase in the void space between the particles, and therefore, is 
labelled the Combined Continuum and Discrete Model (CCDM) \cite{multiphase.Tsuji93, 
multiphase.Hoomans96, multiphase.Xu97, multiphase.Xu98}. Due to a discrete description
of the solid phase, constitutive relations are omitted, and therefore, leads to 
a better understanding of the fundamentals. This was also concluded by 
Zhu et al.~\cite{multiphase.Zhu07b} and Zhu et al.~\cite{multiphase.Zhu08} during 
a review on particulate flows modelled with the CCDM approach. It has seen a 
major development in 
last two decades and describes motion of the solid phase by the Discrete Element 
Method (DEM) on an individual particle scale and the remaining phases
are treated by the Navier-Stokes equations. Thus, the method is recognized as
an effective tool to investigate into the interaction between a particulate and
fluid phase as reviewed by Yu and Xu \cite{multiphase.Yu03}, Feng and Yu 
\cite{multiphase.Feng04} and Deen et al.~\cite{multiphase.Deen07}.

Within a major review on particulate flows modelled with the CCDM approach 
Zhu et al.~\cite{multiphase.Zhu07b} and Zhu et al.~\cite{multiphase.Zhu08} concluded 
that the methodology is well suited to understand the fundamental 
physics of these flows. However, current models including software should be
 extended to
multi-phase flow behaviour and to particle shapes other than spherical
geometries to meet engineering needs. These efforts should lead to a general link
between continuum and discrete approaches so that results are quantified for
process modelling.

Initially, 
such studies are limited to simple flow configurations 
\cite{multiphase.Hoomans96, multiphase.Tsuji93}, however, Chu and Yu 
\cite{multiphase.Chu08} demonstrated that the method could be applied to a 
complex flow configuration consisting of a fluidised bed, conveyor belt and
a cyclone. Similarly, Zhou et al.~\cite{multiphase.Zhou11} applied the CCDM
approach to the complex geometry of fuel-rich/lean burner for pulverised
coal combustion in a plant and Chu et al.~\cite{multiphase.Chu09} modelled the complex flow
of air, water, coal and magnetite particles of different sizes in a dense 
medium cyclone (DMC). For both cases remarkably good agreement between 
experimental data and predictions was achieved. 

The CCDM approach has also been applied to fluidised beds as reviewed
by Rowe and Nienow \cite{multiphase.Rowe76} and Feng and Yu 
\cite{multiphase.Feng04} and applied by Feng and Yu 
\cite{multiphase.Feng08} to the 
chaotic motion of particles of different sizes in a gas fluidised bed.
Kafuia et al.~\cite{multiphase.Kafuia02} describe discrete particle-continuum fluid modelling of gas-solid fluidised beds.

A similar development has been seen for modelling of blast furnaces. 
Computational Fluid Dynamics (CFD) as a tool for 
continuous flow modelling has been applied with success to a large extend 
as reviewed by Chattopahyay et al.~\cite{multiphase.Chattopadhyay10a, multiphase.Chattopadhyay10b}. 
The flow of the solid phase consisting of particles is to be modelled
by a discrete approach as suggested by Dong et al.~\cite{multiphase.Dong07}
and reviewed by Yu \cite{multiphase.Yu05} for several engineering applications.
Simsek et al. \cite{pyrolysis.Simsek09} predicted grate firing systems by 
the CCDM method, but obtained only qualitatively reasonable results highlighting
the fact that more research is required.

However, current CCDM approaches should be extended to a truly 
multi-phase flow behaviour as opposed to the Volume-of-Fluid method and 
the multi-phase mixture model \cite{multiphase.Wang98}. Furthermore, particle 
shapes other than spherical geometries have to be taken into account to meet 
engineering needs according to Zhu et al.~\cite{multiphase.Zhu07b} and 
Zhu et al.~\cite{multiphase.Zhu08}. These efforts should ideally be complemented 
by poly-disperse particle systems since all derivations have done for mono-sized 
particles as stated by Feng and Yu \cite{multiphase.Feng04}. All these efforts 
should contribute to a general link between continuum and discrete approaches so 
that results are quantified for process modelling.

Although the CCDM methodology has been  established over the past decade
\cite{multiphase.Tsuji93, multiphase.Xu97}, prediction of heat transfer is 
still in its infancy. Kaneko et al.~\cite{multiphase.Kaneko99} predicted 
heat transfer for polymerisation reactions in gas-fluidised beds by the 
Ranz-Marshall correlation \cite{multiphase.Ranz52}, however, excluding 
conduction. Swasdisevi et al.~\cite{multiphase.Swasdisevi05} predicted heat 
transfer in a two-
dimensional spouted bed by convective transfer solely. Conduction between
particles as a mode of heat transfer was considered by Li and Mason
\cite{multiphase.Li00, multiphase.Li02, multiphase.Li03} for gas-solid 
transport in horizontal pipes. Zhou et 
al.~\cite{multiphase.Zhou04a, multiphase.Zhou04b} modelled coal combustion in a 
gas-fluidised bed including
both convective and conductive heat transfer. Although, Wang et 
al.~\cite{multiphase.Wang2011} used the two fluid model to 
predict the gas-solid flow in a high-density circulating fluidised bed, Malone
and Xu \cite{multiphase.Malone08} predicted heat transfer in liquid-fluidised 
beds by the CCDM method and stressed the fact that deeper investigations into
heat transfer is required. 

Although Xiang 
\cite{multiphase.Xiang09} investigated into
the effect of air on the packing structure of fine particles, it is not feasible
for large structures due to limited computational resources. 
A recent review of Zhou et al.~\cite{multiphase.Zhu07b} shows that a lot of 
approaches concentrate on flow solely without heat or mass transfer. They anticipate
that future requirements will have to concentrate on the following issues 
\cite{multiphase.Zhu07b}:

\begin{itemize}
\item "Microscale: To develop a more comprehensive theory and 
  experimental techniques to study and quantify the interaction
  forces between particles, and between particle and fluid under 
  various conditions, generating a more concrete basis for
  particle scale simulation." \cite{multiphase.Zhu07b}
\item "Macroscale: To develop a general theory to link the discrete
  and continuum approaches, so that particle scale information, 
  generated from DEM or DEM-based simulation, can
  be quantified in terms of (macroscopic) governing equations,
  constitutive relations and boundary conditions that can be
  implemented in continuum-based process modelling." \cite{multiphase.Zhu07b}
\item "Application: To develop more robust models and efficient
  computer codes so that the capability of particle scale simulation 
  can be extended, say, from two-phase to multiphase
  and/or from simple spherical to complicated non-spherical
  particle system, which is important to transfer the present 
  phenomenon simulation to process simulation and hence meet
  real engineering needs." \cite{multiphase.Zhu07b}
\end{itemize} 

These findings are confirmed by Yu and Xu \cite{multiphase.Yu03} who see the 
difficulties more associated with the solid phase than to the fluid phase. Hence,
a coupling between DEM and CFD is more promissing due to its computational 
convenience than DNS- or LB-DEM interfaces.  Yu and Xu \cite{multiphase.Yu03}, 
Zhu et al.~\cite{multiphase.Zhu08} and Zhu et al.~\cite{multiphase.Zhu07b}
gave a major review on particulate flows modelled with the CCDM approach and 
concluded that it is well suited to understand the fundamental physics
of these flows.

\subsection{Chemical Reaction, Heat and Mass Transfer}

The theoretical foundation for the 
Extended Discrete Element Method (XDEM) was developed in 1999 by 
\cite{Peters99a}, who described incineration of a wooden moving bed on a forward 
acting grate \cite{Peters02a} for which the particulate phase was resolved by an 
individual model describing various conversion processes. The concept was later 
also employed by \cite{pyrolysis.Simsek09} to predict the furnace process of a 
grate firing system. 

Initial efforts within an Euler-Lagrange framework including the thermodynamic 
state of the particulate phase were proposed by Peters \cite{Peters02a},  This 
approach is also reffered to as the Combined 
Continuum and Discrete Model (CCDM) of which major reviews are found in Zhu et 
al.~\cite{multiphase.Zhu07b}, Zhu et al.~\cite{multiphase.Zhu08}
Yu and Xu \cite{multiphase.Yu03}, Feng and Yu \cite{multiphase.Feng04} and 
Deen et al.~\cite{multiphase.Deen07}.

Exchange of heat mass, and momentum between the 
particulate and a fluid phase are described by empirical correlations as 
applied by Miltner et al. [20] for baled biomass. Similarly, Mehrabian et 
al.~\cite{multiphase.Mehrabian2012} and Gomez et 
al.~\cite{multiphase.Gomez2015} employed this approach for packed bed combustion. 
Thermal conversion of solid fuels on a forward acting grate were investigated 
through a coupling of CFD/DEM by Simsek et al.~\cite{pyrolysis.Simsek09}. 
Wiese et al.~\cite{multiphase.Wiese2016} also predicted the 
performance of a pellet stove by CFD/DEM coupling and employing reduced order 
modelling for the particulate phase. However, no details on the ROM approach 
were provided. Further investigations with the coupled CFD/DEM technique within 
the same research group include a variety of applications such as drying in a 
rotary kiln \cite{multiphase.Sudbrock2015,multiphase.Scherer2016}, municipal 
solid waste incineration \cite{multiphase.Wissing2016}, 
heat transfer \cite{multiphase.Oschmann2016} and calcination 
\cite{multiphase.Krause2015}. Kloos et al.~\cite{multiphase.Kloss2012} developed 
a CFD/DEM by coupling the open-source codes Lammps and OpenFoam to analyse heat 
transfer in fluidised beds \cite{multiphase.Radl2015}.
Kaneko et al.~\cite{multiphase.Kaneko99} predicted 
heat transfer for polymerisation reactions in gas-fluidised beds by the 
Ranz-Marshall correlation \cite{multiphase.Ranz52}, however, excluding 
conduction. Swasdisevi et al.~\cite{multiphase.Swasdisevi05} predicted heat 
transfer in a two-
dimensional spouted bed by convective transfer solely. Conduction between
particles as a mode of heat transfer was considered by Li and Mason
\cite{multiphase.Li00, multiphase.Li02, multiphase.Li03} for gas-solid 
transport in horizontal pipes. Zhou et 
al.~\cite{multiphase.Zhou04a, multiphase.Zhou04b} modelled coal combustion in 
a gas-fluidised bed including
both convective and conductive heat transfer. Although, Wang et 
al.~\cite{multiphase.Wang2011} used the two fluid model to 
predict the gas-solid flow in a high-density circulating fluidised bed, Malone
and Xu \cite{multiphase.Malone08} predicted heat transfer in liquid-fluidised 
beds by the CCDM method and stressed the fact that deeper investigations into
heat transfer is required.
\added[id=BP]{However, fluidised beds including thermal conversion are likely to experience temperatures above 1000 K \cite{HT:Chen2005,HT:Baillis2000,HT:Decker1983,HT:Gloski1984,HT:Goshayeshi1986} and therefore, heat transfer through radiation becomes dominant.
Toschkoff et al. ~\cite{HT:Toschkoff2015} applied the Discrete Transfer Radiation Model (DTRM) stressed the feasibility of this approach in conjunction with DEM models, however, becomes computationally increasingly expensive with larger numbers of rays \cite{HT:Amberger2013}.
A less expensive and feasible approach was developed by Peters \cite{Peters03b,Peters02a} who employed view factors to estimate radiative transfer between a particle and its neighbours. This method was also applied by Fogber and Radl \cite{HT:Fogber2018} whereby the view factor is estimated based on per-particle corresponding solid angles and treating the particle surface as a black surface. Although a comparison with analytical, Monte-Carlo and OpenFoam solutions agreed well, it is expected that the method still requires a significant CPU time for larger particle arrangements.}

A further detailed review on approaches, recent advances and applications is
given by Zhong et al.~\cite{mpPIC:Zhong2016}

\section{Technological Concept}

Numerous challenges in engineering exist and evolve, that include a continuous 
and discrete phase simultaneously, and therefore, cannot be solved accurately by 
continuous or discrete approaches, only. Problems that involve both a continuous and a discrete phase  
are important in applications as diverse as
pharmaceutical industry e.g.~drug production, agriculture food and processing industry,
mining, construction and agricultural machinery, metals manufacturing, energy production 
and systems biology. Some predominant examples are coffee, corn flakes, nuts, coal,
sand, renewable fuels e.g.~biomass for energy production and fertilizer.
Therefore, the Extended Discrete Element Method (XDEM) provides a platform, that 
couples  discrete and continuous phases for a large number of engineering applications.

The XDEM-suite roots on the Extended Discrete Element Method (XDEM) that is a 
recently evolved numerical technique. It extends the dynamic properties of the 
classical discrete element method by additional properties such as thermodynamic 
state, stress/strain or electro-magnetic field for each discrete entity. \replaced[id=BP]{Although continuum mechanics approaches are equally applicable, a lack of information is inherent due to averaging of the discrete phase. This loss is usually compensated by additional correlations such as distribution of porosity. However, solutions based on continuum mechanics have the advantage to deal with macro-scale engineering challenges. Methodologies based on DEM-CFD coupling require large computer resources and therefore, deal with applications covering micro- to meso-scales. An analysis of results on smaller scales unveils underlying physics that could be fed into continuum mechanics approaches. }{Unlike
continuum mechanics approaches of which averaging of the discrete phase causes a 
significant loss of information and introduces additional correlations XDEM 
resolves the particulate phase with its relevant properties associated with
each entity.}

Interaction of the discrete phase with continuous fields via heat, mass and 
momentum exchange is achieved by a generic interface to the Finite Volume Method (FVM)  
\deleted[id=BP]{and Finite Element (FEM) as mentioned above. Hence, the interface to the
classical finite element method allows evaluating the mechanical load on 
structures such as hoppers, chutes and machinery dealing with transport and 
storage of granular media due to its impact}. A coupling to computational
fluid dynamics is based on the software platform of OpenFoam \cite{CFD.OpenFoamextend} \deleted[id=BP]{, from 
which a number of solvers has been extended and tailored} to handle efficiently
data transfer between the discrete and continuum phase. 

%

\subsection{Fluid Phase Governing Equations} 
\label{s:equations}
\added[id=MB1]{In this regard, the fluid phases are treated as a continuum 
on macroscopic level using \textit{Eulerian volumetric average} known as multi-fluid model in which the conservation of
momentum, mass and energy are solved for each phase separately \cite{Faghri}.The generic governing equation for the $k^{th}$ phase can be written as Eq.(\ref{e:generic}).}

 \begin{equation}
\underbrace{\frac{\partial(\epsilon_k \rho_k \phi_k)}{\partial t}}_{transient} + \underbrace {\nabla \cdot (\epsilon_k \rho_k \vec{v_k} \phi_k)}_{convection} = \underbrace{\nabla \cdot  \Gamma (\nabla \cdot \epsilon_k \phi_k)}_{diffusion} + \underbrace{S(\phi_k)}_{source / sink} \label{e:generic}
\end{equation}
\added[id=MB1]{
Where the conservation of mass, momentum and energy can be derived by replacing the values in Table \ref{t:values}.}

\begin{table}[h]
   \centering
    \caption{Values for conservation equations for each phase}
    \label{t:values}
\begin{tabular}{| l | l l  l l l|}

\hline
      &  $\phi$    && $\Gamma $ && $S(\phi)$ \\ \hline
    \textbf{continuity} &  1 && 0 && $m'$ \\ \hline
    \textbf{momentum} & $v$ && $\mu$ && $-\nabla p + F$ \\ \hline
    \textbf{energy} & $h$ && ---  && $-\nabla \cdot q+Q$ \\
\hline
    \end{tabular}
\end{table}
\added[id=MB1]{
The term $m'$ is the source term of the continuity equation, represents the mass transfer between phases while the $\mu$ and
$F$ in the momentum equation are the viscosity and momentum transfer between phases. In the energy equation the diffusivity is described as a term of  $q = -\lambda \nabla T$ 
and $\lambda$ is the thermal conductivity. The other source term ($Q$) in the energy equation is any kind of heat transfer between phases. 
 The coupling between the fluid phases and particulate phase is through the $source/sink$ term $S(\phi)$ where the positive 
values represent mass, momentum and energy transfer from the solid particles to the fluid phases and vice versa. 
}

\subsection{Governing Equations for the Particle}
\added[id=MB2]{One-dimensional and transient conservation equations characterize energy and mass transport within each particle.
Conservation of mass for gas within the pore volume of a porous particle writes as follows:}
\begin{equation}
\frac{\partial}{\partial t} \big( \varepsilon_p \rho_g \big) + 
\label{gasCont}
\end{equation} 
\added[id=MB2]{where $S_{mass}$ is the summation of the individual species mass production or consumption rates created by chemical reactions. $\varepsilon_p$ denotes particle porosity and $u_g$ advective
velocity. Moreover, the fluid's intrinsic density $\rho_g$ is given by the sum of partial densities of species present in the gas phase as $\rho_g = \sum\limits_{i} \rho_{i,g}$. 
Transport of gaseous species within the pore space of the particle is considered to obey Darcy’s law:}
\begin{equation}
    - \frac{ \partial p} { \partial r}=\frac{\mu_g \epsilon_p }{K} \left\langle u_g \right\rangle  \label{e:1Dpartdarcy}
\end{equation}
\added[id=MB2]{\noindent where $p$ is pressure and $\mu_g$ denotes the dynamic viscosity and $K$ represents the permeability of the porous particle.
The balance of an individual specie~$i$ within the pores of a given particle is described by}
\begin{equation}
    \frac{\partial({\epsilon_p \rho_{i,g}})}{\partial{t}}   +\frac{1}{r^n}\frac{\partial}{\partial{r}} (r^n \epsilon_p\rho_{i,g} u_g)=\frac{1}{r^n}\frac{\partial}{\partial{r}}(r^n D_{i}\epsilon_p\frac{\partial}{\partial{r}}\rho_{i,g})+\epsilon_p\sum_k\dot{\omega}_{k,i,g} \label{e:species}
\end{equation}
\added[id=MB2]{where $\dot{\omega}_{k,i}$ denotes the source term accounting for the consumption or generation of species~$i$ from reaction $k$ and $D_{i}$ is the molecular diffusion coefficient.}
\added[id=MB2]{The distribution of temperature within a single particle is accounted for by the energy equation as:}
\begin{equation} \label{eq:energyyy}
 \frac{\partial}{\partial t} \left( \rho h  \right) = \frac{1}{r^n}\frac{\partial}{\partial r}\left( r^n \lambda_{\rm eff} \frac{\partial T}{\partial r} \right) + Q'
 \end{equation}
\begin{equation}
       Q' = 
        \begin{cases}
            \sum_{k=1}^{l}\dot{\omega}_k H_k  & \text{in case of chemical reaction}\\
            - m'h_{l,m} - m' L_{f}  & \text{in case of melting}
        \end{cases}
 \end{equation}
\added[id=MB2]{where $\lambda_{\rm eff}$ is the effective thermal conductivity and $h_{l,m}$, $L_{f}$ and $H_k$ donate the enthalpy of the liquid at melting temperature,
latent heat of fusion, and the enthalpy of reaction, respectively.
$m'$ is the melting rate which will be defined later.} 
\added[id=MB2]{To solve the equations, the following boundary conditions are needed:}
\begin{align}
	\left. -\lambda_{\rm eff} \frac{\partial T}{\partial r} \right|_{r=0}    &= 0 \label{TBCatC}\\
	\left. -D_{i} \frac{\partial \rho_{i,g}}{\partial r} \right|_{r=0} &= 0 \label{MBCatC} 
\end{align}
\begin{align}
	\left. -\lambda_{\rm eff} \frac{\partial T}{\partial r} \right|_{r=R} &= \alpha (T_{s} - T_{\infty}) + \dot{q}''_{\rm rad} + \dot{q}''_{\rm cond} \label{Tbc} \\
	\left. -D_{i} \frac{\partial \rho_{i,g}}{\partial r} \right|_{r=R} &= \beta_i (\rho_{i,s} - \rho_{i,\infty}) + \dot{m}''_{i,g}\label{Mbc} 
\end{align}
\added[id=MB2]{where $\alpha$ and $\beta$ denote the heat and mass transfer coefficients, $T_{\infty}$, $\rho_{i,\infty}$ represent
the ambient gas temperature and the 
gas partial density of the species $i$ in ambient gas. $T_{s}$, $\rho_{i,s}$ donate the particle temperature and the partial density of
species $i$ at the surface of the particle, respectively. $\dot{m}''_{i,g}$ accounts for the mass fluxes from the environment. 
Moreover, $\dot{q}''_{\rm rad}$ and $\dot{q}''_{\rm cond} $ account for conductive heat and radiation transport through  
physical contact with the wall and/or particles and are calculated as follow:}
\begin{align}
	\dot{q}''_{p,\rm rad}  &= \sum\limits_{j=1}^M F_{p\rightarrow j} \sigma \left(T_p^4-T_j^4\right)     \label{e:nradiation} \\
	\dot{q}''_{p,\rm cond} &= \sum\limits_{j=1}^N \frac{1}{\frac{1}{\lambda_p} + \frac{1}{\lambda_j}  }  \frac{T_p-T_j}{\Delta x_{p,j}} \label{e:nconduction}
\end{align}
\added[id=MB2]{where $F_{p\rightarrow j}$ is the view factor between particle $p$ and its neighbor $j$ and ${\lambda}$ is the thermal conductivities of the different particles.
The equations are conveniently written to be represented on different coordinates systems. Thus, the particle geometry can be chosen to be an infinite plate $n=0$, 
infinite cylinder $n=1$ or a sphere $n=2$.}

\added[id=MB2]{The discrete element method based on the soft
sphere model is applied to the dynamic module of the XDEM-suite.
The movement of each particle is tracked using the equations of classical mechanics. Newton’s 
and Euler's second law for translation and rotation of each particle are integrated over time and 
the particle positions, orientations and velocity are updated accordingly during time integration.}

\begin{equation}
 m_i \frac{d \vec{v_i}}{d t} = m_i \frac{d^2 \vec{x_i}}{dt^2} = \vec{F}_{i}^c + \vec{F}_{i}^g + \vec{F}_{i}^{ext}
\end{equation}

\begin{equation}
 I_i \frac{d \omega_i}{d t} = \sum_{j=1}^n M_{i,j}
\end{equation}

\added[id=MB2]{\noindent where $\vec{F}_{i}^c$, $M_{i.j}$,  $\vec{F}_{i}^{ext}$, $v_i$, $\omega_i$, $I_i$ are contact forces, torques and external forces acting on particle i,
linear velocity, angular velocity and moment of inertia, respectively. For more description the reader is referred to \cite{Hoffmann_thesis, Mahmoudi_thesis, Mark_thesis, Samiei_thesis}}

\deleted[id=BP]{One of the outstanding predictive capabilities of the XDEM-suite are the thermal
conversion mechanisms that could be attached to the discrete phase. Different
reaction types are available that may be attached to a single reaction or an
extensive reaction mechanism of an arbitrary number. Depending on the reaction
type chosen morphological changes or shrinking of for example burning particles 
is taken into account. Futhermore, phase changes between gas - liquid - solid 
are treated within the current approach:} 


%
%
%

Transport coefficients, thermodynamic and other properties are spatially resolved
to include dependencies such as temperature and are available from an extensive 
data base including NASA polynomials for solid, liquid and gaseous species.

\deleted[id=BP]{Thus, the XDEM suite covers a large range of applications as diverse as 
powder metallurgy, material science, iron making, snow research, thermal
conversion of biomass to promote renewable energy, machinery for handling and
transport of granular material. These efforts are enhanced by a graphical
user interface (GUI), that allows users to set-up, run and post-process results.
It includes also a an extensive property and chemical reaction data base to
assign desired reactions e.g.~mechanisms to the particulate phase. A detailed
analysis of multi-phase results reveals the underlying and complex physics of
such scenarios and thus, supports improved design and operation for engineering
applications.}

\section{Engineering Applications}

The following section addresses relevant and validated applications 
that emphasise the predictive capabilities of the XDEM-suite.

\subsection{Generic Multi-phase Flow Solver} \label{multiphase}

The multi-phase solver of the XDEM-suite is applicable to any system including several fluid phases flowing through 
packed bed of solid particles, which exist in broad spectrum of engineering disciplines as was mentioned before.
The frequently used reactors of these types are packed bed, trickle bed, fluidized bed reactors which can be classified as 
counter-current, co-current and cross-current based on the direction of the fluid phases \cite{Mederos2009}.

The flow behaviour in packed bed reactors is very complex and depends not only on hydrodynamics but also on the mass and 
heat transfer between all phases \cite{Schwidder2012}. The size parameter and geometry of the particulate phase have a pronounced effect 
on the flow distribution of fluid phases \cite{Atta2007, Schwidder2012, Jiang20021}. In order to predict the performance 
of these reactors, the local interaction between all phases such as fluid-fluid, fluid-solid and solid-solid interactions 
must be taken into account. The Eulerian-Lagrangian characteristics of the XDEM-suite allow including all these interactions 
for any particulate system.

\begin{figure}
     \centering
     \includegraphics[trim = 0mm 5mm 0mm 0mm, clip, angle=0, width=0.9\linewidth]{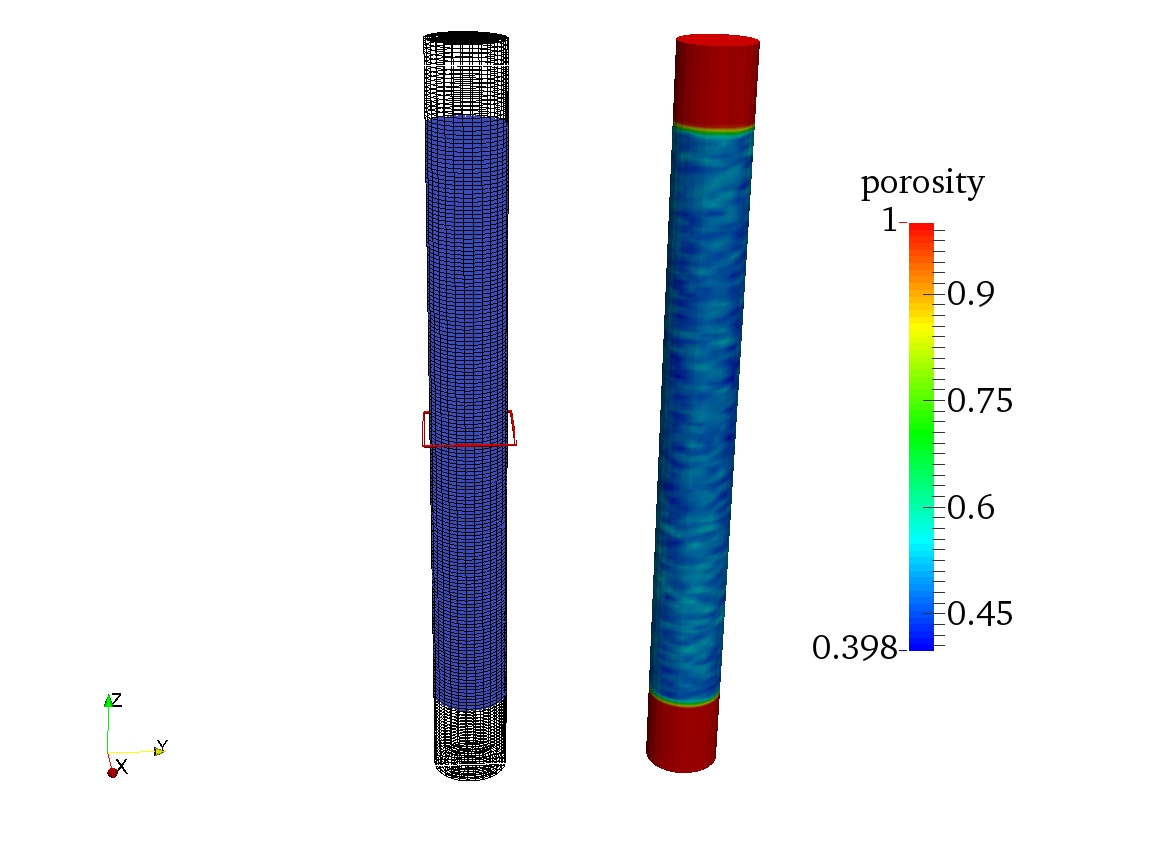}
         \caption{Computational domain and porosity distribution}
         \label{f:porositydistibution}
 \end{figure}

The generic solver was validated using experimental data of Gunjal et al. \cite{Gunjal2005} for flow of \deleted[id=MB1]{gas and liquid}\added[id=MB1]{air and water at room temperature} 
through packed bed of solid particles.
An experiment was carried out in a vertical pipe filled with $6mm$ spherical particles to represent trickle bed reactors. The geometry of the case as well as the 
position of the particles are shown in fig.~\ref{f:porositydistibution} that includes also the distribution of
porosity  \deleted[id=BP]{of the domain computed by XDEM-suite, is also demonstrated}. The 
XDEM-suite calculate the porosity of each CFD cell based on the algorithm proposed by Xiao et al. \cite{xiao2011}. 

\begin{figure}[!tbp]
  \centering
  \begin{minipage}[b]{0.45\textwidth}
    \includegraphics[trim = 0mm 0mm 0mm 0mm, clip, angle=0, width=1.1\linewidth]{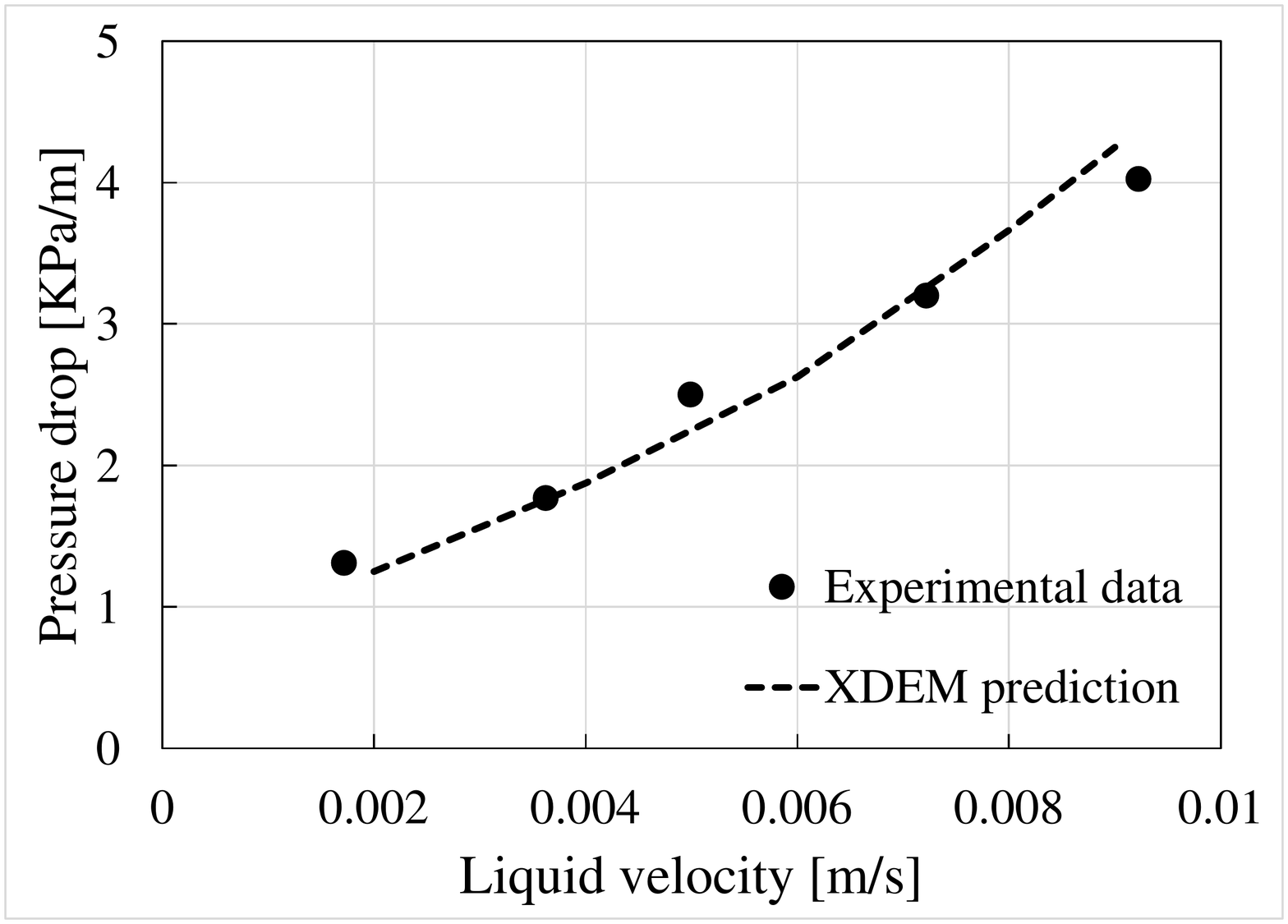}
         \caption{The XDEM prediction for pressure drop vs experimental data of Gunjal et al. \cite{Gunjal2005}}
         \label{f:pressureDrop}  \end{minipage}
  \hfill
  \begin{minipage}[b]{0.45\textwidth}
    \includegraphics[trim = 0mm 0mm 0mm 0mm, clip, angle=0, width=1.1\linewidth]{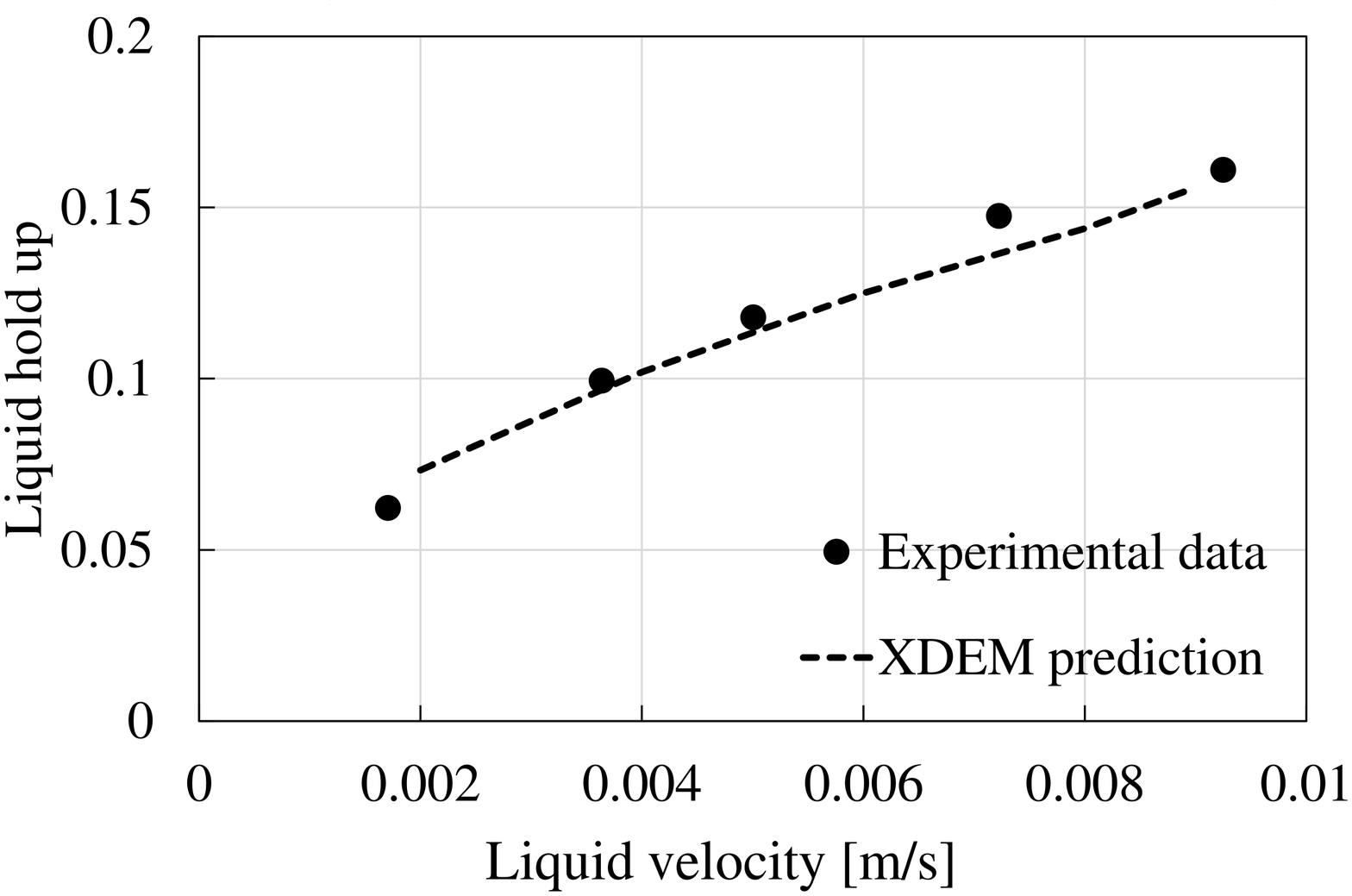}
         \caption{The XDEM prediction for liquid saturation vs experimental data of Gunjal et al. \cite{Gunjal2005}}
         \label{f:saturation}
  \end{minipage}
\end{figure}

The results were validated by comparing predicted results versus experimental data for pressure drop and liquid hold up. In fig.~\ref{f:pressureDrop}, the pressure drop for different 
liquid velocities is shown, which confirms that by increasing the liquid velocities, the pressure 
drop increases as well. In fig.~\ref{f:saturation}, the liquid hold up against liquid velocity is depicted, which shows that higher liquid hold up occurs when higher liquid velocity is introduced. 
Both figures show a very good agreement with experimental data.

\begin{figure}
     \centering
     \includegraphics[trim = 0mm 5mm 0mm 0mm, clip, angle=0, width=1\linewidth]{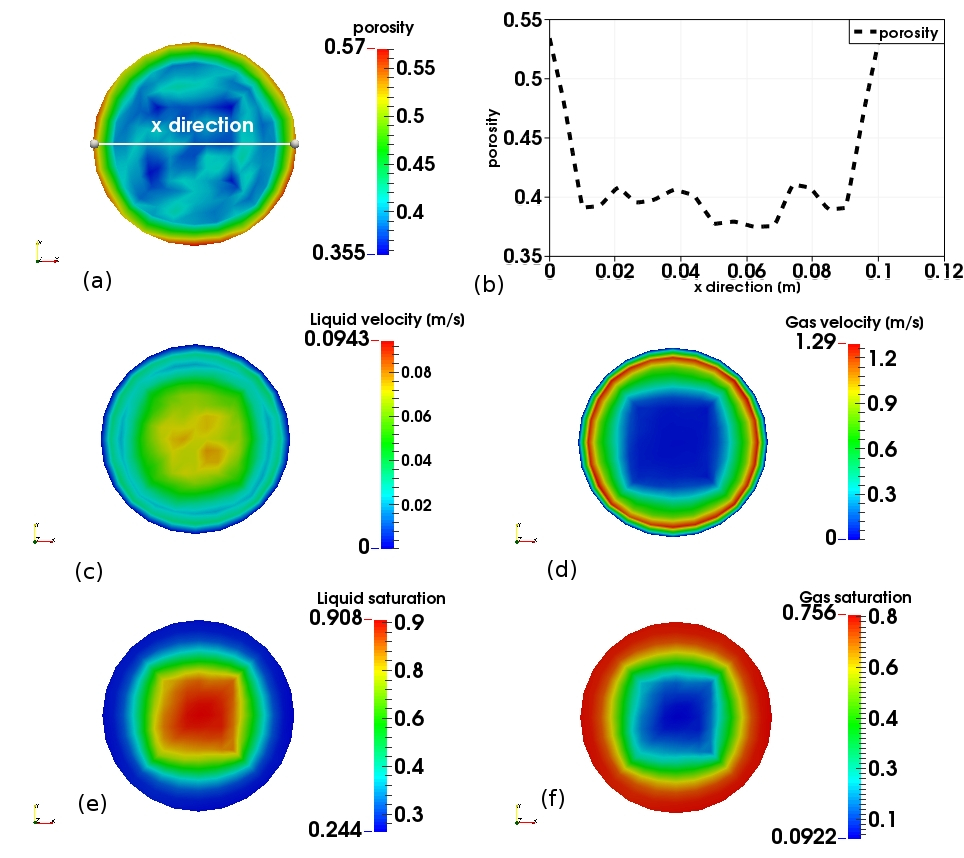}
         \caption{(a) porosity distribusion, (b) porosity distribusion over a line depicted in part (a), (c) liquid phase velocity, (d) gas phase velocity, 
         (e) liquid phase saturation, (f) gas phase saturation}
         \label{f:all}
 \end{figure}

In order to visualize other hydrodynamic parameters such as velocities and saturations, a slice at \replaced[id=BP]{height}{hight} of $0.5~m$ (shown in fig.~\ref{f:porositydistibution}) for a specific case 
was chosen, which 
are displayed in fig.~\ref{f:all}. The porosity distribution for this slice is depicted in fig.~\ref{f:all}(a) and it is drawn over a line in fig.~\ref{f:all}(b) in order to 
show the wall effects. As it is clear in part (b), higher porosities are observed at the vicinity of the wall, which highly effect the hydrodynamic parameters as compared to a constant porosity distribution. 
The liquid and gas phase velocities are shown in part (c) and (d) respectively, while phase saturations are depicted in part (e) and (f). Higher velocities are in respect to higher 
saturation zones. The non-uniform velocity profiles represent the effect of equally non-uniform porosity distribution estimated by the XDEM-suite which leads to 
more accurate and precise prediction of hydrodynamic parameters. More details about the multiphase solver of the XDEM suit can be found in \cite{Baniasasi2017, Baniasadi2018, MeBaniasadiCES2017}. 

\subsection{Melting of particles as Phase Change Including Heat and Mass Transfer}

Transport phenomena including a phase change such as evaporation, melting and solidification play a key role in a 
wide range of engineering applications among which melting is of particular interest and addressed in this section.

A melting process is found in many industrial processes such as metal processing, environmental engineering 
and thermal energy storage systems where three phases of solid, liquid and gas could exist simultaneously. These systems are 
difficult to model with numerical analysis due to the complex interaction between the different 
fluid phases and the granular phase. As was discussed before, the Lagrangian-Eulerian framework is closer to the real 
physical processes rather than the Eulerian-Eulerian model for granular flows which includes particle - particle interaction.
These kind of process can be modelled using the XDEM-suite, Since the XDEM-suite is an Eulerian-Lagrangian platform which \replaced[id=BP]{includes}{include} 
heat, mass and momentum transfer between the fluid phases and solid particles.

\noindent where $h_{sl}$, $m'$ and $h_{lm}$ are latent heat of melting, the melting rate and specific enthalpy of melt, respectively.
The melting rate $m'$ in the energy equation can be estimated based on a the energy balance. It requires that the 
latent heat of melting is related to heat transfer to particle. Heat available above the melting temperature is consumed by 
melting process \cite{MeBaniasadiCES2017, MeBaniasadiAIP}. Therefore, a melting rate is defined as:

    \begin{equation}
       m' = 
        \begin{cases}
            \frac{\rho(h - h_m)}{L_{f}\Delta t}  & h \geq h_m\\
            0  &h < h_m
        \end{cases}
    \end{equation}
    
\added[id=MB2]{where $h$ and $h_m$ are the enthalpy of particle at the particle's temperature and melting temperature, respectively.} The melting rate $m'$ and $m' h_{lm}$ are transferred to the CFD field by introducing a source term in the 
liquid continuity and energy equations for the fluid phases the governing equations of the multiphase solver discussed in the previous section. 

The new radius of the particle is estimated based on the 
mass loss at the surface in each time step. It is assumed that the particle retains its sphericity 
in the whole process. 

\begin{figure}[!tbp]
  \centering
  \begin{minipage}[b]{0.4\textwidth}
    \includegraphics[width=\textwidth]{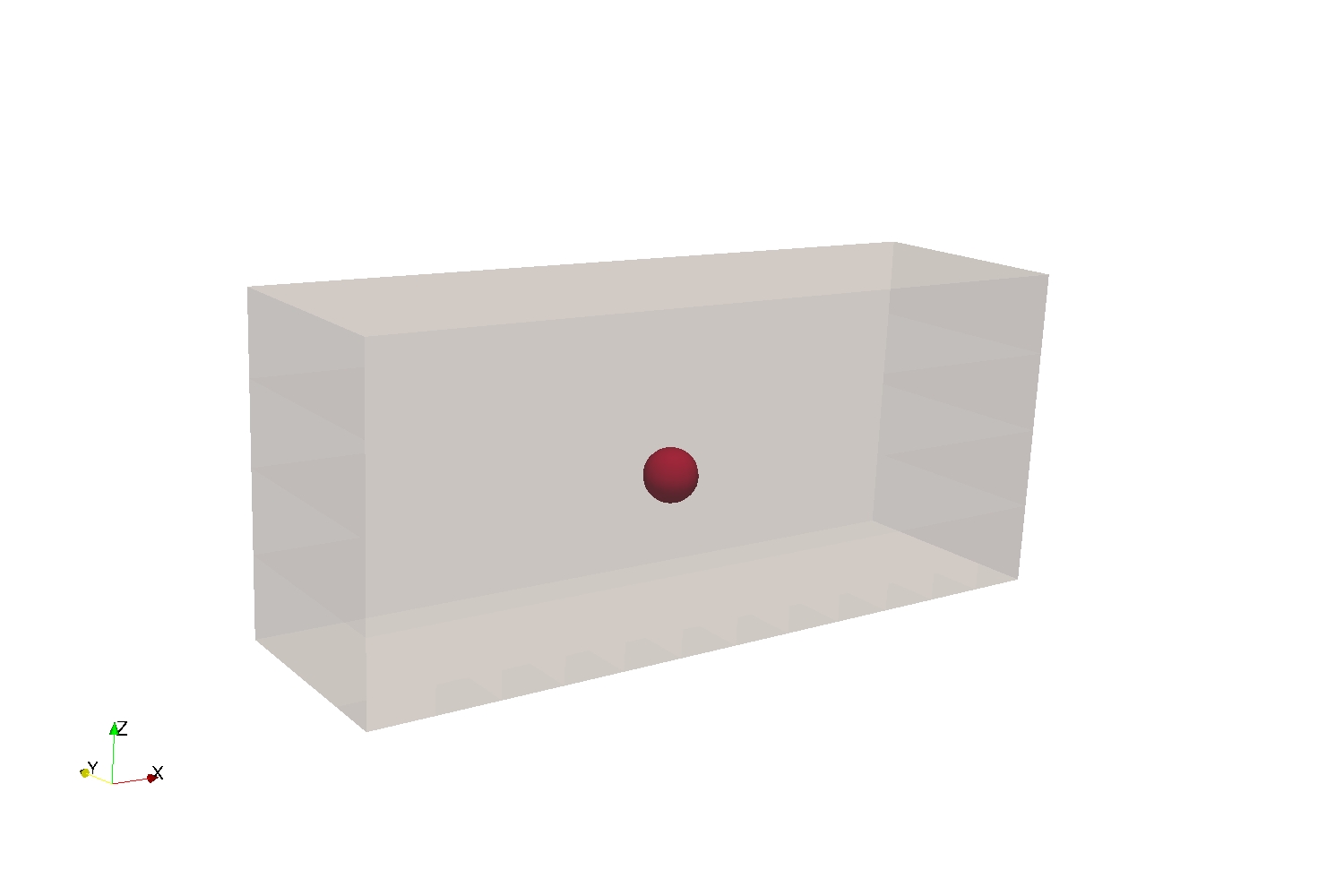}
    \caption{Experimental setup for a single particle in a free stream of water.}
    \label{f:geo}
  \end{minipage}
  \hfill
  \begin{minipage}[b]{0.55\textwidth}
    \includegraphics[width=\textwidth]{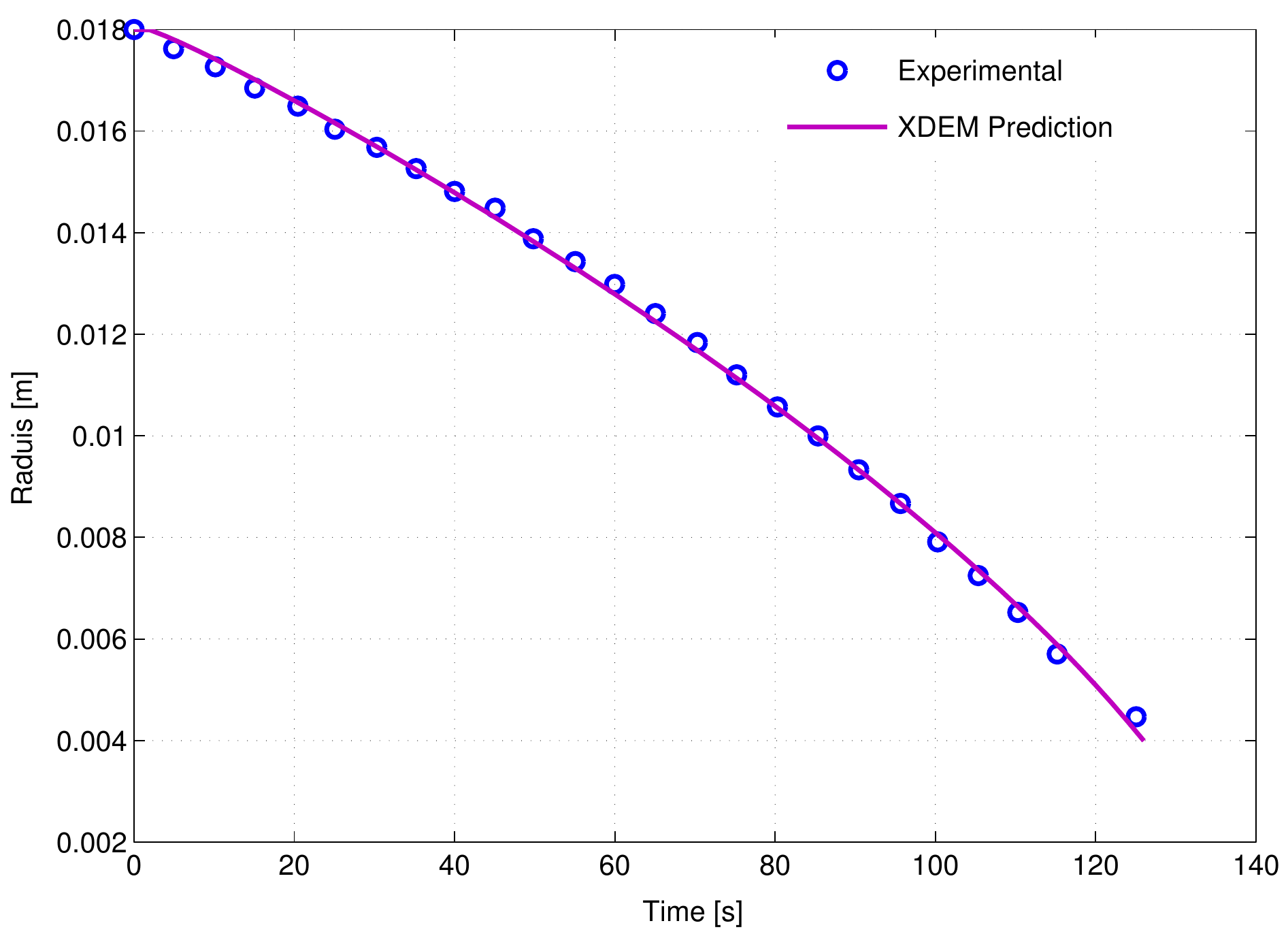}
    \caption{Time history of the radius of the single ice particle.}
    \label{f:plot}
  \end{minipage}
\end{figure}

Validation was performed employing experimental study conducted by Shukla et al. \cite{Shukla2011}.
 The case was considered according to the experimental study under the influence of forced convection \replaced[id=BP]{for which}{and} the set up is shown in fig.~\ref{f:geo}. 
 The experiment was carried on a single 
 sphere ice particle with an initial diameter of $0.036~m$ fixed in a open channel with dimensions of 
 500, 152, 216 mm. Water with a velocity of 0.06 m/s passed the ice sphere horizontally and temperature of the surrounding water, the melting temperature and initial temperature of 
 the ice particle were set to $299.15~K$, $273.15~K$ and $257.15~K$,  respectively. Latent heat of melting of ice is considered
 to be $334~ \frac{KJ}{Kg}$. 

\begin{figure}[h]
\centering
\includegraphics[width=14cm]{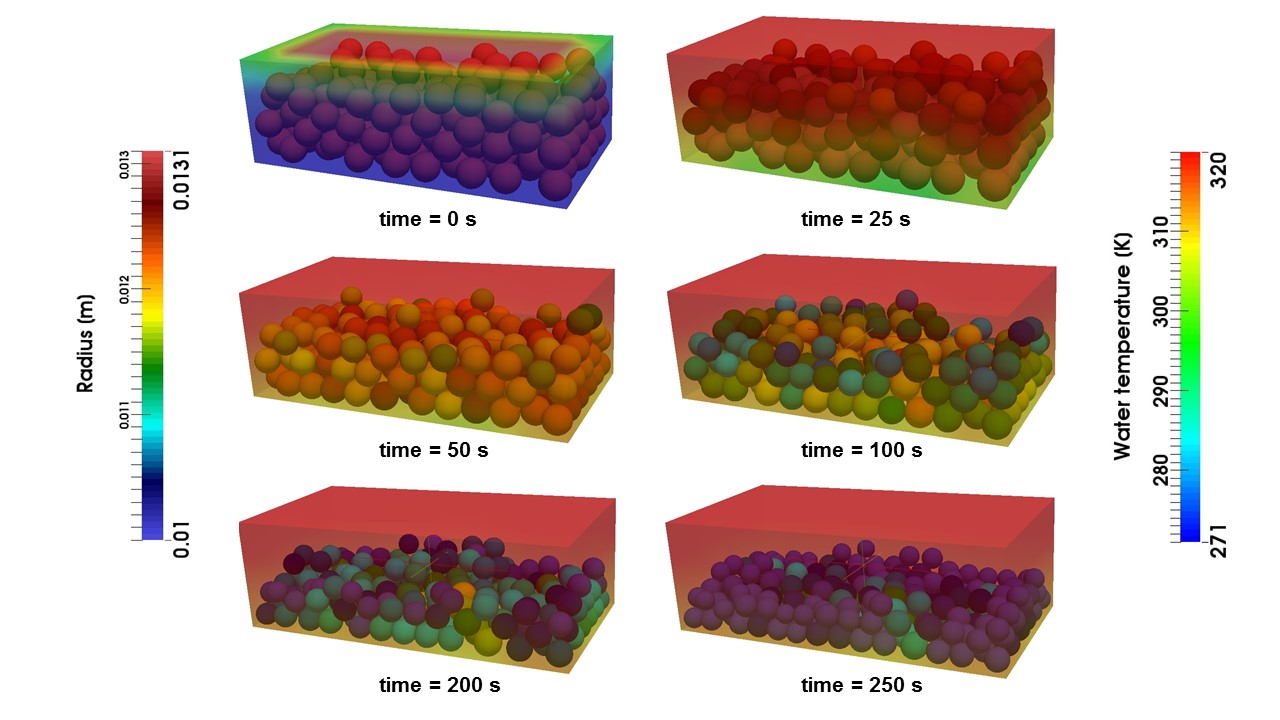}
\caption{Particles size and temperature at different times.}
\label{f:difftime}
\end{figure}

 Fig.~\ref{f:plot} depicts the comparison between experimental data and the predicted results for the ice sphere’s radius versus time 
 \replaced[id=BP]{which}{whci} shows excellent agreement.
 A time step of 0.005 second was used in calculations. Lower values of a time step produced similar result. 
 In addition, the model has been applied for melting of a packed bed of ice particles. The water with the velocity of $0.01 ~m/s$ and the temperature of $320.15~K$
 enters from  the top of a box with dimensions of $300~mm$ in
width, $152 ~mm$ in depth and $100~mm$ in height, and flows through an assemble of particles. 
The water temperature distribution and the size of particles are presented in fig.~\ref{f:difftime} for different instances. 
 
The results include temperature distributions in both \replaced[id=BP]{fluid}{flid} and individual particles,
that additionally experience forces due to collisions and buoyancy. Thus, the 
model framework within the XDEM-suite takes into account all relevant physics
and predicts accurate results. 

\subsection{Powder Metallurgy}
Mining, forming and cutting tools as well as wear resistant tools are some of the 
most important products of the hard metal industry. One of the most employed 
materials in the hard metal industry is tungsten carbide (WC). Tungsten carbide 
is a hard composite that ensures resistance to wear, deformation and fracture to 
the demanding applications. During sintering of WC small grains may dissolve and 
larger grains grow resulting in an overall increase of the grain size. According 
to the sintering conditions, the grains tend to grow differently and some 
individual grains may even grow to an uncommonly large size compared to the 
average grain size. Abnormal large grains negatively impact most the 
mechanical properties of tungsten carbides, because they may act as initiation 
points for breakage. The WC grain size distribution is essentially determined by 
the grain size distribution of the employed tungsten powder \cite{Lassnerbook}. 
Thus, one of the major challenges in carbide technology is to control the grain 
size of tungsten powder effectively and accurately.

Tungsten powders are industrially produced by reduction of tungsten oxides, 
almost exclusively, via hydrogen reduction of tungsten blue oxide (TBO)\cite{ITIA_13}. 
The tungsten blue oxide is not a standard defined tungsten oxide with a chemical 
composition that depends on the production conditions from its precursor APT \cite{Lassnerbook}. 
The size of the reduced tungsten powder is governed by the prescribed humidity, 
temperature of the process, height and morphology of the tungsten oxides powder 
beds, the flow of the feedstock, the direction and flow of the reducing agents \added[id=BP]{and} 
the dew point \cite{Lassnerbook,ITIA_13,WuXiang_09}. The reducing hydrogen-flow 
does not only drive the reduction process, but it also removes the formed water 
vapour. The retention of water vapour within the powder is one of the most important 
factors to control because water vapour can react with the existing oxides forming 
the volatile compound $\rm WO_2(OH)_2$, which is transported and deposited in 
different localities \cite{Lassnerbook}; influencing directly the widening of the 
tungsten grain size distribution including abnormally large grains. Thus, controlling 
the water vapour production and retention during the reduction process is of pivotal 
interest and immediate industrial impact. In order to study this process, scientists 
have developed models to gain understanding and to try to control the phenomena. 
Nowadays, literature offers empirical models that rely on theoretical diffusion 
equations described in terms of temperature, time, diffusion-path lengths, and 
oxide bulk density to predict the extent of heterogeneous reactions. However, 
due to the lack of a comprehensive and accurate representation, industrial 
production today is still relies on empirical knowledge  \cite{Estupinan_thesis}.

In this section the XDEM-suite is employed to describe the tungsten powder 
production under industrial like conditions. Hydrogen reduction of $\rm WO_3$ to W-powder 
is considered a well-established process \cite{Lassnerbook,Luidold_07,Haubner_83a}. 
The chemical conversion of $\rm WO_3$ powder is represented by a staged-mechanism, 
reactions (8) to (11) \cite{Estupinan_14}, where oxygen is removed out of the system 
in the form of water vapour. 

\begin{eqnarray}
WO_{3}(s) + 0.1 H_2(g) &\rightleftarrows&   WO_{2.9}(s)+ 0.1 H_2O 
\label{e:reaction_700} \\
WO_{2.9}(s) + 0.18 H_2(g) &\rightleftarrows&   WO_{2.72}(s)+ 0.18  H_2O 
\label{e:reaction_710} \\
WO_{2.72}(s) + 0.72 H_2(g) &\rightleftarrows&   WO_{2}(s)+ 0.72  H_2O 
\label{e:reaction_720} \\
WO_{2}(s) + 2 H_2(g) &\rightleftarrows&  W(s)+ 2 H_2O 
\label{e:reaction_730}
\end{eqnarray} 

\noindent Each of the above-inserted reversible bi-molecular reactions can be 
treated as equilibrium bi-molecular and finite rate reactions. For such a reaction 
model the reaction rate, or the concentration rate of a species $k$, involved in 
an equilibrium reaction at temperature $T$ is represented by the following 
differential relation
\begin{equation}\label{e:dSdt}
\frac{{dc_k}}{{dt}} =  k_f(T) \cdot \left(
\nu_{k}' \cdot \prod\limits_{i = 1}^N { c_{R_i}^{\nu_{i}' } }  - \frac{\nu_{k}''}{K_{\text{eq,c}}(T)} \cdot \prod\limits_{j = 1}^M { c_{P_j}^{\nu_{j}'' } }\right)
\end{equation}
\noindent where $N$ denotes the number of reactants $R_i$, $M$ denotes the number of products $P_j$,  $\nu_{i/j}$ represents the absolute values of the corresponding stoichiometric coefficients and $k_f(T)$ denotes the Arrhenius coefficient ~\cite{Barrow_96}.

The $\rm WO_2$ powder employed for this study was prepared prior to experimentation. The selected tungsten oxide reduction step is the dominant part occurring during the industrial production of tungsten powders. Table~\ref{t:setup} summarizes the experimental as well as the numerical setup of the above-mentioned experiments.

\begin{table}[h]	
	\caption{Numerical properties selected during $\rm WO_2$ reduction simulation} \label{t:setup}
	\centering
	\begin{tabular}{l l l }%
		Powder\\
		\hline
		$\rm WO_{2}$				& 98		&\%			\\
		Size			& 35	&$\rm \mu m$		\\
		Total mass		& 100	&g		\\
		\\
		Reducing gas\\		
		\hline
		$\rm H_{2}$				& 99.9		&\%			\\	
		Volumetric flow		& 15		& $\rm Nm^3/s$		\\
		\\
		Powder bed\\
		\hline
		Height				& 10		& mm		\\
	\end{tabular}
\end{table}

\begin{figure}[ht]
	\centering
	\includegraphics[trim = 0mm 0mm 0mm 0mm, clip, angle=0, width=1\linewidth]{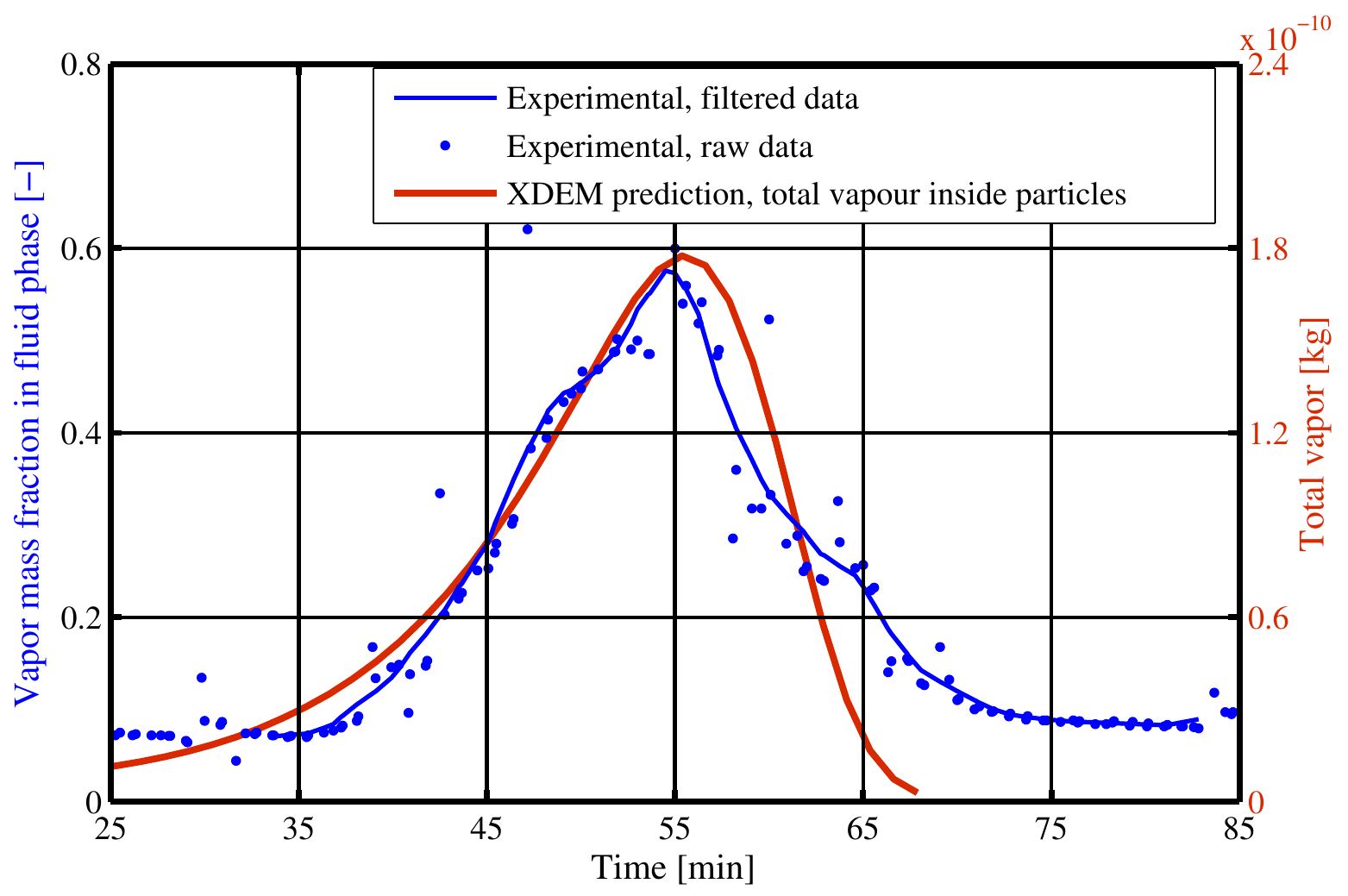}
	\caption{Comparison of water vapour predicted by XDEM-suite and experimental measurements in gas outlet. \added[id=AE]{The prediction for total water vapour in the powder bed is plotted in red and blue lines describe water vapour mass fraction measured at the outlet of the furnace.} }
	\label{f:wo2 validation}
\end{figure}

In order to show the predictive capabilities of the method, XDEM predictions are 
compared to measurements of \added[id=AE]{water} vapour. Fig.~\ref{f:wo2 validation} successfully 
validates the presented approach for the industrial reduction of tungsten oxides. 
\added[id=AE]
{In the figure, the experimental data of the water vapour mass fraction was taken from continuous humidity measurements in the outlet gases. The filtered data,  was obtained after passing the signal measurement through a simple moving-average filtering algorithm. In this case study, water vapour is a product of the hydrogen reduction of $\rm WO_2$. Hydrogen is applied in excess to drive the reaction progress and evacuate the produced water vapour. Consequently, and as described in eq.~\ref{e:reaction_730}, the concentration of water vapour indicates the oxygen removal, and therefore, the reaction progress.
	  Due to the high gas flow, industry assumes that outlet measurements are highly representative of the water concentration inside the furnace, and consequently, indicates the reactions progress. 
The high correlation between measurements and predictions not only validates the numerical predictions, but also indicates that outlet measurements are consistent with the intra-furnace reduction process.}

\noindent \replaced[id=BP]{These}{This} results as well as previous validations documented in 
\cite{Estupinan_thesis,Estupinan_14,Estupinan_15c} anticipate that employing the 
XDEM methodology combined with experimental data provides important breakthroughs 
for the powder metallurgy industry.

\subsection{Thermal Treatment of Biomass}

\begin{figure}[h]
\centering
\includegraphics[width=8cm, angle =0, trim={2cm 4cm 2cm 3.5cm}]{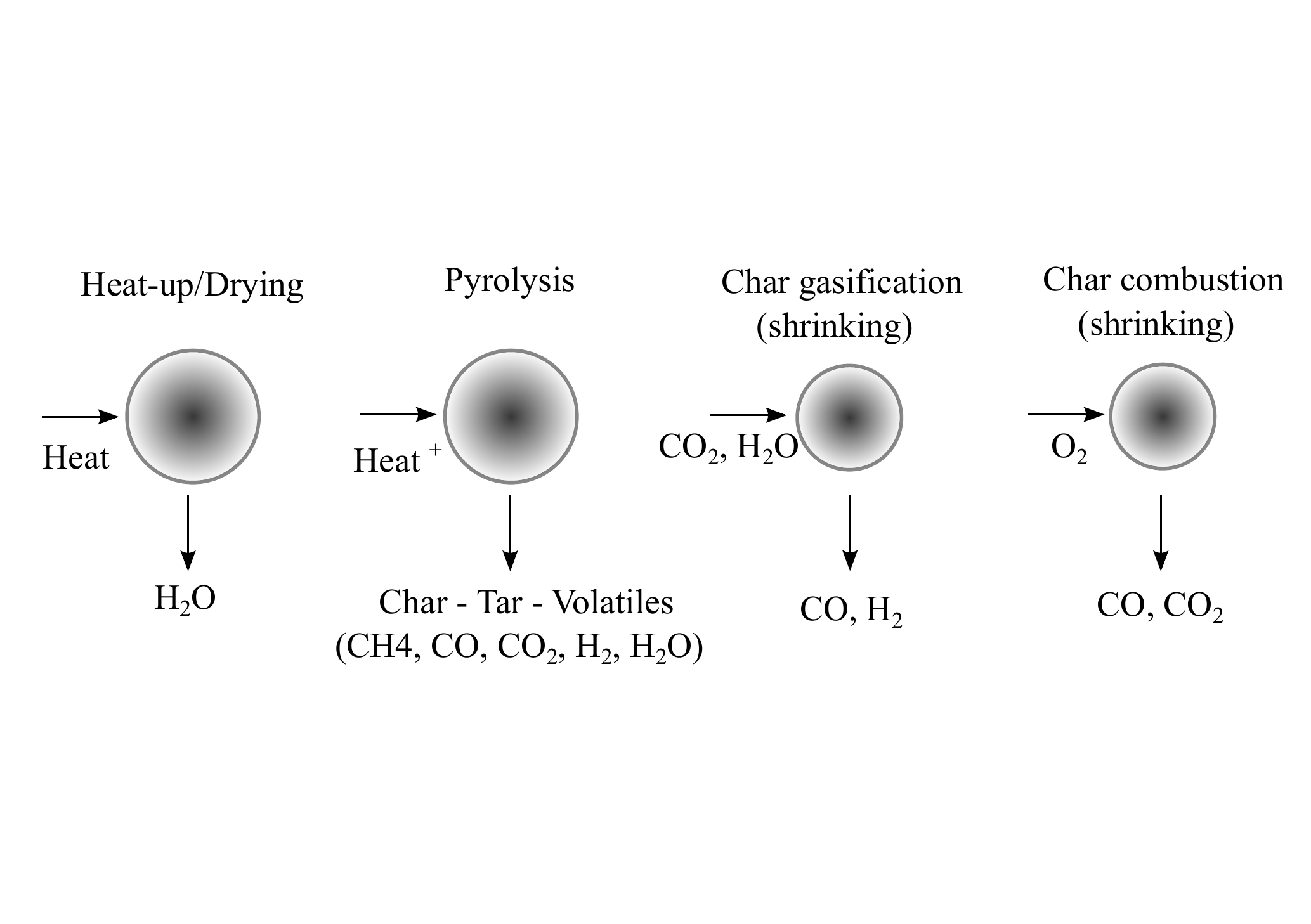} 
\caption{Main steps of conversion process}
\label{BiomassCombustionMechanism}
\end{figure}

Thermal treatment of biomass refers to the conversion of biomass when the essential heat is available in the system.
In this case, physical and chemical processes shown in fig.~\ref{BiomassCombustionMechanism} as heat-up, drying,
pyrolysis, combustion and gasification take place. 

Heat-up is warming biomass to increase the temperature above the ambient temperature.
When the surface temperature of biomass reaches the saturation status (around $100^\circ~C$ at $1~bar$),
the water contents of biomass start to vaporize whereby the drying process begins. After removing the water content and with increasing heat,
the biomass starts to decompose thermally to solid char, liquid
tar and volatile gases ($CH_4, CO, CO_2, H_2, H_2O$). This stage is pyrolysis or devolatilization process.
The released heat from reactions between the volatile gases and oxygen is used for drying and pyrolysis of new feeding biomass 
fuel to the reactor and producing further volatiles so that this cycle continues as far as the oxygen is provided 
to the system. The remaining char from thermal decomposition is gasified via reacting with $CO_2$ and $H_2O$ to generate 
useful gases of $CO$ and $H_2$ which are highly combustible. Also char is oxidized at high temperature above $700^\circ~C$ denoting 
the combustion process when the solid char reacts with oxygen. \cite{Mohseni2017}

Here the XDEM-suite is applied to validate the drying of a single particle with the experiment carried out
by Looi et al. \cite{Looi}. The particle is considered as a spherical wet particle including $60~\%$ of moisture content 
and diameters $10,12~mm$ defined as cases \textbf{A} and \textbf{B}.
The experiments are performed at the pressure of $2.4~bar$, and the superheated steam temperature at 
$170 ^\circ~C$, as well as the steam velocity at $2.7~m/s$. 

\begin{figure}[h]
\centering
\includegraphics[width=11.5cm, trim={0 2cm 0 2cm}]{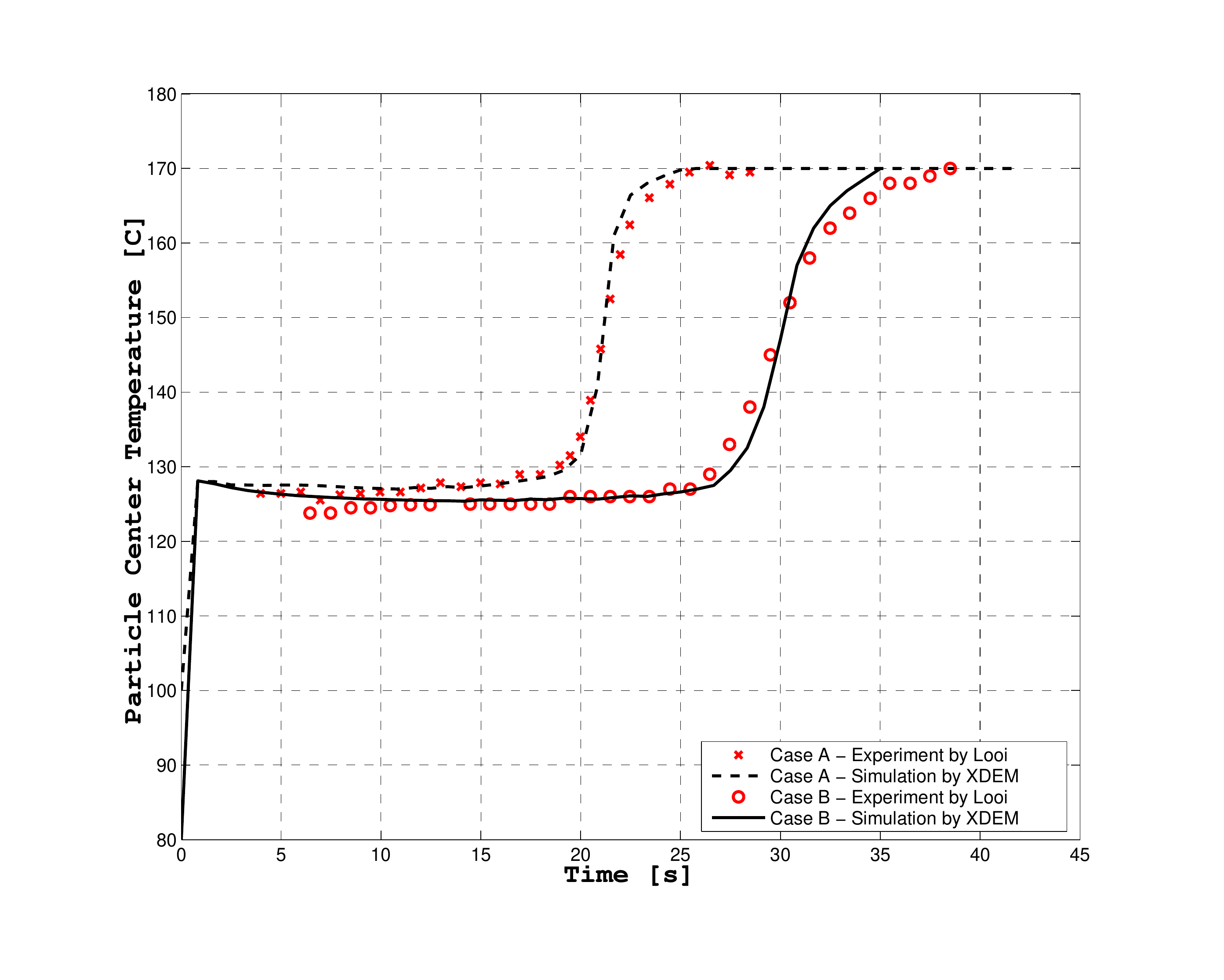}  
\caption{Particle size effect on drying of a coal particle---
Case A: $d_p=10mm$, $P=2.4bar$, $T_a=170^\circ C$, $v_g=2.7m/s$; 
Case B: $d_p=12mm$, $P=2.4bar$, $T_a=170^\circ C$, $v_g=2.7m/s$ }
\label{validation_07_10}
\end{figure}

The prediction of particle core temperature compared to experiment is shown in fig.~\ref{validation_07_10}
that shows a very good agreement.

\begin{figure}[h]
\centering
\includegraphics[width=11.5cm, trim={0 0.3cm 0 1.5cm}]{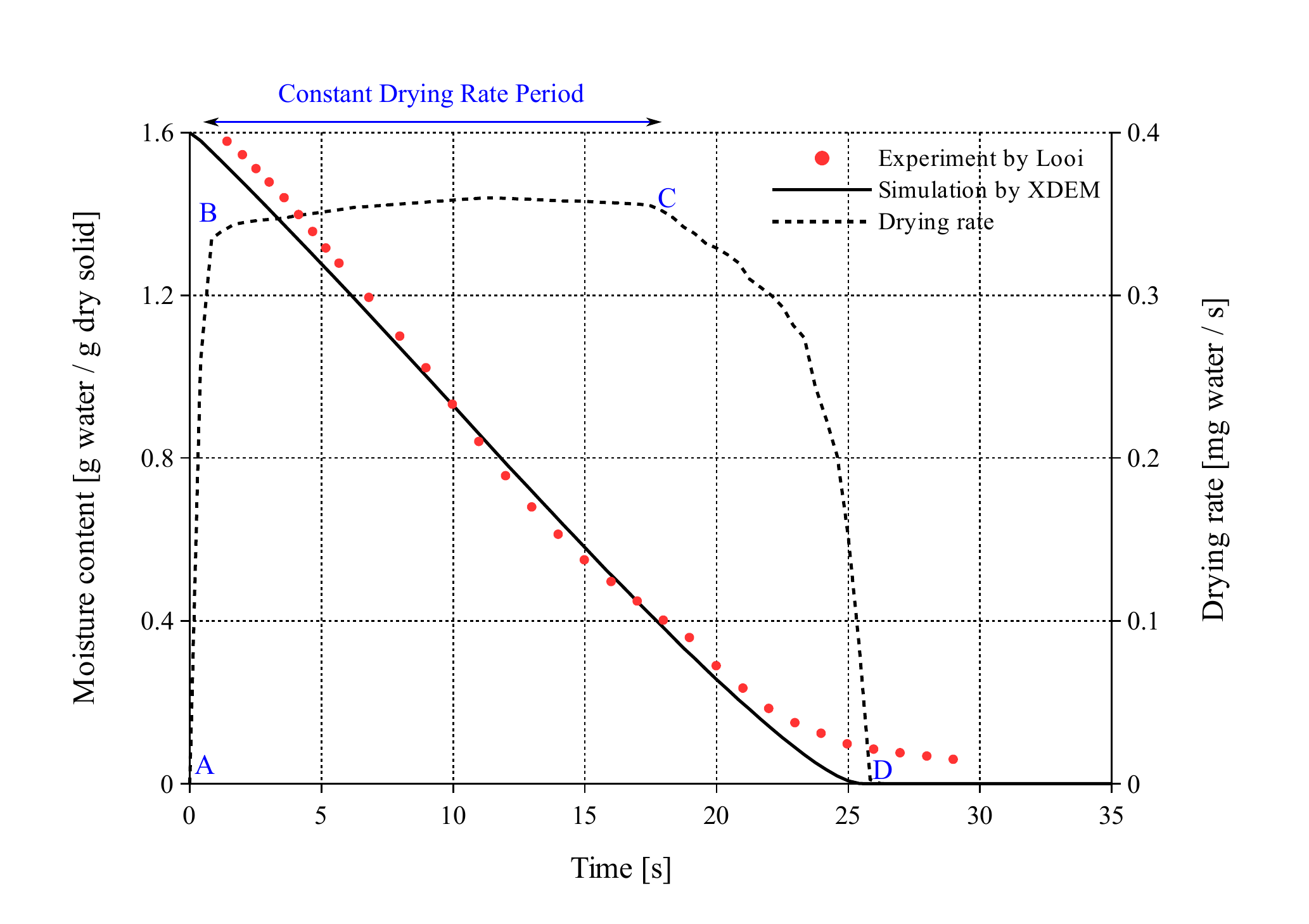}  
\caption{Drying rate and moisture content vs. time---
case \textbf{B}: $d_p=12mm$, $P=2.4bar$, $T_a=170^\circ C$, $v_g=2.7m/s$ }
\label{Exp_dryingRate}
\end{figure}

In case \textbf{A}, the temperature rises from saturation after $20~s$ while in case \textbf{B} it rises after $27~s$ which
states that the small particle is dried faster. The reason is in the small particle, heat transfer from the surface
to core of particle is faster. However, the temperature of particles in both cases reaches the gas
temperature ($170 ^\circ~C$) at the end of drying. 
In addition, fig.~\ref{Exp_dryingRate} shows the behaviour of drying rate based on the moisture content for
case \textbf{B}.
The \textit{AB} part on the curve represents a warming-up period of the particle. The \textit{BC} part is the 
constant-rate period where the particle temperature is constant during this period since heat transfer into 
the surface is constant and the heat is consumed just for water evaporation. When the free water is finished, the 
temperature starts to rise. This point is shown as point \textit{C}, where the constant-rate ends and the 
drying rate begins falling that is termed the critical-moisture content. The curved portion \textit{CD} is 
termed the falling-rate period and is typified by a continuously changing rate throughout the remainder of 
the drying cycle. In addition, fig.~\ref{Exp_dryingRate} depicts the predicted residual moisture content
which agrees well with experimental data \cite{CES_MM}.

\clearpage

\subsection{\deleted[id=GP]{Turbulent Particle-laden Flows}Dual-grid multiscale simulations of three-phase flows}

\added[id=GP]{The dual-grid multiscale approach for three phase flows was introduced in~\cite{PozzettiIJMF,pozzettiicnaam2016}
and applied to several flow configurations ranging from conventional process engineering~\cite{pozzettipowdermet}
to additive manufacturing~\cite{peters2017Flow10993-31734}. It consists in the identification of two length-scales: a bulk scale
where the coupling between CFD and DEM domain is preformed, and  a fluid fine scale at which the fluid equations
are resolved. One CFD grid is associated to each scale, making the grid used for the solution of the fluid flow equations
independent from the particles' characteristic dimension.
In~\cite{PozzettiIJMF} was shown how, by adopting a correct interpolation strategy between the two grids, the multiscale
approach can produce accurate, grid-convergent results when the standard DEM-VOF method cannot.
}
\subsubsection{\added[id=GP]{Laminar three-phase dam-break}}
\begin{figure}[ht!]  
\centering
\includegraphics[width=0.5\textwidth]{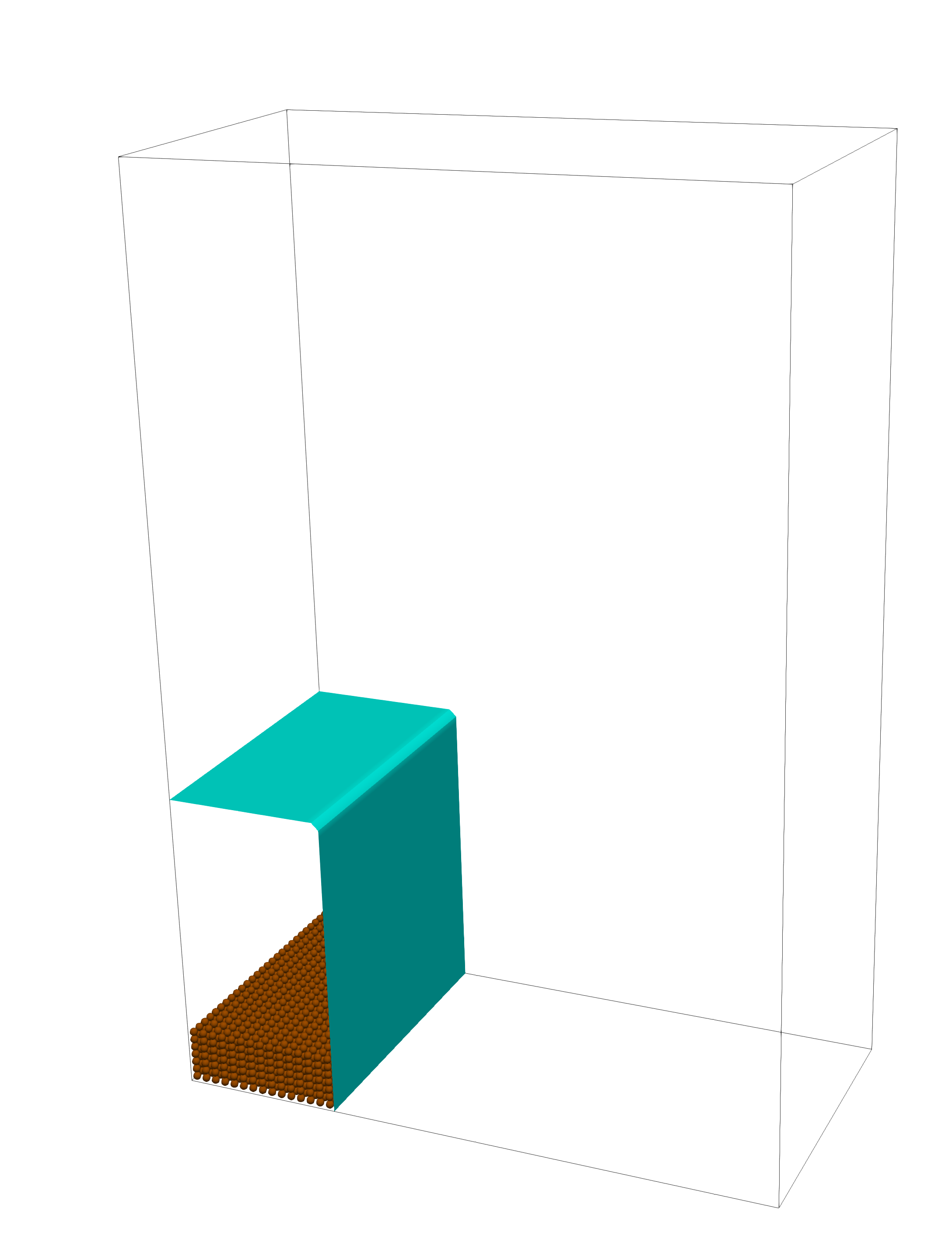} 
 \caption{\label{DBLaminar_setup}Dual-grid multiscale simulation of three-phase flows: Laminar dam-break setup as in~\cite{PozzettiIJMF} .}
\end{figure}
\begin{figure}[ht!]  
\centering
\includegraphics[width=1\textwidth]{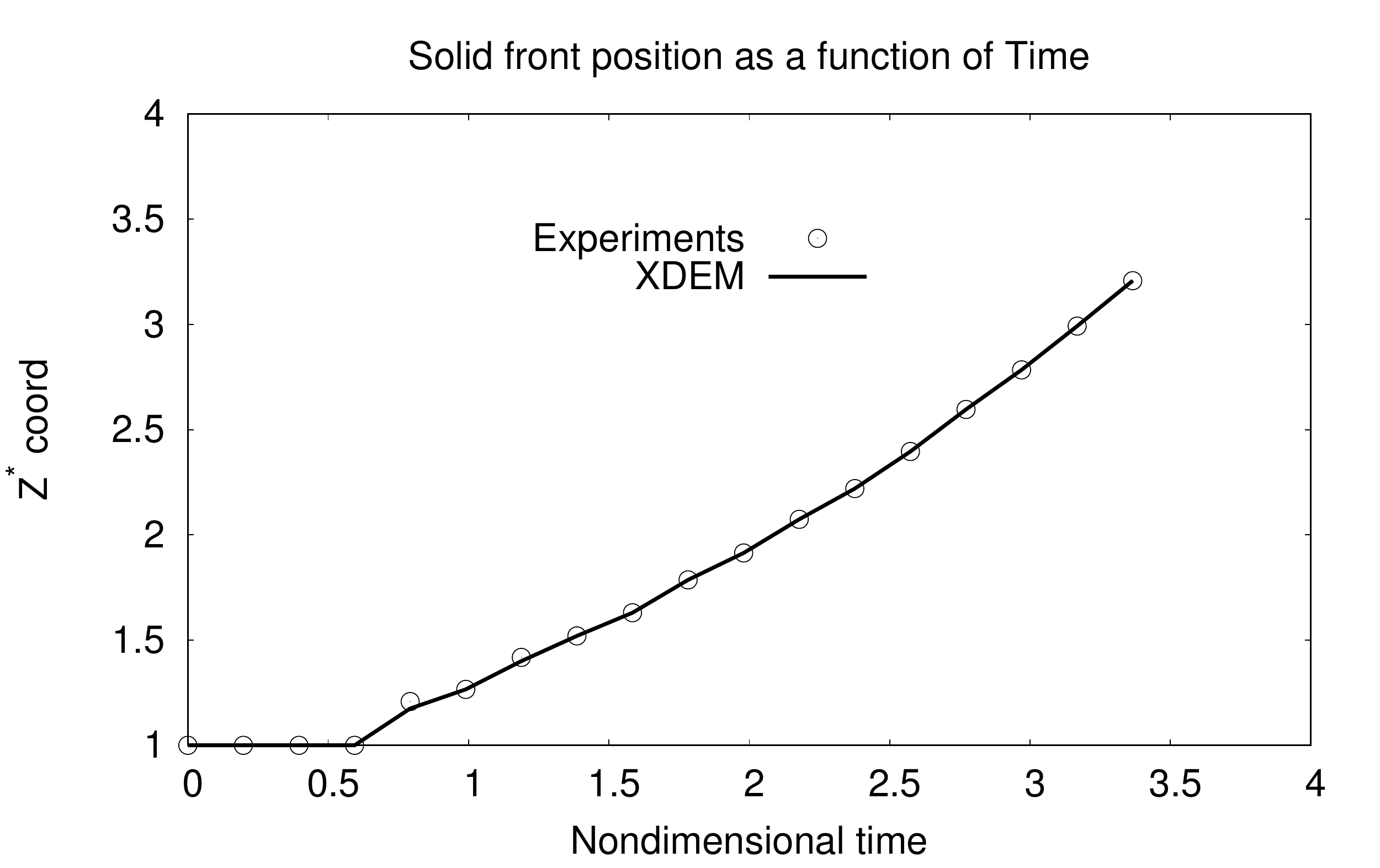} 
\includegraphics[width=1\textwidth]{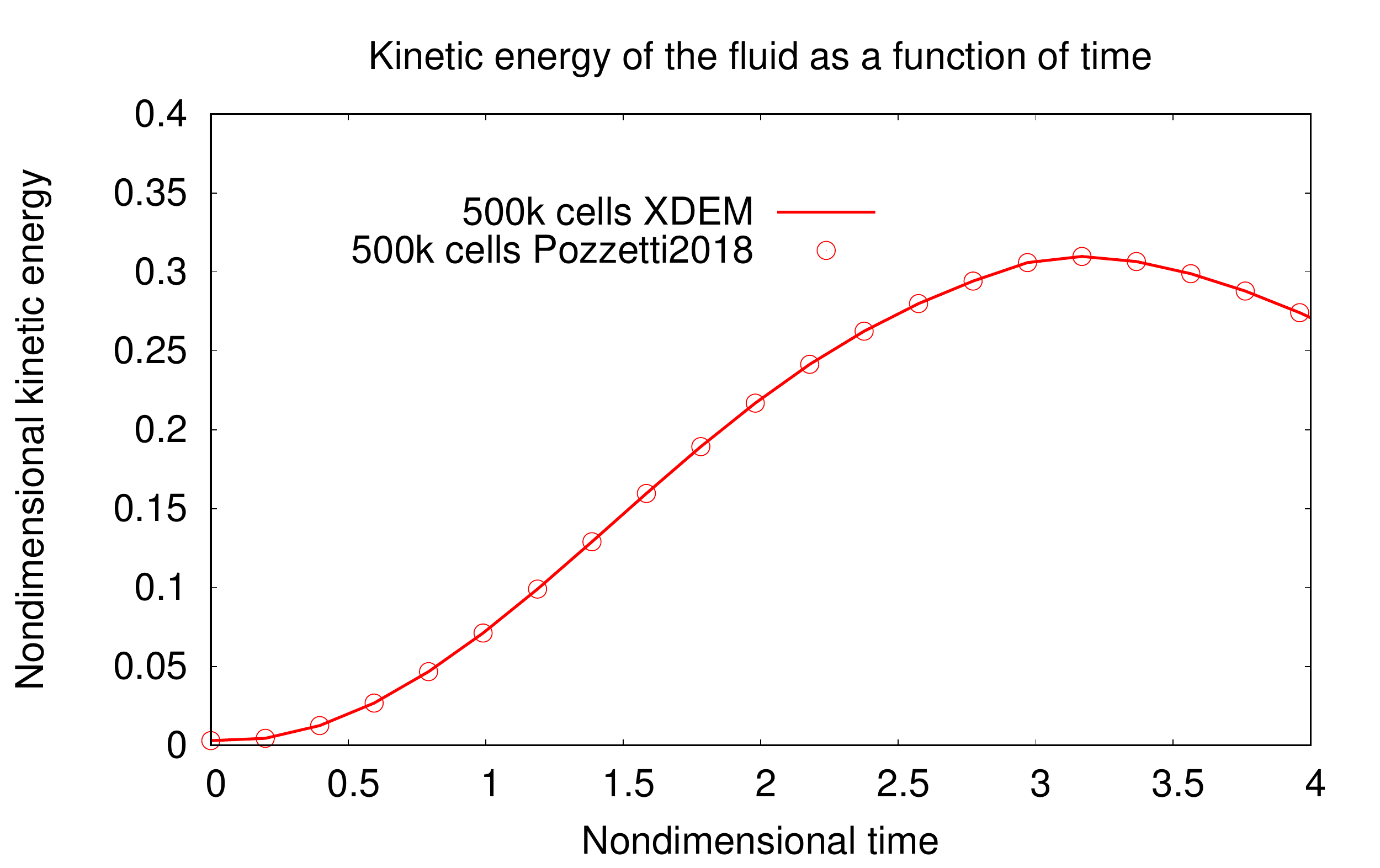}
 \caption{\label{ConfrontIJMF}Dual-grid multiscale simulation of three-phase flows: Comparison between 
the reference results in~\cite{PozzettiIJMF} and the implementation within the XDEM platform.}
\end{figure}
\added[id=GP]{The dam break is a famous benchmark for multiphase flows, that was proposed in various configurations by
different authors~\cite{KLEEFSMAN}. In particular its extension to three-phase flows has been of great interest for civil/chemical engineering.}
\added[id=GP]{We here re-propose a dam-break benchmark as introduced in~\cite{PozzettiIJMF} and citations within,
resolved with the dual-grid multiscale DEM-VOF solver of the XDEM-platform.
The benchmark features a box of dimensions  $0.2 m$ x $0.1 m$ x $0.3 m$,
where column of water of $0.05 m$ x $0.1 m$ x $0.1 m$, and an uniformly 
distributed bed of spherical particles at the bottom,
 breaks and stabilizes.
The spheres have a diameter of $2.7 \, mm$.
As already done in~\cite{PozzettiIJMF} we refer to and adimensional kinetic energy: }
\begin{equation}
\mathbf{E_{kin}}=\frac{\int_D \mathbf{u}_f \cdot \mathbf{u}_f \, \rho_f 
\frac{1}{2} 
dV}{\int_{Z_{col}} A_{in} \rho_h g dZ}.
\end{equation}
The nondimensional time is defined as 
\begin{equation}
{t^* =t  \left(\frac{2 g}{a}\right)^{\frac{1}{2}},}
\end{equation}
with {$a=0.05m$}.
\added[id=GP]{In figure~\ref{ConfrontIJMF}, one can observe how the benchmark is satisfied by the XDEM-platform in terms of accuracy
in the prediction of the experiments, and of convergence to the reference value of the kinetic energy.}

\subsubsection{\added[id=GP]{LES approach to turbulent flows}}
\added[id=GP]{
Industrially-significant flows often have to deal with high velocities.
This represents a major problem for the numerical approach to such flows as 
the non-linearity of the Navier-Stokes equations can trigger physical and numerical instabilities~\cite{POPE2000}.
Different numerical approaches have been proposed to cope with this problems, for a detailed description of which
the reader is referred to~\cite{POPE2000} and citations within.
For a numerical treatment of turbulent phenomena within multiphase flow,
the Large Eddy Simulation (LES) approach have been found to hold interesting advantages~\cite{toutant2008877}.
}
The LES approach aims to compute exactly the large, energy-carrying structures 
of the 
flow, while modeling the smaller scales whose behaviour is accepted to be less 
local \cite{POPE2000}.
This is obtained by a filtering operation applied on every field variable
\begin{eqnarray}
\tilde{\mathbf{y}}(\mathbf{x})=\int_A \mathbf{y}(\mathbf{x}') \Phi 
(\mathbf{x},\mathbf{x}',\tilde{\Delta}) d\mathbf{x}',
\label{LESfiltering}
\end{eqnarray}
where $A$ is the domain of the simulation, $\Phi $ a generic filter, with
$\tilde{\Delta}$  the filter width.
This allows to ideally divide the velocity vector field as
\begin{eqnarray}
\mathbf{u}_f(\mathbf{x})=\tilde{\mathbf{u}_f}(\mathbf{x})+\mathbf{u}_f'(\mathbf{x}),
\end{eqnarray}
with $\tilde{\mathbf{u}_f}(\mathbf{x})$ the filtered term as in \ref{LESfiltering}, 
and $\mathbf{u}_f'(\mathbf{x})$ the remaining contribute whose effects need to be 
modelled.

Classical turbulence models effectively succeeded in describing the effects of 
unresolved structures on the resolved ones. If a dispersed phase is present in 
the flow the effect of the unresolved structures on the particle may be addressed 
by reconstructing a stochastic 
unresolved velocity field using information about the turbulent kinetic energy
\cite{:/content/aip/journal/pof2/11/10/10.1063/1.870162,
:/content/aip/journal/pof2/8/5/10.1063/1.868911,Zhou20044193,Burgener}.
The XDEM platform allows to perform coupling between a discrete phase and  a 
LES solution both considering only the resolved structures as relevant for the 
flow,
and therefore using the filtered velocity as coupling variable, or taking into 
account an average 
effect of the unresolved scales by partially reconstructing a sub-grid velocity 
field.

The partially reconstructed velocity field can be used in order to calculate 
the drag force in the form 
\begin{eqnarray}
\tilde{\mathbf{u}_f'}=&\tilde{\mathbf{u}_f'}(\mathbf{k}), \\
\tilde{\mathbf{u}_f} =& <\mathbf{u}_f>+\tilde{\mathbf{u}_f'},\\
F_{drag} =& F_{drag}(\tilde{\mathbf{u}_f}),
\end{eqnarray}
with $\mathbf{k}$ the turbulent kinetic energy
One of the simplest reconstruction procedure can be obtained by
\begin{eqnarray}
\tilde{\mathbf{u}_f}_i'=& \sqrt{\frac{2}{3}\mathbf{k}} \,  \mathbf{\psi}_i \, 
for \,i=1,..,3,
\end{eqnarray}
with $\psi_i$ being a white noise. This assumes the subgrid-scales to be 
perfectly isotropic and the different components of $\tilde{\mathbf{u}_f}$ to be 
uncorrelated. 

\begin{figure*}[!htb]
\centering
\includegraphics[width=0.3\textwidth]
{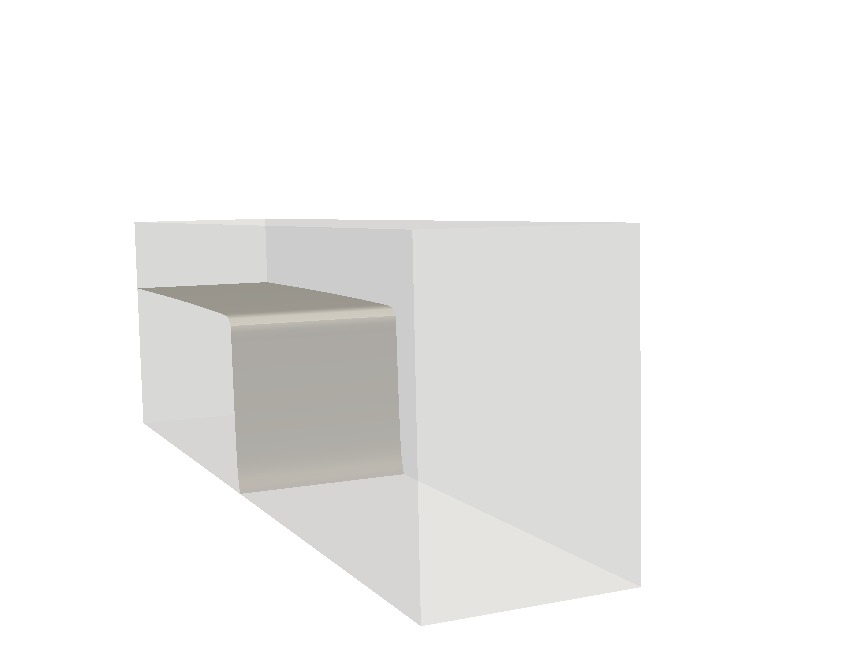}   
\includegraphics[width=0.3\textwidth]
{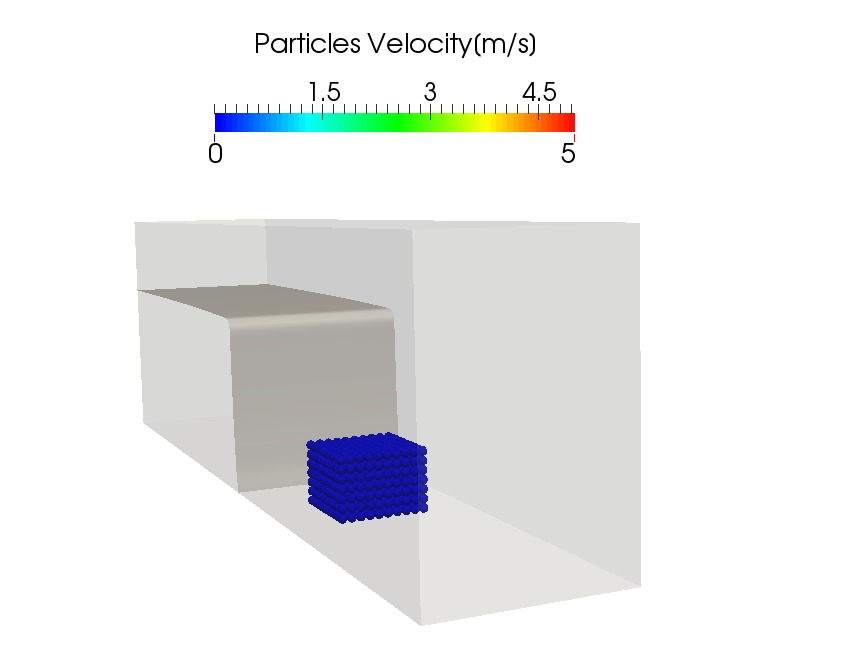}   
\includegraphics[width=0.3\textwidth]
{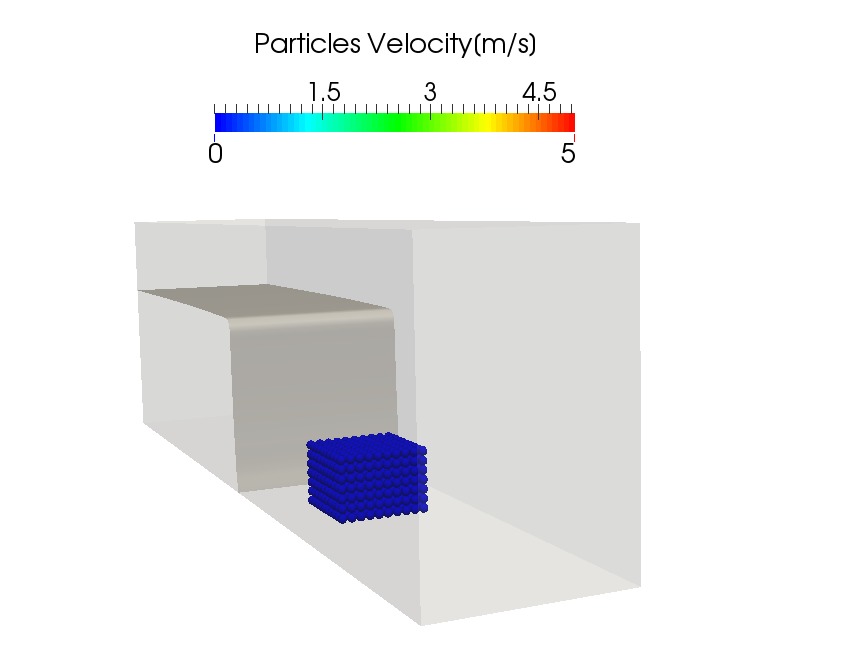} \\ 
\includegraphics[width=0.3\textwidth]
{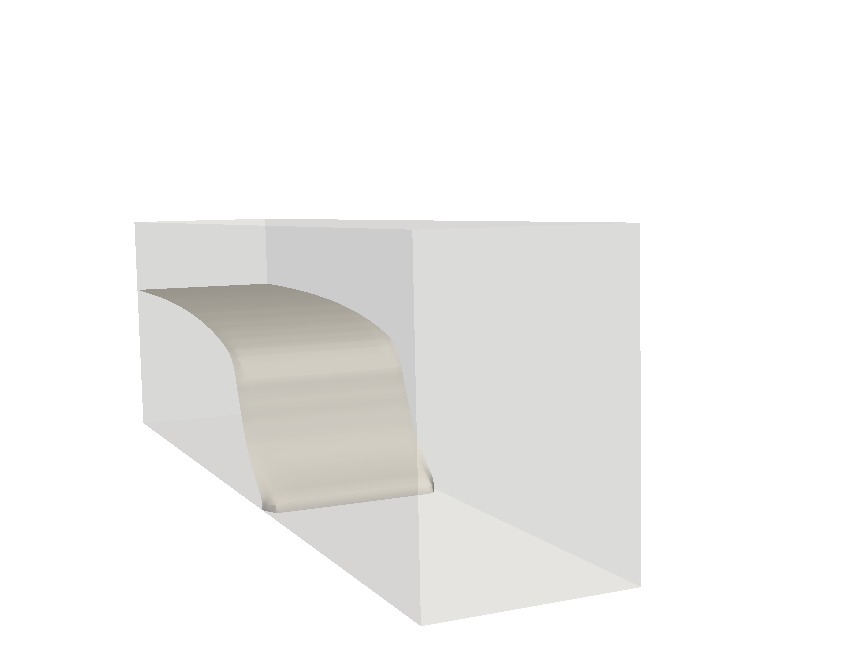} 
\includegraphics[width=0.3\textwidth]
{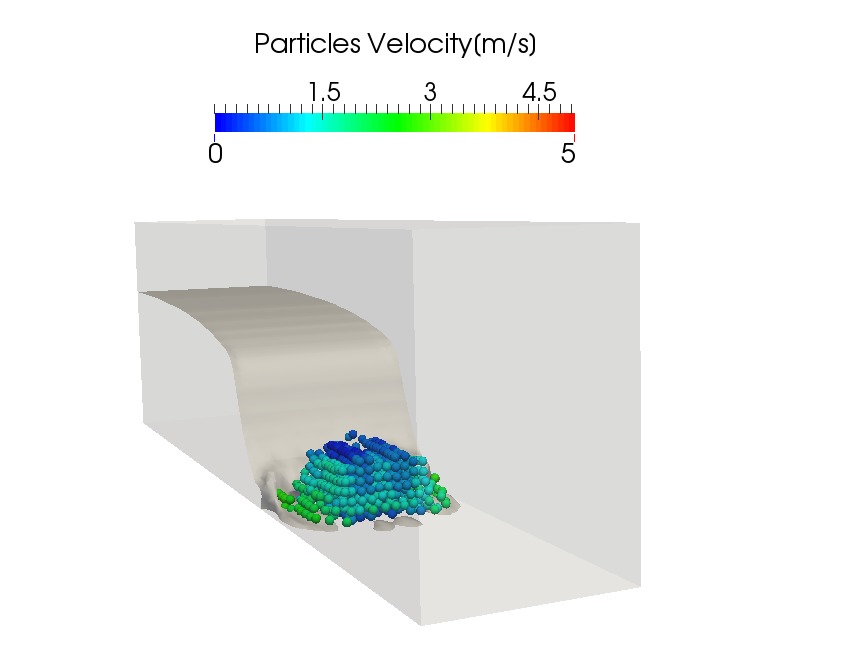}  
\includegraphics[width=0.3\textwidth]
{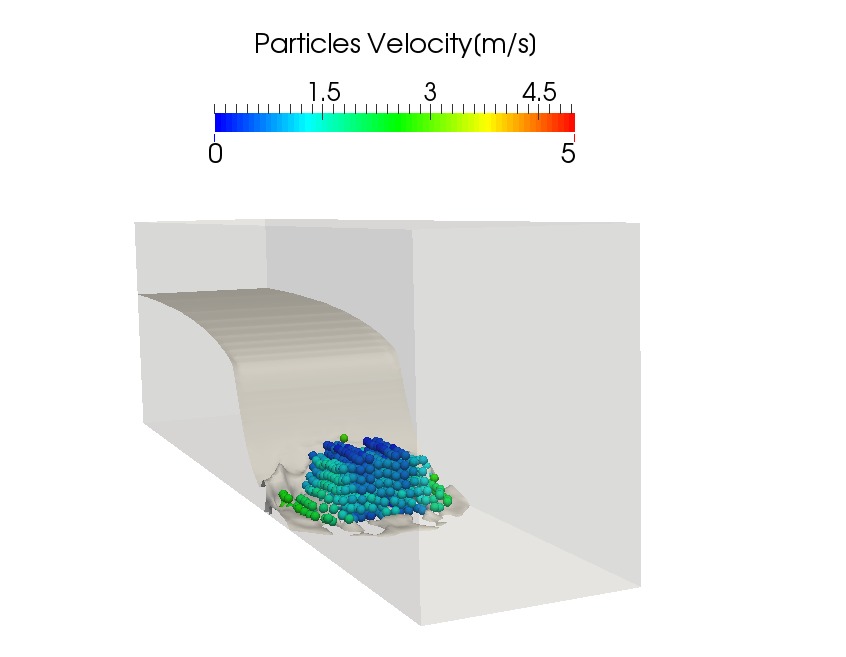} \\ 
\includegraphics[width=0.3\textwidth]
{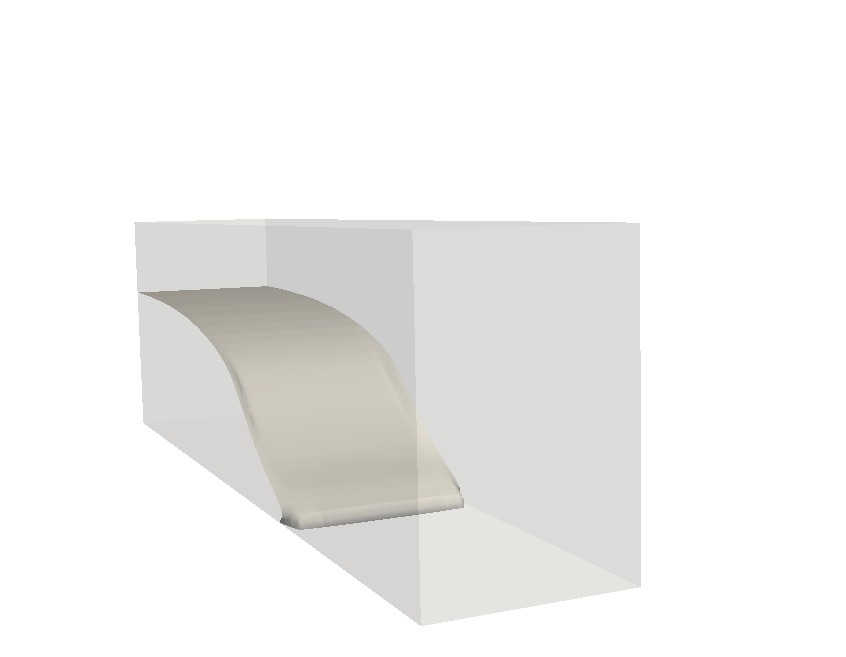} 
\includegraphics[width=0.3\textwidth]
{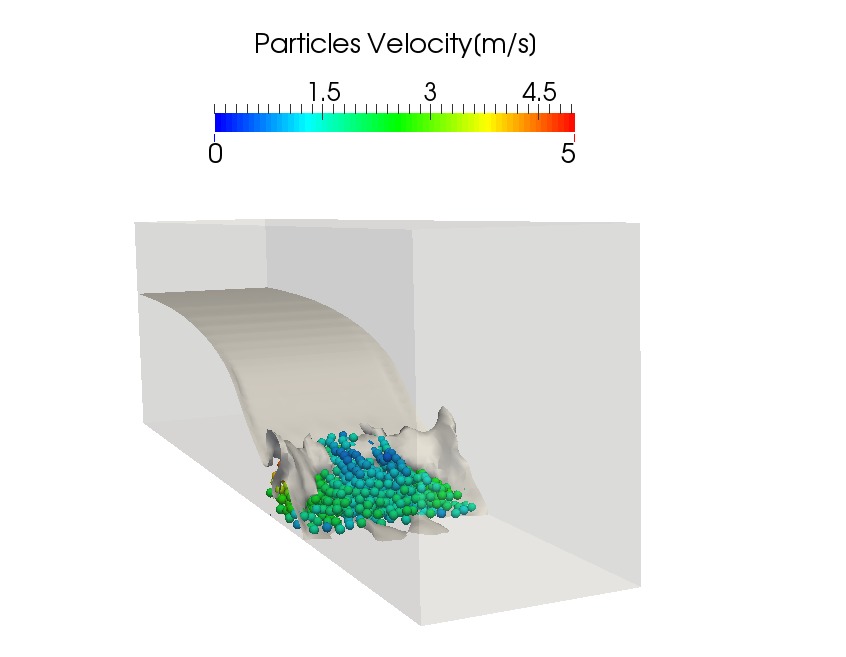} 
\includegraphics[width=0.3\textwidth]
{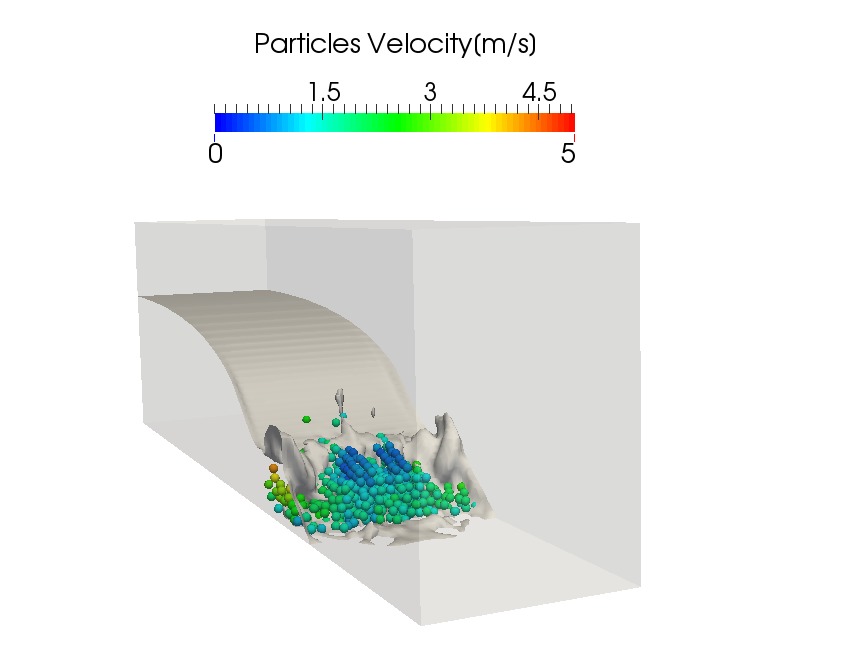}\\  
\includegraphics[width=0.3\textwidth]
{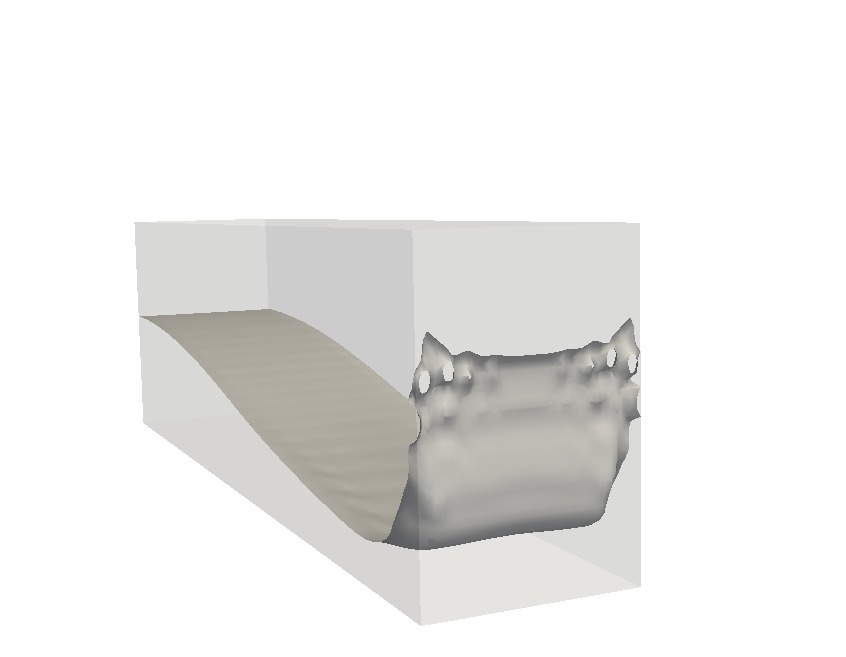}  
\includegraphics[width=0.3\textwidth]
{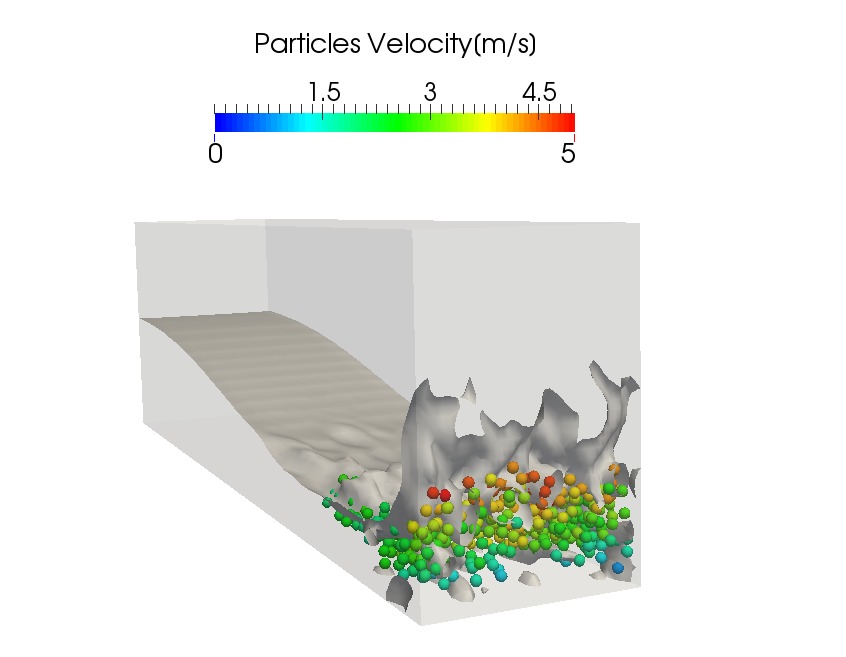} 
\includegraphics[width=0.3\textwidth]
{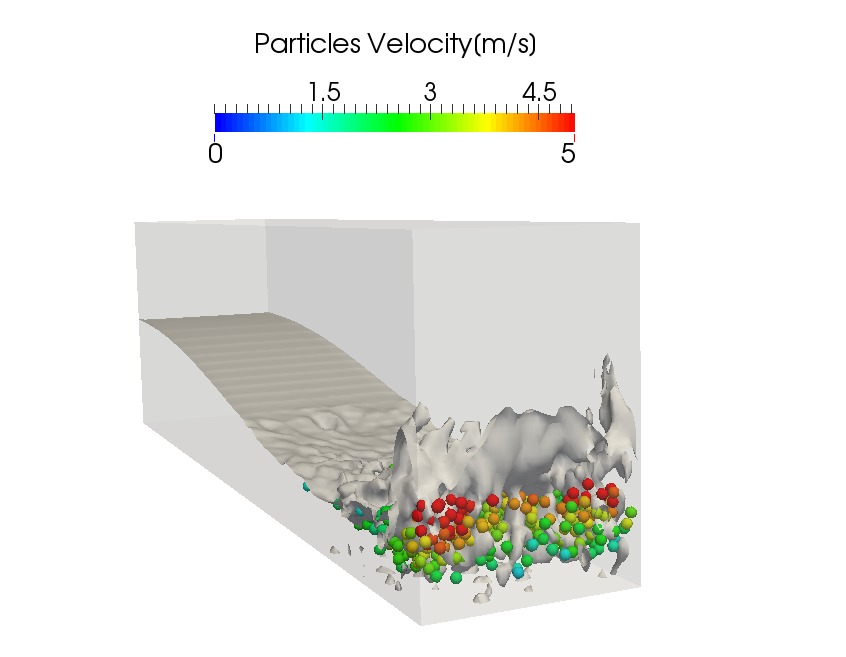}
 \caption{\label{DamBreakAgainstPBIG} Dam breaking against a pile of particles 
significant time step. Comparison between a Dam without obstacles (left), and 
with obstacle and subdomain discretization of $150k$ elements 
(middle) and $500k$ elements(right).
From the top to the bottom are shown times of $0.1s$, $0.35s$, $0.50s$, $0.85s$.}
\end{figure*}

The standard two-phase dam break usually involves an obstacle against which the 
water column is breaking.
We propose a similar benchmark test by substituting the obstacle with a pile of 
particles.
The water column is breaking against the pile of particles, lifting them, and 
transporting them in a 
slurry flow that impacts on the wall.
The setup for the simulations include a domain size of  $10 \, m \,$x $\, 2 \, m\,$ x $\, 3\, m$  initially the 
water is at $ 5 \, m\,$x $\, 2 \, m\,$ x $\, 2\, m$.
A block of $700$ particles of $10\, cm$ diameter, consisting in seven layers of 
$10$ x $10$ reticulates is posed at $1 \,m$  of distance from the dam.
The particles density is fixed at $800 \, kg/m^3$. 
Considering the actual dimensions of the domain the problem could not be solved
with laminar assumptions. The LES scheme described in the previous section 
is used.
We propose results for a sub-domain discretisation of $150k$ and $500k$ 
hexahedral piecewise constant elements.
In figure \ref{DamBreakAgainstPBIG} the evolution of the dam break is shown.

\begin{figure}[ht!]  
\centering
\includegraphics[width=1.0\textwidth]{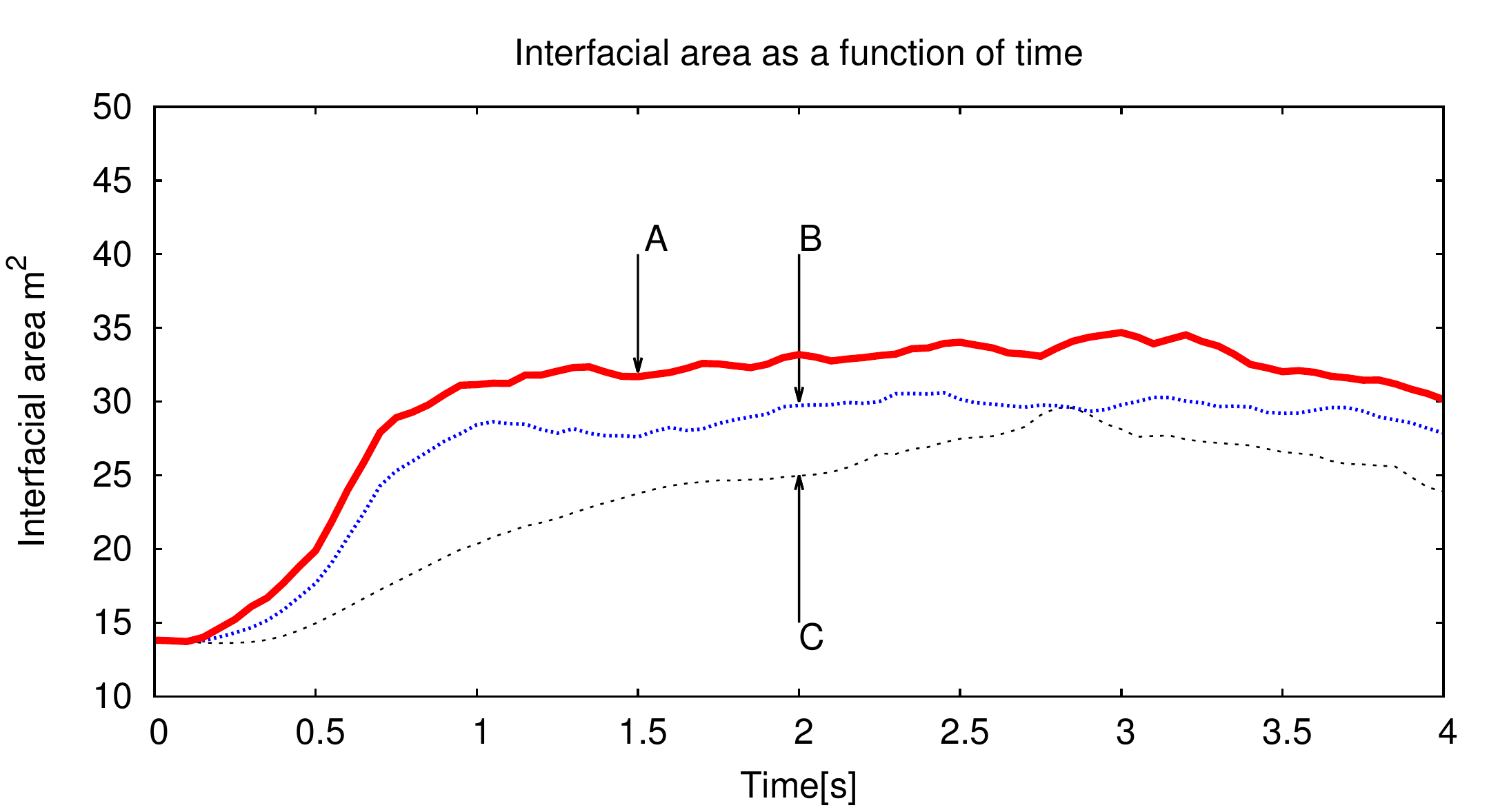} 
 \caption{\label{DAM_Obstacle_area}Turbulent Dam Break: Interface area as a 
function of time, Comparison between 
a sub-domain discretisation of $500k$ elements (A), $150k$ elements(B), 
and a dam without obstacle (C).}
\end{figure}

Three configurations are compared: the dam in absence of particles, and at 
the presence of particles  with different domain discretisation.
It can be noticed how the flow configuration changes completely at the presence 
of the particle,
in particular the water partially filters through the bed while lifting the 
particles.
One can observe how the discretisation of the domain changes not only the 
resolution of the liquid-gas interface, but as well the  velocity distribution 
of the particles.
In fig.~\ref{DAM_Obstacle_area} the interface area as a function of time is 
shown for the three cases.

\begin{figure}[ht] 
\centering
\includegraphics[width=1.0\textwidth]{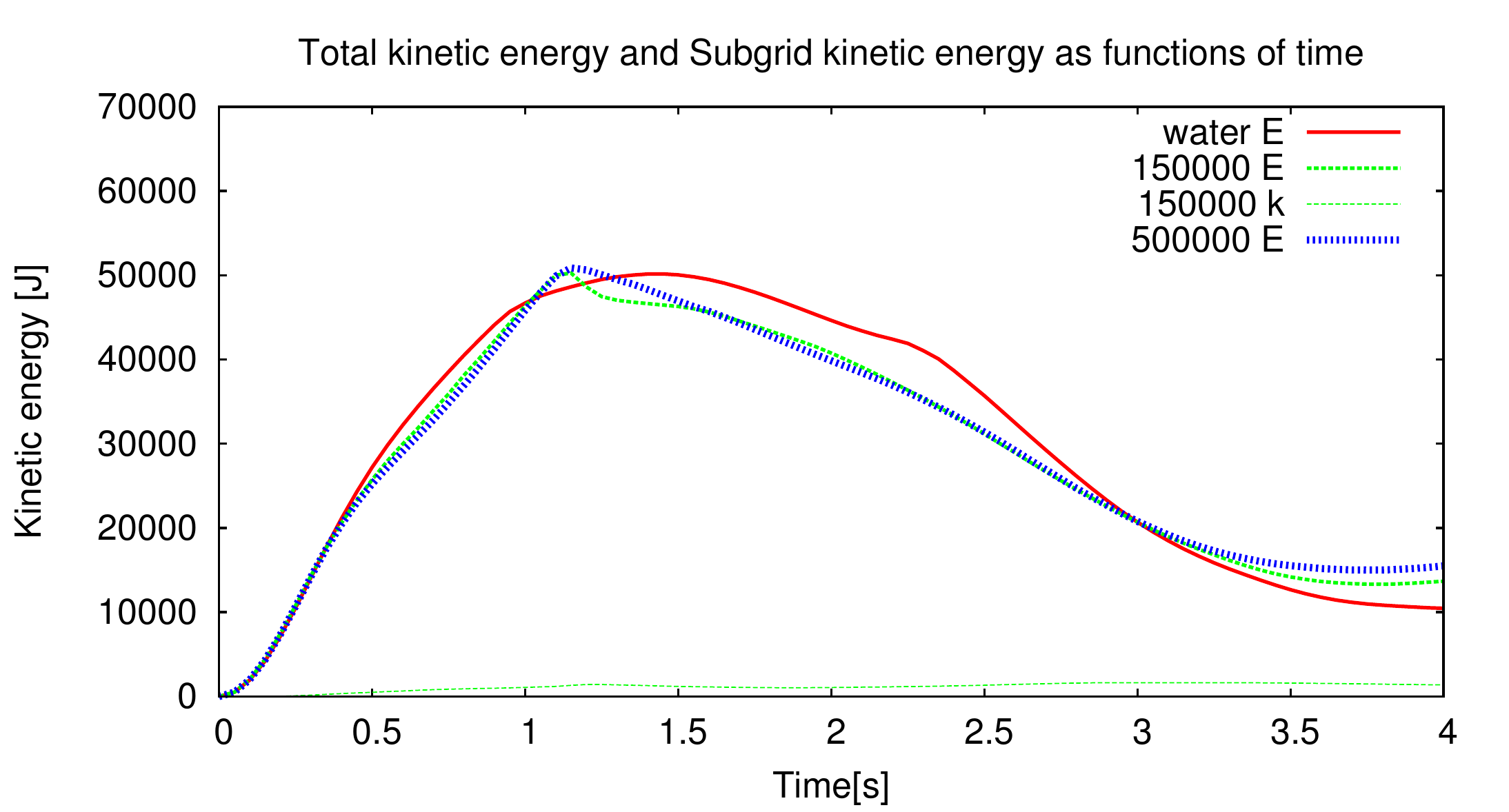} 
 \caption{\label{DAM_Obstacle_Turbulence}Turbulent Dam Break. Total kinetic 
energy (E) as a 
function of time, Comparison between 
a subdomain discretization of $500k$ elements (500000), $150k$ 
elements(1500.000), 
and a dam without obstacle (water). The sub-filter kinetic energy (k) for the 
discretization of $150k$  elements is shown as  a comparison term.}
\end{figure}

\begin{figure}[ht]  
\centering
\includegraphics[width=1.0\textwidth]{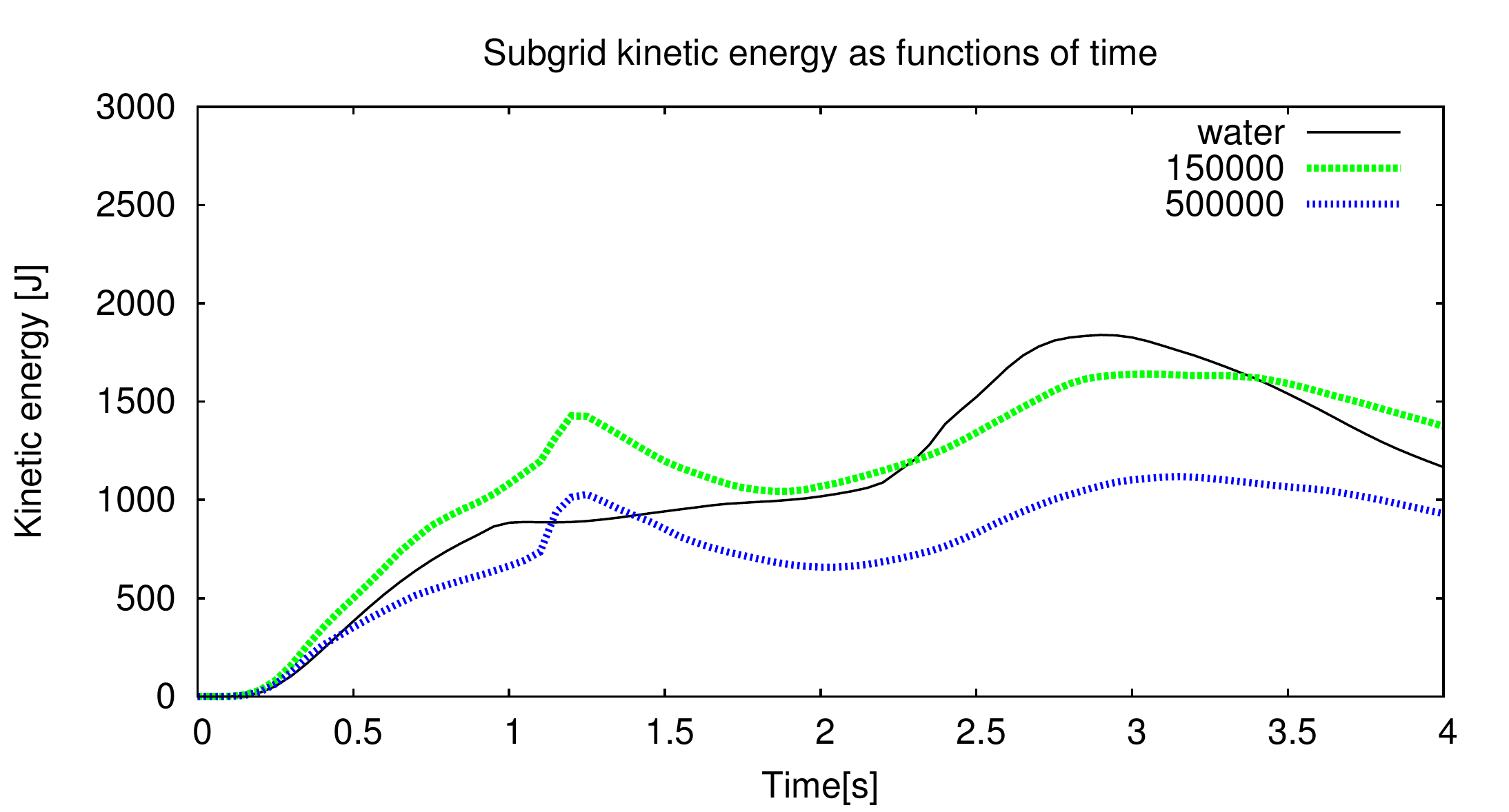} 
 \caption{\label{DAM_Obstacle_k_turb}Turbulent Dam Break. Total 
sub-filter kinetic energy (k) as a 
function of time, Comparison between 
a subdomain discretization of $500k$ elements (500000), $150k$ 
elements(150000), 
and a dam without obstacle (water).}
\end{figure}

It can be noticed how the presence of the particle obstacle significantly 
changes the qualitative behaviour of the dam breaking, causing a premature 
rupture of the interface (approximatively around $t=0.5 s$) with consequent 
increment of the interface area.
At the same time the usage of a finer discretisation allows to capture smaller 
structures and therefore leads to an higher interface area prediction.
In figure \ref{DAM_Obstacle_Turbulence} and  \ref{DAM_Obstacle_k_turb} the 
profiles for the total kinetic 
energy and the sub-filter kinetic energy of the system are presented.

It can be noticed how the total kinetic energy of the fluid system is 
influenced by the particle presence. In 
particular areas where the particles are storing kinetic energy from the fluid
(and the profiles are are under the curve relative to the water at $t=0.5-0.9$ 
and $t=1.2-2.7$) and areas  
where the particles are releasing kinetic energy into the fluid 
(approximatively for times $t=1.1-1.3s$ and $t=3-4s$) can be observed. 
With reference to fig.~\ref{DAM_Obstacle_Turbulence} the turbulent kinetic 
energy appears correctly bounded. Looking at fig.~\ref{DAM_Obstacle_k_turb} one 
can notice how the presence of the particle 
influences the turbulence kinetic energy profile.
Comparing the behaviour of the system with only water and the one with 
particles and a discretisation of  $150000$ element (that is also the one 
adopted for the dam without particles), a qualitatively different time 
distribution of the sub-filter kinetic energy is observed. In particular during 
the impact with the wall at $\mathbf{x}=10m$ a higher subgrid kinetic energy 
can be observed when particles are present in the system. In reverse during the 
second impact with the wall at $\mathbf{x}=0$ the sub-filter kinetic energy is 
higher for the pure fluid, while the presence of the particles gives now to 
the fluid a more viscous behaviour.
By comparison between the two different discretisation of the fluid-particle 
system,
 one can notice how the contribution of the sub-filter kinetic energy decrease 
when a finer 
discretisation is adopted, while the qualitative behaviour remains similar. 
This is in accordance to the general LES theory 
since a finer discretisation allows to resolve more scales, and therefore a 
bigger part of the total kinetic energy.

\clearpage

\section{Summary}

This contribution reviews relevant literature addressing multi-phase flow
with a solid phase as particulate material present. In addition, the article 
introduces the predictive capabilities of the XDEM-suite applied to a
solid phase that is closely linked to the fluid phase by an intensive heat, mass
and momentum transfer. The interface between continuum and discrete 
numerical approaches includes a selection of fluid dynamic solvers and a
variety of different and fundamental reaction mechanism that cover a large 
number of numerical approaches to describe thermal conversion of a particulate 
phase. Approaches presented are validated with experimental data,
that establishes an accurate framework for investigating a variety of 
engineering applications. Results obtained exhibit a large degree of detail,
in particular on smallest scales. A thorough analysis of the results obtained 
allows unveiling the underlying physics, that enables engineers to improve both
design and performance.

\section{Acknowledgements}
The authors would like to acknowledge the support of Fond National de la 
Recherche Luxembourg, Europe FP7 IAPP program (Grant IAP-GA-2012-323526-AMST) and the
HPC centre of the University of Luxembourg.

\section*{References}


\begin{thebibliography}{100}
\expandafter\ifx\csname url\endcsname\relax
  \def\url#1{\texttt{#1}}\fi
\expandafter\ifx\csname urlprefix\endcsname\relax\def\urlprefix{URL }\fi
\expandafter\ifx\csname href\endcsname\relax
  \def\href#1#2{#2} \def\path#1{#1}\fi

\bibitem{blevins84}
R.~D. Blevins, {Applied Fluid Dynamics Handbook}, Krieger Publishing Company,
  Malabar, Florida, 1984.

\bibitem{Peric96}
J.~H. Ferzinger, M.~Peric, {Computation Methods for Fluid Dynamics}, Springer
  Verlag, Heidelberg, 1996.

\bibitem{Bathe96}
K.~J. Bathe, {Finite Element Procedures}, Prentice Hall, 1996.

\bibitem{Zienkiewicz84}
O.~Zienkiewicz, {Methode der {F}initen {E}lemente}, Carl Hanser Verlag, 1984.

\bibitem{multiphase.Gidaspow94}
D.~Gidaspow, Multiphase flow and Fluidisation, Academic Press, 1994.

\bibitem{Peters03c}
B.~Peters, Thermal Conversion of Solid Fuels, WIT Press, Southampton, 2003.

\bibitem{comb.Spliethoff10}
Power Generation from Solid Fuels, Springer-Verlag Berlin Heidelberg, 2010.

\bibitem{multiphase.Omori87}
Y.~Omori, Blast Furnace Phenomena and Modelling, Elsevier, London, UK, 1987.

\bibitem{multiphase.Szekely79}
J.~Szekely, Y.~Kajiwara, A mathematical representation of spatially
  non-uniform, counter-current flow of gases and liquids in packed beds-of
  relevance to flow phenomena in the bosh of iron blast furnaces, Trans Iron
  Steel Inst. 19.

\bibitem{multiphase.Sugiyama87b}
T.~Sugiyama, M.~Sugata, Development of two-dimensional mathematical model for
  the blast-furnace, Tech. Rep. Nippon Steel Technical Report No. 35, Nippon
  Steel, 32-42 (October 1987).

\bibitem{multiphase.Austin97b}
P.~R. Austin, Modelling of the blast furnace based on the multifluid concept
  with applications to advanced operations, Ph.D. thesis, Tohoku University,
  Japan (1997).

\bibitem{multiphase.Ohno88}
Y.~Ohno, M.~Schneider, Effect of horizontal gas flow on liquid dropping flow in
  two dimensional packed bed, Tetsu-to-Hagane 74 (1988) 1923--1930.

\bibitem{multiphase.Wang97a}
G.~X. Wang, S.~J. Chew, A.~B. Yu, P.~Zulli, Modeling the discontinuous liquid
  flow in a blast furnace, Metallurgical and Materials Transactions B 28B
  (1997) 333--342.

\bibitem{multiphase.Wang97b}
G.~X. Wang, D.~Y. Liu, J.~D. Litster, A.~B. Yu, S.~J. Chew, P.~Zulli,
  Experimental and numerical simulation of discrete liquid flow in a packed
  bed, Chemical Engineering Science 52 (1997) 4013--4019.

\bibitem{multiphase.Liu02}
D.~Y. Liu, G.~X. Wang, J.~D. Litster, Unsaturated liquid percolation flow
  through nonwetted packed beds, AIChE Journal 48 (2002) 953--962.

\bibitem{multiphase.Eto93}
Y.~Eto, K.~Takeda, S.~Miyagawa, Experiments and simulation of the liquid flow
  in the dropping zone of a blast furnace, Iron and Steel Institute of Japan
  (ISIJ-International) 33 (1993) 681.

\bibitem{multiphase.Wang91}
J.~Wang, R.~Takahashi, J.~Yagi, Simulation model of the gas-liquid flows in the
  packed bed, Tetsu-to-Hagane 77 (1991) 1585--1592.

\bibitem{multiphase.Takahashi89}
H.~Takahashi, K.~Kushima, T.~Takeuchi, Two dimensional analysis of burden flow
  in blast furnace based on plasticity theory, Iron and Steel Institute of
  Japan (ISIJ-International) 29 (1989) 117--124.

\bibitem{multiphase.Takahashi93}
H.~Takahashi, N.~Komatsu, Cold model study on burden behaviour in the lower
  part of blast furnace, Iron and Steel Institute of Japan (ISIJ-International)
  33 (1993) 655--663.

\bibitem{multiphase.Kuwabara91}
M.~Kuwabara, S.~Takane, K.~Sekido, I.~Muchi, Mathmatical two-dimensional model
  of the blast furnace process, Tetsu-to-Hagane 77 (1991) 1593--1600.

\bibitem{multiphase.Chen93}
J.~Chen, T.~Akiyama, H.~Nogami, J.~Yagi, H.~Takahashi, Modeling of solid flow
  in moving beds, Iron and Steel Institute of Japan (ISIJ-International) 33
  (1993) 664--671.

\bibitem{multiphase.Takatani94}
K.~Takatani, T.~Inada, Y.~Ujisawa, 3-dimensional dynamic mathematical simulator
  of blast furnace, CAMP-Iron and Steel Institute of Japan 7 (1994) 50--53.

\bibitem{multiphase.Kajiwara79}
Y.~Kajiwara, S.~J., Interraction between gas and liquid flow in a simulated
  blast furnace, J. Metall Trans. 10B.

\bibitem{multiphase.Yagi93}
J.~Yagi, Mathematical modeling of the flow of four fluids in a packed bed,
  Journal of Iron And Steel Research, International 33 (1993) 619--639.

\bibitem{multiphase.Castro00}
J.~Castro, H.~Nogami, J.~Yagi, Transient mathematical model of blast furnace
  based on multi-fluid concept with application to high pci operation, Iron and
  Steel Institute of Japan (ISIJ-International) 40 (2000) 637--646.

\bibitem{multiphase.Sugiyama87a}
T.~Sugiyama, A.~Nakagawa, H.~Shibaike, Analysis on liquid flow in the dripping
  zone of blast furnace, Tetsu-to-Hagane 75.

\bibitem{multiphase.Zhang98}
S.~J. Zhang, A.~B. Yu, P.~Zulli, B.~Wright, U.~Tuzun, Modelling of the solids
  flow in a blast furnace, Iron and Steel Institute of Japan
  (ISIJ-International) 38 (1998) 1311--1319.

\bibitem{multiphase.Austin98}
P.~R. Austin, H.~Nogami, J.~Yagi, Computational investigation of scrap charging
  to the blast furnace, Journal of Iron And Steel Research, International 38.

\bibitem{multiphase.Zaimi00}
S.~A. Zaimi, T.~Akiyama, J.~B. Guillot, J.~Yagi, Validation of a blast furnace
  solid flow model using reliable 3-d experimental results, Iron and Steel
  Institute of Japan (ISIJ-International) 40 (2000) 332--341.

\bibitem{multiphase.Castro02a}
J.~A. Castro, H.~Nogami, J.~Yagi, Three-dimensional multi-phase mathematical
  modeling of the blast furnace based on the multifluid model, Journal of Iron
  And Steel Research, International 42.

\bibitem{multiphase.Castro02b}
J.~A. Castro, H.~Nogami, J.~Yagi, Numerical investigation of simultaneous
  injection of pulverized coal and natural gas with oxygen enrichment to the
  blast furnace, Journal of Iron And Steel Research, International 42.

\bibitem{multiphase.Chu02}
M.~Chu, H.~Nogami, J.~Yagi, Numerical analysis on blast furnace performance
  under operation with top gas recycling and carbon composite agglomerates
  charging, Journal of Iron And Steel Research, International 44.

\bibitem{multiphase.Li01}
M.~Li, Y.~Banda, T.~Tsuge, Analysis of liquid distribution in non-uniformly
  packed trickle bed with single phase flow, Chemical Engineering Science 56.

\bibitem{multiphase.Wen01}
X.~Wen, Y.~Shu, K.~Nandakumar, Profile in randomly packed beds from computer
  simulation, AIChE Journal 47.

\bibitem{multiphase.Wang08}
C.-s. Wang, M.~Xiao-jing, Z.~Shao-bo, X.~Xing-guo, W.~Wen-zhong, A new
  mathematical model for description of the liquid discrete flow within a
  packed bed, Journal of Iron And Steel Research, International 15 (2008)
  16--23.

\bibitem{multiphase.Wang09}
C.-S. Wang, X.-J. Mu, A amthematical model capable of describing the liquid
  flow mainly in a blast furnace, International Journal of Minerals, Metallurgy
  and Materials 16 (2009) 505.

\bibitem{multiphase.Zhang02}
S.~J. Zhang, A.~B. Yu, P.~Zulli, B.~Wright, P.~Austin, Numerical simulation of
  solids flow in a blast furnace, Applied Mathematical Modelling 26 (2002)
  141--154.

\bibitem{multiphase.Takahashi96}
H.~Takahashi, M.~Tanno, J.~Katayama, Burden descending behavior with renewal of
  deadman in a two dimensional cold model of blast furnace, Iron and Steel
  Institute of Japan (ISIJ-International) 36~(11) (1996) 1354--1359.

\bibitem{multiphase.Das2011}
S.~K. Das, A.~Kumari, D.~Bandopadhay, S.~A. Akbar, G.~K. Mondal, A mathematical
  model to characterize effects of liquid hold-up on bosh silicon transport in
  the dripping zone of a blast furnace, Appl. Math. Modelling
  10.1016/j.apm.2011.02.045.

\bibitem{multiphase.Tsuchiya76}
N.~Tsuchiya, M.~Tokuda, M.~Ohtani, The transfer of silica from the gas phase to
  molten iron in the blast furnace, Metallurgical Transactions B 7B (1976)
  315--320.

\bibitem{multiphase.Wang97c}
G.~Wang, S.~Chew, A.~Yu, P.~Zulli, Model study of liquid flow in blast furnace
  lower zone, Iron and Steel Institute of Japan (ISIJ-International) 37 (1997)
  573--582.

\bibitem{multiphase.Austin98b}
P.~R. Austin, H.~Nogami, J.~Yagi, Analysis of actual blast furnace operations
  and evaluation of static liquid holdup effects by the four fluid model,
  Journal of Iron And Steel Research, International 38.

\bibitem{multiphase.Jin10a}
H.~Jin, S.~Choi, J.-I. Yagi, J.~Chung, Dripping liquid metal flow in the lower
  part of a blast furnace, Iron and Steel Institute of Japan
  (ISIJ-International) 50 (2010) 1023--1031.

\bibitem{multiphase.Jin10b}
H.~J. Jin, S.~Choi, Numerical analysis of isothermal flow in lower part of
  blast furnace considering effect of cohesive zone, Ironmaking and Steelmaking
  37~(2) (2010) 89--97.

\bibitem{multiphase.Danloy2011}
G.~Danloy, J.~Mignon, R.~Munnix, G.~Dauwels, L.~Bonte, A blast furnace model to
  optimize the burden distribution, Tech. rep., Centre for Research in
  Metallurgy (CRM), Li{\'e}ge, Belgium, www.crm-eur.com (2009).

\bibitem{multiphase.Hamilius78}
A.~Hamilius, M.~Deroo, G.~Monteyne, R.~Bekaert, R.~D'hondt, Blast furnace
  practice with stave- coolers and with a rotating chute for burden
  distribution, Ironmaking Proceedings, Chicago 37 (1978) 160--168.

\bibitem{multiphase.Nogami06}
H.~Nogami, M.~Chu, J.~Yagi, Numerical analysis on blast furnace performance
  with novel feed material by multi-dimensional simulator based on multi-fluid
  theory, Applied Mathematical Modelling 30 (2006) 1212--1228.

\bibitem{multiphase.Takahashi84}
R.~e.~a. Takahashi, Operation and simulation of pressurized shaft furnace for
  direct reduction, Ironmaking Proc. 43 (1984) 485--500.

\bibitem{multiphase.Fun70}
F.~Fun, Rates and mechanisms of feo reduction from slags, Metall. Trans. 1
  (1970) 2537--2541.

\bibitem{general.Levenspiel76}
O.~Levenspiel, Chemical Reaction Engineering, Wiley, New York, 2nd Edition,
  1976.

\bibitem{multiphase.Turkdogan80}
E.~Turkdogan, G.~J.~W. Kor, R.~J. Fruehan, Iron Steelmaker 7 (1980) 268.

\bibitem{multiphase.Inoue87}
H.~Inoue, T.~Terui, Y.~Jeng, Y.~Omori, M.~Ohtani, Sio generation kinetics in
  the reaction of sic with carbon monoxide in the temperature range 1900-2000
  °c, Bull. Res. Inst. Min. Dressing Metall. 43 (1987) 43--52.

\bibitem{multiphase.Ozturk86}
B.~Ozturk, R.~J. Fruehan, Silicon transfer in blast furnace, Process Technol.
  Proc. 6 (1986) 959--966.

\bibitem{multiphase.Yu04}
A.~B. Yu, G.~X. Wang, S.~J. Chew, P.~Zulli, Modelling the gas-liquid flow in an
  ironmaking blast furnace, Progress in Computational Fluid Dynamics 4~(1)
  (2004) 29--38.

\bibitem{multiphase.Gupta94}
G.~S. Gupta, J.~D. Litster, V.~R. Rudolph, E.~T. White, A cold model study of
  liquid flow in the blast furnace lower zone, in: 6th AusIIM Extractive
  Metallurgy Conference, Brisbane, 3-6 July 1994, 1994, pp. 295--301.

\bibitem{multiphase.Gupta96}
G.~Gupta, J.~D. Litster, V.~R. Rudolph, Model studies of liquid flow in the
  blast furnace lower zone, Journal of Iron And Steel Research, International
  36 (1996) 32--39.

\bibitem{multiphase.Gupta97}
G.~Gupta, J.~D. Litster, E.~T. White, V.~R. Rudolph, Nonwetting flow of a
  liquid through a packed bed with gas cross-flow, Metallurgical and Materials
  Transactions B 28B (1997) 597--604.

\bibitem{multiphase.Mackey73}
P.~Mackey, N.~Warner, Studies in the vaporization of mercury in irrigated
  packed beds, Chemical Engineering Science 28 (1973) 2141--2154.

\bibitem{multiphase.Standish68}
N.~Standish, Dynamic holdup in liquid metal irrigated packed beds, Chemical
  Engineering Science 23 (1968) 51--56.

\bibitem{multiphase.Warner59}
N.~A. Warner, Liquid metal irrigation of a packed bed, Chemical Engineering
  Science 11 (1959) 149--160.

\bibitem{multiphase.Liu98}
D.~Y. Liu, S.~Wijeratne, J.~D. Litster, Scand. J. Metall. 26 (1998) 79.

\bibitem{granular.Bridgewater94}
J.~Bridgewater, Granular Matter - An Interdisciplinary Apporach, Springer,
  1994, Ch. Mixing and segregation mechanisms in particle flow, pp. 161--194.

\bibitem{multiphase.Wang00}
G.~X. Wang, J.~D. Litster, A.~B. Yu, Simulation of gas-liquid flow in dripping
  zone of blast furnace involving impermeable fused layers, Iron and Steel
  Institute of Japan (ISIJ-International) 40~(7) (2000) 627--636.

\bibitem{multiphase.Xu00}
B.~H. Xu, A.~B. Yu, S.~J. Chew, P.~Zulli, Simulation of the gas-solid flow in a
  bed with lateral gas blasting, Powder Technol. 109 (2000) 14--27.

\bibitem{multiphase.Chew01a}
S.~J. Chew, P.~Zulli, A.~B. Yu, Modeling of liquid flow in the blast furnace,
  Journal of Iron And Steel Research, International 41 (2001) 1112.

\bibitem{multiphase.Chew01b}
S.~J. Chew, P.~Zulli, A.~B. Yu, Modelling of liquid flow in the blast furnace:
  Theoretical analysis of the effects of gas, liquid and packing properties,
  Iron and Steel Institute of Japan (ISIJ-International) 41 (2001) 1112--1121.

\bibitem{multiphase.Chew01c}
S.~J. Chew, P.~Zulli, A.~B. Yu, Modelling of liquid flow in the blast furnace:
  Application in a comprehensive blast furnace model, Iron and Steel Institute
  of Japan (ISIJ-International) 41 (2001) 1122--1130.

\bibitem{multiphase.Singh06}
V.~Singh, G.~S. Gupta, A discrete model for non-wetting liquid flow from a
  point source in a packed bed under the influence of gas flow, Chemical
  Engineering Science 61 (2006) 6855--6866.

\bibitem{cohesivezone.Chew97}
S.~J. Chew, G.~X. Wang, A.~B. Yu, P.~Zulli, Experimental study of liquid flow
  in the blast furnace cohesive zone, Ironmaking and Steelmaking 24 (1997)
  392--400.

\bibitem{multiphase.Aussillous04}
P.~Aussillous, D.~Quere, Shapes of rolling liquid drops, Journal of Fluid
  Mechanics 512 (2004) 133--155.

\bibitem{multiphase.Dimitrakopoulos97}
P.~Dimitrakopoulos, J.~J.~L. Higdon, Displacement of fluid droplets from solid
  surfaces in low reynolds-number shear flows, Journal of Fluid Mechanics 336
  (1997) 351--378.

\bibitem{multiphase.Durbin88}
D.~P. A., On the wind force needed to dislodge a drop adhered to a surface,
  Journal of Fluid Mechanics 196 (1988) 205--222.

\bibitem{multiphase.Dussan87}
V.~E.~B. Dussan, On the stability of drops or bubbles to stick to non-horizontl
  surfaces of solids part 3: The influence of motion of surrounding fluid on
  dislodging drops, Journal of Fluid Mechanics 174 (1987) 381--397.

\bibitem{multiphase.Dussan74}
V.~E.~B. Dussan, S.~S. Davis, On motion of a fluid-fluid interfacealong a solid
  surface, Journal of Fluid Mechanics 65 (1974) 71--95.

\bibitem{multiphase.King93}
A.~King, E.~Tuck, Thin liquid layers supported by steady air-flow surface
  traction, Journal of Fluid Mechanics 251 (1993) 709--718.

\bibitem{multiphase.Wilson88}
S.~D.~R. Wilson, The slow dripping of a viscous fluid, Journal of Fluid
  Mechanics 190 (1988) 561--570.

\bibitem{multiphase.Shibata91}
K.~Shibata, M.~Shimizu, S.~Inaba, R.~Takahashi, J.~Yagi, Pressure loss and
  hold-up powders for gas-powder two phase flow in packed beds, Iron and Steel
  Institute of Japan (ISIJ-International) 31 (1991) 434.

\bibitem{multiphase.Chen94}
J.~Chen, H.~Nogami, T.~Akiyama, T.~R., J.~Yagi, Behavior of powders in a packed
  bed with lateral inlets, Iron and Steel Institute of Japan
  (ISIJ-International) 34 (1994) 133.

\bibitem{multiphase.Austin97a}
P.~Austin, H.~Nogami, J.~Yagi, A mathematical model of four phase motion and
  heat transfer in the blast furnace, Iron and Steel Institute of Japan
  (ISIJ-International) 37 (1997) 458--467.

\bibitem{multiphase.Austin97c}
P.~R. Austin, H.~Nogami, J.~Yagi, A mathematical model for blast furnace
  reaction analysis based on the four fluid model, Iron and Steel Institute of
  Japan (ISIJ-International) 37 (1997) 748--755.

\bibitem{multiphase.Castro02}
J.~Castro, H.~Nogami, J.~Yagi, Numerical investigation of simultaneous
  injection of pulverized coal and natural gas with oxygen enrichment to the
  blast furnace, Iron and Steel Institute of Japan (ISIJ-International) 42
  (2002) 1203 -- 1211.

\bibitem{multiphase.Nogami04}
H.~Nogami, P.~Austin, J.~Yagi, K.~Yamaguchi, Numerical investigation on effects
  of deadman structure and powder properties on gasand powder flows in lower
  part of blast furnace, Iron and Steel Institute of Japan (ISIJ-International)
  44 (2004) 500--509.

\bibitem{multiphase.Pintowantoro04}
S.~Pintowantoro, H.~Nogami, J.~Yagi, Numerical analysis of static holdup of
  fine particles in blast furnace, Iron and Steel Institute of Japan
  (ISIJ-International) 44 (2004) 304--309.

\bibitem{multiphase.Hidaka00}
N.~Hidaka, T.~Matsumoto, K.~Kusakabe, S.~Morooka, Transient accumulation
  behavior of fines in packed beds of pulverized coke particles, J. Chem. Eng.
  Jpn. 33 (2000) 152--159.

\bibitem{multiphase.Dong04a}
X.~F. Dong, D.~Pinson, S.~J. Zhang, A.~B. Yu, P.~Zulli, Gas-powder flow and
  powder accumulation in a packed bed i: Experimental study, Powder Technology
  149 (2004) 1--9.

\bibitem{multiphase.Horio92}
M.~Horio, Transport phenomena in the lower part of blast furnace, Iron and
  Steel Institute of Japan (ISIJ-International) 68.

\bibitem{multiphase.Jackson63}
R.~Jackson, The mechanics of fluidized beds: part i: the stability of
  computation of multiphase-flow phenomena with interphase slip,the state of
  uniform fluidisation, Trans. Inst. Chem. Eng. 41 (1963) 13--21.

\bibitem{multiphase.Soo90}
S.~L. Soo, Multiphase Fluid Dynamics, Science Press, Beijing, 1990.

\bibitem{multiphase.Garg75}
S.~K. Garg, J.~W. Pritchett, Dynamics of gas-fluidized beds, J. Appl. Phys. 46
  (1975) 4493--4500.

\bibitem{multiphase.Gidaspow83}
D.~Gidaspow, B.~Ettehadie, Fluidization in two-dimensional beds with a jet: 2.
  hydrodynamic modeling, Ind. Eng. Chem. Fundam. 22 (1983) 193--201.

\bibitem{multiphase.Kuipers91}
J.~A.~M. Kuipers, W.~Prins, W.~P.~M. Van~Swaajj, Theoretical and experimental
  bubble formation at a single orifice in a two-dimensional gas-fluidized bed,
  Chem. Eng. Sci. 46 (1991) 2881--2894.

\bibitem{multiphase.Levy00}
A.~Levy, Two-fluid approach for plug flow simulations in horizontal pneumatic
  conveying, Powder Technol. 112 (2000) 263--272.

\bibitem{cohesivezone.Dong06}
X.~F. Dong, D.~Pinson, S.~J. Zhang, A.~B. Yu, P.~Zulli, Gas-powder flow in
  blast furnace with different shapes of cohesive zone, Applied Mathematical
  Modelling 30 (2006) 1293--1309.

\bibitem{multiphase.Sugiyama96}
T.~Sugiyama, Analysis on the powder movement and accumulation in the deadman
  and dripping zone of blast furnace by the experiment of the powder-gas
  2-phase flow through the parallel packed bed, CAMP-Iron and Steel Institute
  of Japan 9 (1996) 19--21.

\bibitem{multiphase.Yamaoka86}
H.~Yamaoka, Flow characteristics of gas and fine particles in a two-dimensional
  space of packed bed, Tetsu-to-Hagane 72 (1986) 2194 -- 2201.

\bibitem{multiphase.Ichida92}
M.~Ichida, T.~Nakayama, K.~Tamura, H.~Shiota, K.~Araki, Y.~Sugisaki, Behavior
  of fines in the blast furnace, Iron and Steel Institute of Japan
  (ISIJ-International) 32 (1992) 505--513.

\bibitem{multiphase.Yamaoka91}
H.~Yamaoka, Mechanisms of hanging caused by dust in a shaft furnace, Iron and
  Steel Institute of Japan (ISIJ-International) 31 (1991) 939--946.

\bibitem{multiphase.Hidaka98}
N.~Hidaka, J.~Iyama, T.~Matsumoto, K.~Kusakabe, S.~Morooka, Entrainment of fine
  particles with upward gas flow in a packed bed of coarse particles, Powder
  Technology 95 (1998) 265--271.

\bibitem{multiphase.Steiler91}
J.~M. Steiler, R.~Nicolle, P.~Negro, M.~Helleisen, N.~Jusseau, B.~Metz,
  C.~Thirion, in: Proc. Ironmaking Conf., Vol.~50, ISS, Warrendale, PA, 1991,
  p. 715.

\bibitem{multiphase.Dong09}
X.~F. Dong, A.~B. Yu, J.~M. Burgess, , D.~Pinson, S.~Chew, P.~Zulli, Modelling
  of multiphase flow in ironmaking blast furnace, Ind. Eng. Chem. Res. 48
  (2009) 214--226.

\bibitem{multiphase.Aoki93}
R.~Aoki, H.~Nogami, H.~Tsuge, T.~Miura, T.~Furukawa, Simulation of transport
  phenomena around the raceway zone in the blast furnace with and without
  pulverized coal injection, Iron and Steel Institute of Japan
  (ISIJ-International) 33 (1993) 646.

\bibitem{multiphase.Yuu05}
S.~Yuu, T.~Umekage, T.~Miyahara, Predicition of stable and unstable flows in
  blast furnace raceway using numerical simulation methods for gas and
  particles, Iron and Steel Institute of Japan (ISIJ-International) 45 (2005)
  1406--1415.

\bibitem{multiphase.Takeda96}
K.~Takeda, F.~C. Lockwood, Stochastic model of flow and dispersion of fine
  particles in a packed bed, Tetsu-to-Hagane 82 (1996) 492--497.

\bibitem{multiphase.Takeda97}
K.~Takeda, F.~C. Lockwood, Integrated mathematical model of pulverised coal
  combustion in a blast furnace, Iron and Steel Institute of Japan
  (ISIJ-International) 37 (1997) 432--440.

\bibitem{multiphase.Tsuji93}
Y.~Tsuji, T.~Kawaguchi, T.~Tanaka, Discrete particle simulation of
  two-dimensional fluidized bed, Powder Technol. 77~(79).

\bibitem{multiphase.Hoomans96}
B.~P.~B. Hoomans, J.~A.~M. Kuipers, W.~J. Briels, W.~P.~M. Van~Swaaij, Discrete
  particle simulation of bubble and slug formation in a two-dimensional
  gas-fluidized bed: A hard-sphere approach, Chem. Eng. Sci. 51.

\bibitem{multiphase.Xu97}
B.~H. Xu, A.~B. Yu, Numerical simulation of the gas-solid flow in a fluidized
  bed by combining discrete particle method with computational fluidd dynamics,
  Chemical Engineering Science 52 (1997) 2785.

\bibitem{multiphase.Xu98}
B.~H. Xu, A.~B. Yu, Comments on the paper “numerical simulation of the
  gas-solid flow in a fluidized bed by combining discrete particle method with
  computational fluid dynamics-reply, Chemical Engineering Science 53 (1998)
  2646--2647.

\bibitem{multiphase.Zhu07b}
H.~P. Zhu, Z.~Y. Zhou, R.~Y. Yang, A.~B. Yu, Discrete particle simulation of
  particulate systems: Theoretical developments, Chemical Engineering Science
  62 (2007) 3378 -- 3396.

\bibitem{multiphase.Zhu08}
H.~P. Zhu, Z.~Y. Zhou, R.~Y. Yang, A.~B. Yu, Discrete particle simulation of
  particulate systems: A review of major applications and findings, Chemical
  Engineering Science 63 (2008) 5728--5770.

\bibitem{multiphase.Yu03}
A.~B. Yu, B.~H. Xu, Particle-scale modelling of gas-solid flow in fluidisation,
  Journal of Chemical Technology and Biotechnology 78~(2-3) (2003) 111--121.

\bibitem{multiphase.Feng04}
Y.~Q. Feng, A.~B. Yu, Assessment of model formulations in the discrete particle
  simulation of gas-solid flow, Industrial \& Engineering Chemistry Research 43
  (2004) 8378--8390.

\bibitem{multiphase.Deen07}
N.~G. Deen, M.~V.~S. Annaland, M.~A. Van Der~Hoef, J.~A.~M. Kuipers, Review of
  discrete particle modeling of fluidized beds, Chemical Engineering Science 62
  (2007) 28--44.

\bibitem{multiphase.Chu08}
K.~W. Chu, A.~B. Yu, Numerical simulation of complex particle-fluid flows,
  Powder Technology 179 (2008) 104--114.

\bibitem{multiphase.Zhou11}
H.~Zhou, G.~Mo, J.~Zhao, K.~Cen, Dem-cfd simulation of the particle dispersion
  in a gas-solid two-phase flow for a fuel-rich/lean burner, Fuel 90 (2011)
  1584--1590.

\bibitem{multiphase.Chu09}
K.~W. Chu, B.~Wang, A.~B. Yu, A.~Vince, G.~D. Barnett, P.~J. Barnett, Cfd-dem
  study of the effect of particle density distribution on the multiphase flow
  and performance of dense medium cyclone, Minerals Engineering 22 (2009)
  893--909.

\bibitem{multiphase.Rowe76}
P.~N. Rowe, A.~W. Nienow, Particle mixing and segregation in gas fluidized
  beds—a review, Powder Technology 15 (1976) 141--147.

\bibitem{multiphase.Feng08}
Y.~Q. Feng, A.~B. Yu, An analysis of the chaotic motion of particles of
  different sizes in a gas fluidized bed, Particuology 6 (2008) 549--556.

\bibitem{multiphase.Kafuia02}
K.~D. Kafuia, C.~Thornton, M.~J. Adams, Discrete particle-continuum fluid
  modelling of gas-solid fluidised beds, Chemical Engineering Science 57 (2002)
  2395--2410.

\bibitem{multiphase.Chattopadhyay10a}
K.~Chattopadhyay, M.~Isac, R.~I.~L. Guthrie, Applications of computational
  fluid dynamics (cfd) in iron- and steelmaking: Part 1, Ironmaking and
  Steelmaking 37~(8) (2010) 554--561.

\bibitem{multiphase.Chattopadhyay10b}
K.~Chattopadhyay, M.~Isac, R.~I.~L. Guthrie, Applications of computational
  fluid dynamics (cfd) in iron- and steelmaking: Part 2, Ironmaking and
  Steelmaking 37~(8) (2010) 562--569.

\bibitem{multiphase.Dong07}
X.~Dong, A.~Yu, J.~ichiro Yagi, P.~Zulli, Modelling of multiphase flow in a
  blast furnace: Recent developments and future work, ISIJ International
  47~(11) (2007) 1553--1570, 3.

\bibitem{multiphase.Yu05}
A.~B. Yu, Encyclopedea of Condensed Matter Physics, Elsevier, 2005, Ch. Powder
  Processing: Models and Simulations, pp. 401--414.

\bibitem{pyrolysis.Simsek09}
E.~Simsek, B.~Brosch, S.~Wirtz, V.~Scherer, F.~Krüll, Numerical simulation of
  grate firing systems using a coupled cfd/discrete element method (dem),
  Powder Technology 193 (2009) 266--273.

\bibitem{multiphase.Wang98}
C.~Y. Wang, Tramsport Phenomena in Porous Media, Oxford Pergamon, 1998, Ch.
  Modelling Multiphase Flow and Transport in Porous Media.

\bibitem{multiphase.Kaneko99}
Y.~Kaneko, T.~Shiojima, M.~Horio, Dem simulation of fluidized beds for
  gas-phase olefin polymerization, Chemical Engineering Science 54 (1999) 5809.

\bibitem{multiphase.Ranz52}
W.~E. Ranz, W.~R. Marshall, Evaporation from drops, Chemical Engineering
  Progress 48 (1952) 141.

\bibitem{multiphase.Swasdisevi05}
T.~Swasdisevi, W.~Tanthapanichakoon, T.~Charinpanitkul, T.~Kawaguchi, T.~Tsuji,
  Prediction of gas-particle dynamics and heat transfer in a two-dimensional
  spouted bed, Advanced Powder Technology 16 (2005) 275.

\bibitem{multiphase.Li00}
J.~T. Li, D.~J. Mason, A computational investigation of transient heat transfer
  in pneumatic transport of granular particles, Powder Technology 112 (2000)
  273.

\bibitem{multiphase.Li02}
J.~T. Li, D.~J. Mason, Application of the discrete element modelling in air
  drying of particulate solids, Drying Technology 20 (2002) 255.

\bibitem{multiphase.Li03}
J.~T. Li, D.~J. Mason, A.~S. Mujumdar, A numerical study of heat transfer
  mechanisms in gas-solids flows through pipes using a coupled cfd and dem
  model, Drying Technology 21 (2003) 1839.

\bibitem{multiphase.Zhou04a}
H.~Zhou, G.~Flamant, D.~Gauthier, Dem-les of coal combustion in a bubbling
  fluidized bed. part i: gas-particle turbulent flow structure, Chemical
  Engineering Science 59 (2004) 4193.

\bibitem{multiphase.Zhou04b}
H.~Zhou, G.~Flamant, D.~Gauthier, Dem-les simulation of coal combustion in a
  bubbling fluidized bed. part ii: coal combustion at the particle level,
  Chemical Engineering Science 59 (2004) 4205.

\bibitem{multiphase.Wang2011}
X.~Wang, F.~Jiang, J.~Lei, J.~Wang, S.~Wang, X.~Xu, Y.~Xiao, A revised drag
  force model and the application for the gas-solid flow in the high-density
  circulating fluidized bed, Applied Thermal Engineering 31~(14-15) (2011)
  2254--2261.

\bibitem{multiphase.Malone08}
K.~F. Malone, B.~H. Xu, Particle-scale simulation of heat transfer in
  liquid-fluidised beds, Powder Technology 184 (2008) 189--204.

\bibitem{multiphase.Xiang09}
J.~Xiang, The effect of air on the packing structure of fine particles, Powder
  Technology 191 (2009) 280--293.

\bibitem{Peters99a}
B.~Peters, Classification of combustion regimes in a packed bed based on the
  relevant time and length scales, Combustion and Flame 116 (1999) 297 -- 301.

\bibitem{Peters02a}
B.~Peters, Measurements and application of a discrete particle model ({DPM}) to
  simulate combustion of a packed bed of individual fuel particles, Combustion
  and Flame 131 (2002) 132--146.

\bibitem{multiphase.Mehrabian2012}
R.~Mehrabian, S.~Zahirovic, R.~Scharler, I.~Obernberger, S.~Kleditzsch,
  S.~Wirtz, A cfdmodel for thermal conversion of thermally thick biomass
  particles, Fuel Process. Technol. 95 (2012) 96--108.

\bibitem{multiphase.Gomez2015}
M.~Gomez, Fast-solving thermally thick model of biomass particles embedded in a
  cfd code for the simulation of fixed-bed burners, Energy Convers. Manag. 105
  (2015) 30–44.

\bibitem{multiphase.Wiese2016}
J.~Wiese, F.~Wissing, D.~Höhner, S.~Wirtz, V.~Scherer, U.~Ley, H.~Behr,
  Dem/cfd modelling of the fuel conversion in a pellet stove, Fuel Processing
  Technology 152 (2016) 223–239.

\bibitem{multiphase.Sudbrock2015}
F.~Sudbrock, H.~Kruggel-Emden, S.~Wirtz, V.~Scherer, Convective drying of
  agitated silica gel and beech wood particle beds - experiments and transient
  dem-cfd simulations, Drying Technology 33 (2015) 15--16.

\bibitem{multiphase.Scherer2016}
V.~Scherer, M.~Mönnigmann, M.~Berner, F.~Sudbrock, Coupled dem-cfd simulation
  of drying wood chips in a rotary drum - baffle design and model reduction,
  Fuel 184 (2016) 896--904.

\bibitem{multiphase.Wissing2016}
F.~Wissing, S.~Wirtz, V.~Scherer, Discrete element modelling of msw
  incineration on grate firing systems: Influence of waste properties, in: 41st
  International Technical Conference on Clean Coal and Fuel Systems, 05.-09.
  Juni 2016, Clearwater, Florida, USA, 2016.

\bibitem{multiphase.Oschmann2016}
T.~Oschmann, S.~Wirtz, H.~Kruggel-Emden, Investigation of heat transfer in
  packed/fluidized beds for spherical particles resolved by an implicit 3d
  finite difference approach, in: Partec 2016, 19.-21. April 2016, Nuremberg,
  Germany, 2016.

\bibitem{multiphase.Krause2015}
B.~Krause, B.~Liedmann, J.~Wiese, S.~Wirtz, V.~Scherer, Coupled three
  dimensional dem-cfd simulation of a lime shaft kiln - calcination, particle
  movement and gas phase flow field, Chemical Engineering Science 134 (2015)
  834--849.

\bibitem{multiphase.Kloss2012}
C.~Kloss, C.~Goniva, A.~Hager, S.~Pirker, Models, algorithms and validation for
  opensource dem and cfd-dem, Progress in Computational Fluid Dynamics 12(2/3)
  (2012) 140--152.

\bibitem{multiphase.Radl2015}
S.~Radl, F.~Krainer, T.~Puffitsch, C.~Kloss, Biot number effects on the local
  heat and mass transfer rate in fixed and fluidized beds, in: AIChE Meeting,
  Salt Lake City, 2015.

\bibitem{HT:Chen2005}
J.~Chen, J.~Grace, M.~Golriz, Heat transfer in fluidized beds: design methods,
  Powder Technol. 150 (2) (2005) 123--132.

\bibitem{HT:Baillis2000}
D.~Baillis, J.-F. Sacadura, Thermal radiation properties of dispersed media:
  theoretical prediction and experimental characterization, J. Quant.
  Spectrosc. Radiat. Transf. 67 (5) (2000) 327--363.

\bibitem{HT:Decker1983}
N.~Decker, L.~Glicksman, Heat transfer in large particle fluidized beds, Int.
  J. Heat Mass Transf. 26 (9) (1983) 1307--1320.

\bibitem{HT:Gloski1984}
D.~Gloski, L.~Glicksman, N.~Decker, Thermal resistance at a surface in contact
  with fluidized bed particles, Int. J. Heat Mass Transf. 27 (4) (1984)
  599--610.

\bibitem{HT:Goshayeshi1986}
A.~Goshayeshi, J.~Welty, R.~Adams, N.~Alavizadeh, Local heat transfer
  coefficients for horizontal tube arrays in high-temperature large-particle
  fluidized beds: an experimental study, J. Heat Transf. 108 (4) (1986)
  907--912.

\bibitem{HT:Toschkoff2015}
G.~Toschkoff, S.~Just, K.~Knop, P.~Kleinebudde, A.~Funke, D.~Djuric,
  G.~Scharrer, J.~Khinast, Modeling of an active tablet coating process, J.
  Pharm. Sci. 104 (12) (2015) 4082--4092.

\bibitem{HT:Amberger2013}
S.~Amberger, S.~Pirker, C.~Kloss, Thermal radiation modeling using ray tracing
  in liggghts, in: 6th International Conference on Discrete Element Methods
  (DEM6), 2013.

\bibitem{Peters03b}
B.~Peters, C.~Bruch, T.~Nussbaumer, Modelling of wood combustion under fixed
  bed conditions, Fuel 82~(6) (2003) 729--738.

\bibitem{HT:Fogber2018}
T.~Fogber, S.~Radl, A novel approach to calculate radiative thermal exchange in
  coupled particle simulations, Powder Technology 323 (2018) 24--44.

\bibitem{mpPIC:Zhong2016}
W.~Zhong, A.~Yu, G.~Zhou, J.~Xie, H.~Zhang,
  \href{http://www.sciencedirect.com/science/article/pii/S0009250915006661}{{CFD}
  simulation of dense particulate reaction system: Approaches, recent advances
  and applications}, Chemical Engineering Science 140 (2016) 16 -- 43.
\newblock \href {http://dx.doi.org/http://dx.doi.org/10.1016/j.ces.2015.09.035}
  {\path{doi:http://dx.doi.org/10.1016/j.ces.2015.09.035}}.
\newline\urlprefix\url{http://www.sciencedirect.com/science/article/pii/S0009250915006661}

\bibitem{CFD.OpenFoamextend}
http://openfoamwiki.net (February 2017).

\bibitem{Faghri}
A.~Faghri, Y.~Zhang, Transport phenomena in multiphase systems, Academic Press,
  2006.

\bibitem{Hoffmann_thesis}
F.~Hoffmann, {Modelling Heteregeneous Reactions in Packed Beds and its
  Application to the Upper Shaft of a Blast Furnace}, Phd. dissertation,
  University of Luxembourg (2014).

\bibitem{Mahmoudi_thesis}
A.~Mahmoudi, {PREDICTION OF HEAT-UP, DRYING AND GASIFICATION OF FIXED AND
  MOVING BEDS BY THE DISCRETE PARTICLE METHOD (DPM)}, Phd. dissertation,
  University of Luxembourg (2016).

\bibitem{Mark_thesis}
M.~M., {A Discrete Approach to Describe the Kinematics between Snow and a Tire
  Tread}, Phd. dissertation, University of Luxembourg (2014).

\bibitem{Samiei_thesis}
S.~K., {Assessment of implicit and explicit algorithms in numerical simulation
  of granular matter}, Phd. dissertation, University of Luxembourg (2012).

\bibitem{Mederos2009}
F.~S. Mederos, J.~Ancheyta, J.~Chen, Review on criteria to ensure ideal
  behaviors in trickle-bed reactors, Applied Catalysis A: General 355~(1–2)
  (2009) 1 -- 19.
\newblock \href {http://dx.doi.org/10.1016/j.apcata.2008.11.018}
  {\path{doi:10.1016/j.apcata.2008.11.018}}.

\bibitem{Schwidder2012}
S.~Schwidder, K.~Schnitzlein, A new model for the design and analysis of
  trickle bed reactors, Chemical Engineering Journal 207–208 (2012) 758 --
  765, 22nd International Symposium on Chemical Reaction Engineering (ISCRE
  22).
\newblock \href {http://dx.doi.org/10.1016/j.cej.2012.07.054}
  {\path{doi:10.1016/j.cej.2012.07.054}}.

\bibitem{Atta2007}
A.~Atta, S.~Roy, K.~D. Nigam, Investigation of liquid maldistribution in
  trickle-bed reactors using porous media concept in \{CFD\}, Chemical
  Engineering Science 62~(24) (2007) 7033 -- 7044, 8th International Conference
  on Gas-Liquid and Gas-Liquid-Solid Reactor Engineering.
\newblock \href {http://dx.doi.org/10.1016/j.ces.2007.07.069}
  {\path{doi:10.1016/j.ces.2007.07.069}}.

\bibitem{Jiang20021}
Y.~Jiang, M.~Khadilkar, M.~Al-Dahhan, M.~Dudukovic, \{CFD\} of multiphase flow
  in packed-bed reactors: I. k-fluid modeling issues, AIChE Journal 48~(4)
  (2002) 701--715.
\newblock \href {http://dx.doi.org/10.1002/aic.690480406}
  {\path{doi:10.1002/aic.690480406}}.

\bibitem{Gunjal2005}
P.~R. Gunjal, M.~N. Kashid, V.~V. Ranade, R.~V. Chaudhari,
  \href{http://dx.doi.org/10.1021/ie0491037}{Hydrodynamics of trickle-bed
  reactors:  experiments and cfd modeling}, Industrial \& Engineering
  Chemistry Research 44~(16) (2005) 6278--6294.
\newblock \href {http://arxiv.org/abs/http://dx.doi.org/10.1021/ie0491037}
  {\path{arXiv:http://dx.doi.org/10.1021/ie0491037}}, \href
  {http://dx.doi.org/10.1021/ie0491037} {\path{doi:10.1021/ie0491037}}.
\newline\urlprefix\url{http://dx.doi.org/10.1021/ie0491037}

\bibitem{xiao2011}
H.~Xiao, J.~Sun, Algorithms in a robust hybrid cfd-dem solver for
  particle-laden flows, Communications in Computational Physics 9~(02) (2011)
  297--323.

\bibitem{Baniasasi2017}
M.~Baniasadi, B.~Peters, Resolving multiphase flow through packed bed of solid
  particles using extended discrete element method with porosity calculation,
  Industrial \& Engineering Chemistry Research 56~(41) (2017) 11996--12008.
\newblock \href
  {http://arxiv.org/abs/http://dx.doi.org/10.1021/acs.iecr.7b02903}
  {\path{arXiv:http://dx.doi.org/10.1021/acs.iecr.7b02903}}, \href
  {http://dx.doi.org/10.1021/acs.iecr.7b02903}
  {\path{doi:10.1021/acs.iecr.7b02903}}.

\bibitem{Baniasadi2018}
M.~Baniasadi, B.~Peters, {Preliminary investigation on the capability of
  eXtended Discrete Element method for treating the dripping zone of a blast
  furnace}, ISIJ International 48~(1).

\bibitem{MeBaniasadiCES2017}
M.~Baniasadi, M.~Baniasadi, B.~Peters, Coupled cfd-dem with heat and mass
  transfer to investigate the melting of a granular packed bed, Chemical
  Engineering Science 178 (2018) 136 -- 145.
\newblock \href {http://dx.doi.org/https://doi.org/10.1016/j.ces.2017.12.044}
  {\path{doi:https://doi.org/10.1016/j.ces.2017.12.044}}.

\bibitem{MeBaniasadiAIP}
M.~Baniasadi, M.~Baniasadi, B.~Peters, Application of the extended discrete
  element method (xdem) in the melting of a single particle, AIP Conference
  Proceedings 1863~(1) (2017) 180003.
\newblock \href
  {http://arxiv.org/abs/http://aip.scitation.org/doi/pdf/10.1063/1.4992363}
  {\path{arXiv:http://aip.scitation.org/doi/pdf/10.1063/1.4992363}}, \href
  {http://dx.doi.org/10.1063/1.4992363} {\path{doi:10.1063/1.4992363}}.

\bibitem{Shukla2011}
A.~K. Shukla, R.~Dmitry, O.~Volkova, P.~R. Scheller, B.~Deo, Cold model
  investigations of melting of ice in a gas-stirred vessel, Metallurgical and
  Materials Transactions B 42~(1) (2011) 224--235.
\newblock \href {http://dx.doi.org/10.1007/s11663-010-9442-9}
  {\path{doi:10.1007/s11663-010-9442-9}}.

\bibitem{Lassnerbook}
E.~Lassner, W.-D. Schubert, {Tungsten: Properties, Chemistry, Technology of the
  Elements, Alloys, and Chemical Compounds}, 1st Edition, Plenum Publishers,
  1999.

\bibitem{ITIA_13}
M.~Weil, W.-D. Schubert, {The Beautiful Colours of Tungsten Oxides}, Tungsten
  (2013) 1--12.

\bibitem{WuXiang_09}
X.-W. Wu, J.-S. Luo, B.-Z. Lu, C.-H. Xie, Z.-M. Pi, M.-z. Hu, T.~Xu, G.-G. Wu,
  Z.-M. Yu, D.-Q. Yi, {Crystal Growth of Tungsten During Hydrogen Reduction of
  Tungsten Oxide at High Temperature}, Transactions of Nonferrous Metals
  Society of China 19 (2009) 785--789.

\bibitem{Estupinan_thesis}
A.~A. {Estupinan Donoso}, {A Discrete-Continuous Approach to Model Powder
  Metallurgy Processes}, Phd, University of Luxembourg (2016).

\bibitem{Luidold_07}
S.~Luidold, H.~Antrekowitsch, {Hydrogen as a Reducing Agent: State-of-the-art
  Science and Technology}, J Miner Met {\&} Mater Soc 2~(June) (2007) 20--27.

\bibitem{Haubner_83a}
R.~Haubner, W.-D. Schubert, E.~Lassner, M.~Schreiner, {Mechanism of Technical
  Reduction of Tungsten: Part 1 Literature Review}, International Journal of
  Refractory Metals and Hard Materials 2~(September 1983) (1983) 108--115.

\bibitem{Estupinan_14}
A.~A.~A. {Estupinan Donoso}, B.~Peters,
  \href{http://authors.elsevier.com/a/1Qj9a{\_}5-MTnPsJ http://
  authors.elsevier.com/a/1Qj9a{\%}7B{\_}{\%}7D5-MTnPsJ}{{XDEM Employed to
  Predict Reduction of Tungsten Oxide in a Dry Hydrogen Atmosphere}},
  International Journal of Refractory Metals and Hard Materials 49~(March)
  (2015) 88--94.
\newblock \href {http://dx.doi.org/10.1016/j.ijrmhm.2014.08.012}
  {\path{doi:10.1016/j.ijrmhm.2014.08.012}}.
\newline\urlprefix\url{http://authors.elsevier.com/a/1Qj9a{\_}5-MTnPsJ http://
  authors.elsevier.com/a/1Qj9a{\%}7B{\_}{\%}7D5-MTnPsJ}

\bibitem{Barrow_96}
G.~M. Barrow, {Physical Chemistry}, 6th Edition, McGraw-Hill, Boston, 1996.

\bibitem{Estupinan_15c}
A.~A. {Estupinan Donoso}, B.~Peters, {Predicting Tungsten Oxide Reduction with
  the Extended Discrete Element Method}, Advances in Powder Metallurgy {\&}
  Particulate Materials (2015) 02.35--02.48.

\bibitem{Mohseni2017}
M.~Mohseni, Drying and conversion analysis of biomass by a dem-cfd coupling
  approach, Ph.D. thesis, University of Luxembourg (2017).

\bibitem{Looi}
A.~Looi, K.~Golonka, M.~Rhodes, Drying kinetics of single porous particles in
  superheated steam under pressure, Chemical Engineering Journal 87 (2002)
  329--338.

\bibitem{CES_MM}
M.~Mohseni, B.~Peters, Effects of particle size distribution on drying
  characteristics in a drum by xdem: A case study, Chemical Engineering Science
  152 (2016) 689--689.

\bibitem{PozzettiIJMF}
G.~Pozzetti, B.~Peters,
  \href{http://www.sciencedirect.com/science/article/pii/S0301932216307327}{{A
  multiscale DEM-VOF method for the simulation of three-phase flows}},
  International Journal of Multiphase Flow\href
  {http://dx.doi.org/10.1016/j.ijmultiphaseflow.2017.10.008}
  {\path{doi:10.1016/j.ijmultiphaseflow.2017.10.008}}.
\newline\urlprefix\url{http://www.sciencedirect.com/science/article/pii/S0301932216307327}

\bibitem{pozzettiicnaam2016}
G.~Pozzetti, B.~Peters, \href{http://orbilu.uni.lu/handle/10993/28863}{{On the
  choice of a phase interchange strategy for a multiscale DEM-VOF Method}}, AIP
  Conference Proceedings 1863.
\newline\urlprefix\url{http://orbilu.uni.lu/handle/10993/28863}

\bibitem{pozzettipowdermet}
G.~Pozzetti, B.~Peters, \href{http://hdl.handle.net/10993/31360}{{Evaluating
  Erosion Patterns in an abrasive water jet cutting using XDEM}}, Advances in
  Powder Metallurgy \& Particulate Materials (2017) 191--205.
\newline\urlprefix\url{http://hdl.handle.net/10993/31360}

\bibitem{peters2017Flow10993-31734}
B.~Peters, G.~Pozzetti, \href{http://hdl.handle.net/10993/31734}{Flow
  characteristics of metallic powder grains for additive manufacturing}, EPJ
  Web of Conferences 13001 (2017) 140.
\newline\urlprefix\url{http://hdl.handle.net/10993/31734}

\bibitem{KLEEFSMAN}
K.~Kleefsman, G.~Fekken, A.~Veldman, B.~Iwanowski, B.~Buchner,
  \href{http://www.sciencedirect.com/science/article/pii/S0021999104005170}{A
  volume-of-fluid based simulation method for wave impact problems}, Journal of
  Computational Physics 206~(1) (2005) 363 -- 393.
\newblock \href {http://dx.doi.org/https://doi.org/10.1016/j.jcp.2004.12.007}
  {\path{doi:https://doi.org/10.1016/j.jcp.2004.12.007}}.
\newline\urlprefix\url{http://www.sciencedirect.com/science/article/pii/S0021999104005170}

\bibitem{POPE2000}
S.~Pope, {Turbulent Flows}, Cambridge Press.

\bibitem{toutant2008877}
A.~Toutant, E.~Labourasse, O.~Lebaigue, O.~Simonin,
  \href{http://www.sciencedirect.com/science/article/pii/S004579300700165X}{{\{DNS\}
  of the interaction between a deformable buoyant bubble and a spatially
  decaying turbulence: A priori tests for \{LES\} two-phase flow modelling}},
  Computers \& Fluids 37~(7) (2008) 877--886, special Issue of the ``Turbulence
  and Interaction-TI2006'' Conference.
\newblock \href {http://dx.doi.org/10.1016/j.compfluid.2007.03.019}
  {\path{doi:10.1016/j.compfluid.2007.03.019}}.
\newline\urlprefix\url{http://www.sciencedirect.com/science/article/pii/S004579300700165X}

\bibitem{:/content/aip/journal/pof2/11/10/10.1063/1.870162}
V.~Armenio, U.~Piomelli, V.~Fiorotto,
  \href{http://scitation.aip.org/content/aip/journal/pof2/11/10/10.1063/1.870162}{{Effect
  of the subgrid scales on particle motion}}, Physics of Fluids 11~(10) (1999)
  3030--3042.
\newblock \href {http://dx.doi.org/10.1063/1.870162}
  {\path{doi:10.1063/1.870162}}.
\newline\urlprefix\url{http://scitation.aip.org/content/aip/journal/pof2/11/10/10.1063/1.870162}

\bibitem{:/content/aip/journal/pof2/8/5/10.1063/1.868911}
Q.~Wang, K.~D. Squires,
  \href{http://scitation.aip.org/content/aip/journal/pof2/8/5/10.1063/1.868911}{{Large
  eddy simulation of particle-laden turbulent channel flow}}, Physics of Fluids
  8~(5) (1996) 1207--1223.
\newblock \href {http://dx.doi.org/10.1063/1.868911}
  {\path{doi:10.1063/1.868911}}.
\newline\urlprefix\url{http://scitation.aip.org/content/aip/journal/pof2/8/5/10.1063/1.868911}

\bibitem{Zhou20044193}
H.~Zhou, G.~Flamant, D.~Gauthier,
  \href{http://www.sciencedirect.com/science/article/pii/S0009250904002933}{{DEM-LES
  of coal combustion in a bubbling fluidized bed. Part I: gas-particle
  turbulent flow structure}}, Chemical Engineering Science 59~(20) (2004)
  4193--4203.
\newblock \href {http://dx.doi.org/10.1016/j.ces.2004.01.069}
  {\path{doi:10.1016/j.ces.2004.01.069}}.
\newline\urlprefix\url{http://www.sciencedirect.com/science/article/pii/S0009250904002933}

\bibitem{Burgener}
T.~Burgener, D.~Kadau, H.~J. Herrmann, {Simulation of particle mixing in
  turbulent channel flow due to intrinsic fluid velocity fluctuation}, Physical
  Review E 83.

\end{thebibliography}

\end{document}